# Asteroids and Comets

Y. R. Fernández[a], J.-Y. Li[b], E. S. Howell[c], L. M. Woodney[d]

[a] Department of Physics, University of Central Florida, 4000 Central Florida Blvd., Orlando, FL 32816-2385 U.S.A., yan@ucf.edu.
[b] Planetary Science Institute, 1700 E. Fort Lowell Rd., Suite 106, Tucson, AZ 85719 U.S.A.
[c] Arecibo Observatory/USRA, HC 3 Box 53995, Arecibo, PR 00612-8346 U.S.A.
[d] Department of Physics, California State University – San Bernardino, 5500 University Pkwy San Bernardino, CA 92407-2397 U.S.A.

**Abstract**

Asteroids and comets are remnants from the era of Solar System formation over 4.5 billion years ago, and therefore allow us to address two fundamental questions in astronomy: what was the nature of our protoplanetary disk, and how did the process of planetary accretion occur? The objects we see today have suffered many geophysically-relevant processes in the intervening eons that have altered their surfaces, interiors, and compositions. In this chapter we review our understanding of the origins and evolution of these bodies, discuss the wealth of science returned from spacecraft missions, and motivate important questions to be addressed in the future.

# 1. Introduction

## 1.1. The Basics

The importance of small bodies – asteroids, comets, and small satellites -- to the overall picture of planetary formation and planetary evolution has been recognized for decades (see reviews by e.g. Festou et al. 2004 and Bottke et al. 2002b). Only since the 1990s, however, have we learned enough about the small bodies and about solar systems in general to start to make detailed connections. Our advances have come mainly thanks to new opportunities in ground- and space-based telescopic observation of both our own Solar System and of other systems. Additionally, the multitude of spacecraft that have now visited a variety of small bodies has provided a wealth of detailed context. The 1990s, 2000s, and 2010s have brought incredible insight into the nature of the small bodies.

In this chapter we will give a review of recent developments in our astronomical understanding of small bodies. To keep the discussion manageable, note that we focus here mainly on comets and asteroids that are not in orbit around major planets; i.e., we are excluding those small satellites that are likely captured comets and asteroids. Such satellites can inform and have informed our understanding of the general comet and asteroid population (see reviews by e.g. Bell et al. 1993, Jewitt and Haghighipour 2007, Turrini et al. 2009).

With this restriction, note that when we refer to "small bodies" in this chapter, we will mean just comets and asteroids. In this first section, we will discuss the historical and current definitions of small bodies, and motivate their study by describing their scientific importance. In section 2, we will describe the bodies' origins, formation scenarios, interaction with planets and with each other, and some of their dynamical processes. In section 3, we discuss the nature of their surfaces, while in section 4 we cover their interiors. Section 5 treats the composition of small bodies, and we finish in section 6 with a summary of some of the big questions in and future directions of small-body research.

## 1.2. Definitions

Comets have been noted in recorded history for millennia, and asteroids have been known for over two hundred years. For most of their observational history, the two groups of small bodies were thought to be clearly separate. More specifically, observational differences were assumed to translate directly to differences in the fundamental nature of the objects. Asteroids were point-like objects in the sky, mostly physically located between Mars and Jupiter in the Asteroid Belt (a.k.a. the Main Asteroid Belt or just the main belt). Comets were extended-sources -- showing comae and tails that are generally much brighter than the solid-body source itself -- and were either on parabolic or highly-elliptical orbits that took them beyond the planetary region of the Solar System, or on short-period orbits that barely (if at all) went past Jupiter. Even the names given to the bodies revealed this dichotomy: asteroids were "minor planets," basically just smaller, fainter versions of the other planets that were seen to wander across the sky. Comets were entirely different beasts, historically not even recognized as astronomical phenomena (as opposed to atmospheric) until Tycho Brahe's parallax studies in the 16th Century. Even the physical explanation for why comets looked extended -- i.e. that they had dust -- wasn't established until the 19th Century by e.g. Bessel (1836) and Bredichin (1877). The early 20th Century saw further characterization of the composition of the two groups, which further reinforced the apparent differences. Cometary spectra showed emission lines from gas species on top of a continuum spectrum, while those of asteroids sometimes showed broad absorption features in the continuum.

One of the major advances since the early 1990s is the realization that comets and asteroids are in fact not clearly separate classes of objects but instead part of a continuum of composition (e.g. Weissman et al. 1989, 2002). Comets are relatively icy and asteroids relatively rocky, but there are many objects that are now recognized to populate everywhere in between. For example: objects in the trans-Neptunian region – nominally called asteroids because of their lack of activity -- are icy and would show cometary activity if they were only closer to the Sun (e.g. Barucci et al. 2002, Prialnik et al. 2008); objects that have orbits that are clearly cometary in nature are seen to be inactive, suggesting that their volatile component is deeply buried or now absent (e.g. Ishiguro et al. 2011); and sustained activity that has been seen to develop from objects in the Asteroid Belt between Mars and Jupiter appears to be driven by volatiles, indicating that ice must still exist somewhere on such objects that have been in the same thermal environment for billions of years (e.g. Hsieh et al. 2010).

Hence, while the terms "asteroid" and "comet" are still commonly used in the small-bodies community, there is broad recognition that these terms no longer necessarily act as shorthand descriptors for their true physical nature. In this chapter we will continue to follow common usage when talking about specific objects that are referred to as comets or asteroids, but it is important to recognize that such terms can occasionally be misleading.

For this chapter, we will limit ourselves to a discussion of objects within a range of diameters. At the low end, we will not be talking about dust in and of itself but only dust as it relates to macroscopic comets and asteroids. In this sense our lower limit to the diameter is about 0.1 m, although in practice the lower limit is a few meters since almost no small bodies on heliocentric orbits have been observed that are smaller than that. At the upper end, we will limit ourselves to objects that have not pulled themselves into a roughly-spherical or spheroidal shape as a result of self-gravity. This is consistent with the IAU definitions of planets and dwarf planets (IAU 2006); we exclude these bodies from consideration here. We will, however, discuss (4) Vesta as it pertains to asteroid phenomena, though it arguably could be called a dwarf planet and has many terrestrial-planet like characteristics (e.g. Keil 2002, Zuber et al. 2011).

Finally, a discussion of terminology is in order. Among comets, short- and long-period comets (SPCs and LPCs) are those that have orbital periods less than and longer than 200 years, respectively. Short-period comets include those of the "Jupiter-family" (JFCs) and of the "Halley-type" (HTCs), the former referring to comets whose orbits tend to be strongly influenced by Jupiter, have aphelia near Jupiter's orbit, are mainly in the ecliptic plane, and generally have orbital periods of about 6 years. Halley-type comets are now thought to be dynamically evolved counterparts to the long-period comets; they can have any inclination. This terminology is the traditional one; Levison (1996) discusses comet classification and his newer system is in use by some comet scientists. That convention divides the population into "ecliptic comets" (ECs) and "nearly-isotropic comets" (NICs). ECs include the Jupiter-family but also a few other inner-Solar System comets that have been decoupled from Jupiter. Furthermore ECs can include the active "Centaurs," which are objects orbiting among the outer planets beyond 5 AU and are in transition from the trans-Neptunian region to the inner Solar System. NICs include the HTCs and LPCs and so dispenses with the arbitrary 200-year boundary.

Among asteroids, most of the over 380,000 numbered asteroids known reside in the Asteroid Belt between Mars and Jupiter. Asteroids are assigned permanent numbers when their orbits are sufficiently well determined to be identified decades in the future without ambiguity. An additional 260,000 unnumbered asteroids are currently catalogued, and may become numbered as new observations are made, and their orbits are better determined. Hundreds more asteroids are discovered in a typical month. Some objects are Mars-crossers, coming closer to the Sun than Mars, and some objects are Earth-crossers, coming closer to the Sun than Earth. Some asteroids cross the orbits of Venus and Mercury as well. A near-Earth asteroid (NEA) is any object with a perihelion distance less than 1.3 AU, and the NEAs are broken down into "Amors," "Apollos," "Atens," and "Atiras". Amors are not Earth-crossers, staying outside Earth's orbit; Apollos and Atens do cross Earth's orbit, with the distinction being whether the orbital semimajor axis is larger or small than 1 AU; Atiras also do not cross Earth's orbit, staying entirely interior to it. While these definitions are still in use, it is important to note that many of these distinctions are no longer meaningful, as we now know that objects can easily move from one dynamical group to another due to both gravitational perturbations and non-gravitational forces.

A traditional dynamical distinction between comets and asteroids is the Tisserand parameter ($T_J$). In a Solar System consisting of just the Sun, Jupiter on a circular orbit, and the small body (called the "restricted three-body problem"), this parameter is nearly constant (i.e. an "integral of the motion" in classical mechanics parlance). In the real Solar System, it is still nearly constant and useful since in general asteroids have $T_J > 3$ and comets have $T_J < 3$. This comes from the fact that short-period comets tend to be dynamically coupled to Jupiter; the Tisserand parameter can be related to the encounter velocity of a small-body with Jupiter. However there are many asteroids with $T_J < 3$ and several comets with $T_J > 3$, so the distinction is not exact.

Some of the classification distinctions can be seen in Figure 1, which shows a plot of the semimajor axis versus eccentricity of many of the asteroids in the inner Solar System. In particular, the curves "EC," "MC," and "JC" indicate the curves that mark which asteroids can cross the orbits of Earth, Mars, and Jupiter, respectively. We will return to this figure in later sections when we discuss various dynamical properties of the small bodies.

Note that in this chapter we will give the complete designation when referring to an asteroid or comet for the first time, but references thereafter will use only the name (unless there is potential cause for confusion).

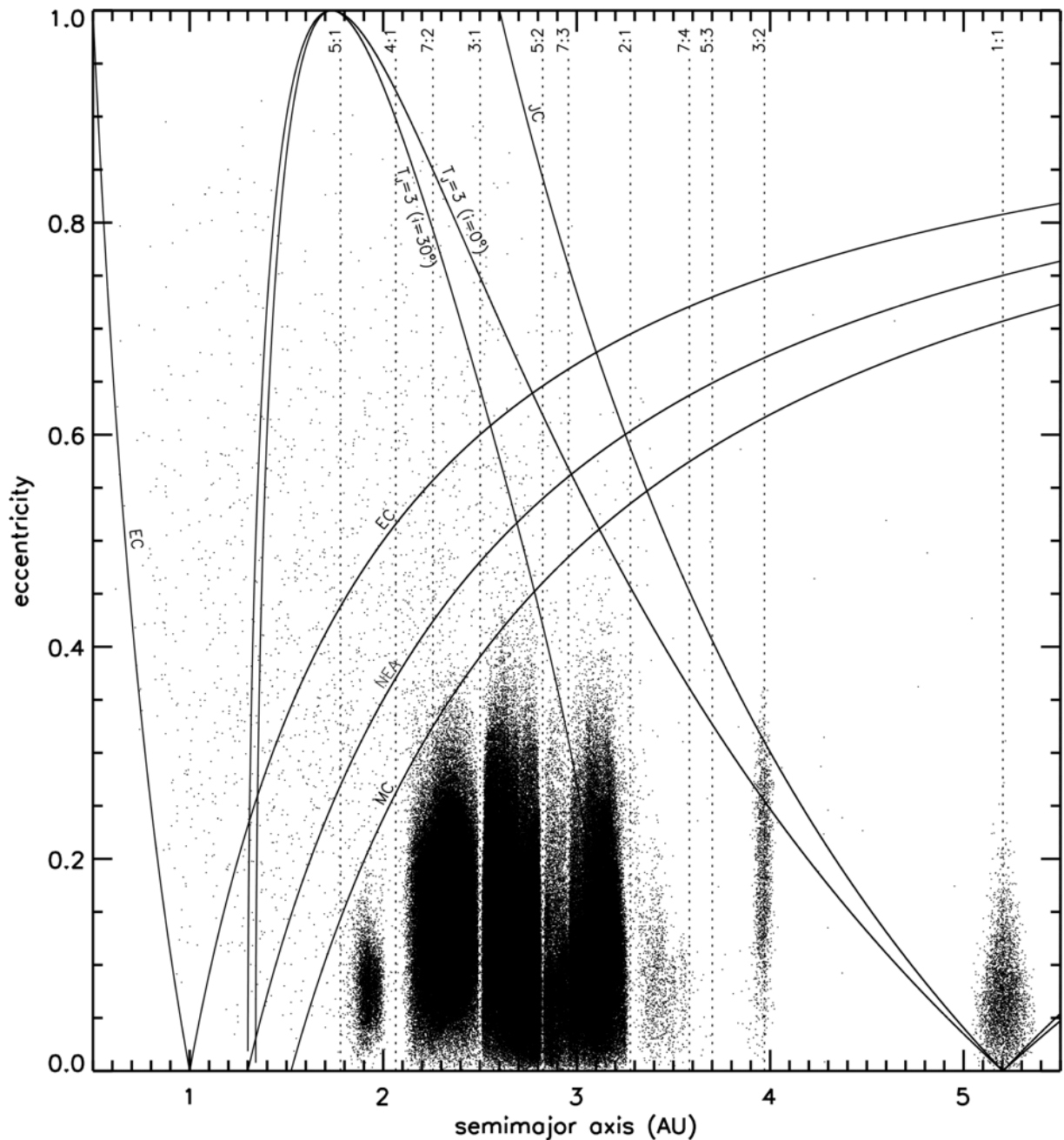

Figure 1. Orbital elements of the over 363000 numbered asteroids (at time of writing), as represented in a-e space, where a is the semimajor axis in AU and e is the orbital eccentricity. Note that 1 AU is currently defined as 149,597,870.7 km exactly and is very nearly identical to the average Sun-Earth distance. There is obvious structure in the Asteroid Belt, and there are regions of orbital element space that are favored. "NEA" indicates the locus where perihelion distance q is 1.3 AU, the defined boundary for near-Earth objects. "EC," "MC," and "JC" indicate the Earth-, Mars- and Jupiter-crossing boundaries, respectively. The two curves marked "$T_J=3$" show where the Tisserand parameter is 3.0 for two possible values of the orbital inclination. Vertical dotted lines indicate mean-motion resonances with Jupiter; the resonance is listed at the top of each line. Note that Jupiter's semimajor axis is 5.20 AU, Mars's is 1.52 AU.

1.3. Dynamical Relationships

Some of the subpopulations of the Solar System's small bodies are dynamically linked. In particular, the JFCs are linked to several other groups. The "scattered disk" of the Kuiper Belt – containing objects beyond Neptune on high-eccentricity orbits – is thought to be the original source of these comets, and the objects progress in a dynamical cascade from beyond Neptune and through the outer-planet region, essentially being handed off from the control of Neptune, to Uranus, to Saturn, and finally to Jupiter (Duncan et al. 2004). Jupiter-controlled comets have aphelia near 5 AU, and these are the Jupiter-family comets we see today. As mentioned above, many members of the transitional population, the Centaurs, have been discovered inhabiting the outer-planet region, and some of the Centaurs show cometary activity. The time scale for an object to exist among the giant planets is only a few million years (Tiscareno and Malhotra 2003, Horner et al. 2004).

Jupiter-family comets share some of their orbital parameter space with NEAs. Current work suggests that up to a few percent of the NEA population could be Jupiter-family comets that have simply lost their surface volatiles (e.g. Bottke et al. 2002a, Fernandez et al. 2005, DeMeo and Binzel 2008).

The connection between Jupiter's Trojan asteroids – objects in 1:1 resonance, but on average about 60 degrees ahead or behind Jupiter in orbital longitude – and the Jupiter-family comets is uncertain. It is possible that there is some leakage from the Trojan swarms into the comet population (Marzari and Scholl 2002), however no one has ever reported an observation of a Trojan having a comet-like coma or tail. Nor is it clear what would be an observational difference between a Jupiter-family comet that originated in the Trojan swarm vs. one that originated in the Kuiper Belt. One possible distinction is that Trojans might be expected to have a lower D/H ratio than objects that formed in the outer Solar System, depending on how much compositional radial mixing has occurred; we discuss D/H ratios in more detail in section 5.

The existence of a link between the Asteroid Belt and the NEAs is well established, although the details of which dynamical process moves which asteroids is still under debate (e.g. Bottke et al. 2002a).

1.4. Relevance to Geology

The most obvious geophysical link between Earth and the small bodies is cratering. Many craters are visible around Earth today, and the only reason we do not see more of them is because Earth's surface is in constant change. Most – though not all! – asteroids have numerous craters, and indeed most of the larger asteroid surfaces are saturated with craters. The study of the abundant craters on other Solar System objects can bring insight into the impact process and how an impact's effects depend on the composition, density, and other bulk parameters of both the impactor and of the target body.

Erosive processes that garden and turnover surfaces occur on most if not all small bodies. The main difference is that small bodies are airless bodies – none has an atmosphere in the same sense that Mars, Venus, Earth (or even Mercury and the Moon for that matter) have atmospheres. The only thing that comes close is the gravitationally-bound coma that sometimes appears around the largest cometary nuclei (e.g. Meech et al. 1997). But in general, no asteroid that we consider here has an atmosphere, and comets only have exospheres. The building up or tearing down of surface structures comes mainly from external effects (e.g. micrometeorite bombardment, impact cratering), though also perhaps from internal processes in the case of comets.

As will be discussed in later sections, the surfaces and interiors of small bodies can be quite different from those of Earth and the other terrestrial planets. Densities, porosities, compositions, collision histories, and surface processes can all be different from what we know on Earth. Yet all of these properties are crucial for understanding the geophysics of small bodies. Complicating our study of these objects is the fact that detailed geomorphology can only be obtained through the use of spacecraft or radar echoes, meaning that many comets and asteroids simply have not been studied in sufficient detail for there to be a true geophysical analysis. One important point to note is that spacecraft flybys of comets and asteroids have made it abundantly clear that we have not yet seen the full diversity of small bodies close up.

1.5. Relevance to Meteorites

We know through laboratory or *in situ* sample analysis that only a small fraction of meteorites originate from the Moon and Mars; the vast majority are from asteroids. The meteorites thus give us important insights into the composition of the solar nebula from which the planets formed. A comprehensive review of meteorite studies is beyond the scope of this chapter, and indeed is the subject of entire volumes (e.g. Lauretta and McSween 2006). We rely heavily on the studies of meteorites to give us an understanding of the composition of asteroids, while at the same time recognizing that the sampling of the Asteroid Belt may be extremely non-uniform. Attempts to link telescope spectra of asteroids with laboratory measurements of meteorites have shown that there is not a simple one-to-one correspondence, and many asteroids do not seem to be represented in the meteorites we have. On the other hand, some links are considered very secure, such as the spectra of howardites, diogenites and eucrites matching the spectrum of asteroid Vesta; these meteorites are considered to be samples of this asteroid (Binzel and Xu 1993, Drake 2001). The identifiable spectrum of Vesta has let us identify collisional fragments of previous impacts in the Asteroid Belt and among the near-Earth asteroids (Binzel et al. 2004). However, other connections of specific asteroids to specific meteorites are not so clear.

Several asteroid taxonomies have been suggested to better understand the relationships between different asteroid spectral types, meteorites, and compositions. Tholen and Barucci (1989) give a thorough review of the most commonly used taxonomic system based on photometry in the visible range (0.35-1.0 microns). Bus and Binzel (2002) used higher resolution spectra over a slightly narrower spectral range (0.45-0.95 microns) to extend the Tholen taxonomy to include more specific spectral features. DeMeo et al. (2009) extended the spectral range from visible to near-infrared (0.4-2.5 microns) and this Bus-DeMeo taxonomy is now the most widely used. Three groups of spectral types, S-, C-, and X-complex, encompass the majority of the asteroids. Each complex includes several related taxonomic classes designated by one or two letters; the specific taxonomic class that an asteroid is assigned depends on the details of the spectral shape. Figure 2 summarizes the spectra of all the taxonomic classes in the Bus-DeMeo system; it shows both the average spectrum of each taxonomic type as well as some actual asteroid spectra. This figure shows the diversity of asteroid reflectances in visible and near-infrared wavelengths among all the taxonomic types. The S-complex asteroids are primarily pyroxene and olivine, and in some cases metallic iron. Ordinary chondrites and other stony but largely undifferentiated meteorites are thought to be samples of S-complex asteroids. The C-complex asteroids are mostly low albedo, flat featureless spectra except for a broad absorption band at 0.7 microns indicative of oxidized iron (Vilas and Gaffey 1989) defining the Ch class. Various carbonaceous chondrite meteorites are thought to be samples of these types of asteroids, although the spectral matches are not perfect. The X-complex asteroids have been linked to many different meteorite groups such as aubrites, enstatite chondrites and achondrites. Links to iron-nickel meteorites are problematic; originally these were thought to come from so-called (in the Tholen system) M-type asteroids, but now it is unclear if any of the large asteroids can be primarily iron-nickel (see e.g. Shepard et al. 2010, 2015, and references therein for detailed discussion). A few additional taxonomic classes such as A, V, and R are unusual, or match specific asteroids. For example, the V-types are most likely related to Vesta, as discussed above. Finally, the D-type asteroids have featureless and very red sloped spectra, and match the outer belt asteroids and Jupiter Trojans, but do not match any meteorites very well, except perhaps Tagish Lake (Hiroi et al. 2001). Interestingly, both of Mars's moons are D-type objects; this is unusual since D-types are not so abundant in the inner part of the Solar System. For convenience in this chapter, and unless otherwise noted, we will refer to broad spectral groups of asteroids using the Bus-DeMeo taxonomy.

The connection between meteorites and comets is very sparse; no meteorite currently in hand is considered to have definitely come from a comet. Cometary meteorites are expected to be rare simply due to high entry speed (Campins and Swindle 1998), although it is possible that some relatively-slow meteor showers may provide meteorites that survive (Brown et al. 2013).

1.6. Relevance to Astronomy

The asteroids and comets we see today are the survivors of the planetary formation process that proceeded during the start of our Solar System. The current dominant hypothesis is that the nebular cloud that formed the Sun also had a protoplanetary disk, from which the first planetesimals, and then planets, formed. The details of what the protoplanetary disk was like, and the actual mechanical process that formed the planets, are both major questions in astronomy that are the subject of significant research. Thus the study of comets and asteroids is tied to some of the "holy grails" of astronomy. The role of comets and asteroids in the history of our Solar System's planets has become even more important since the late 1990s and 2000s with the discovery of many other planetary systems in our Galaxy.

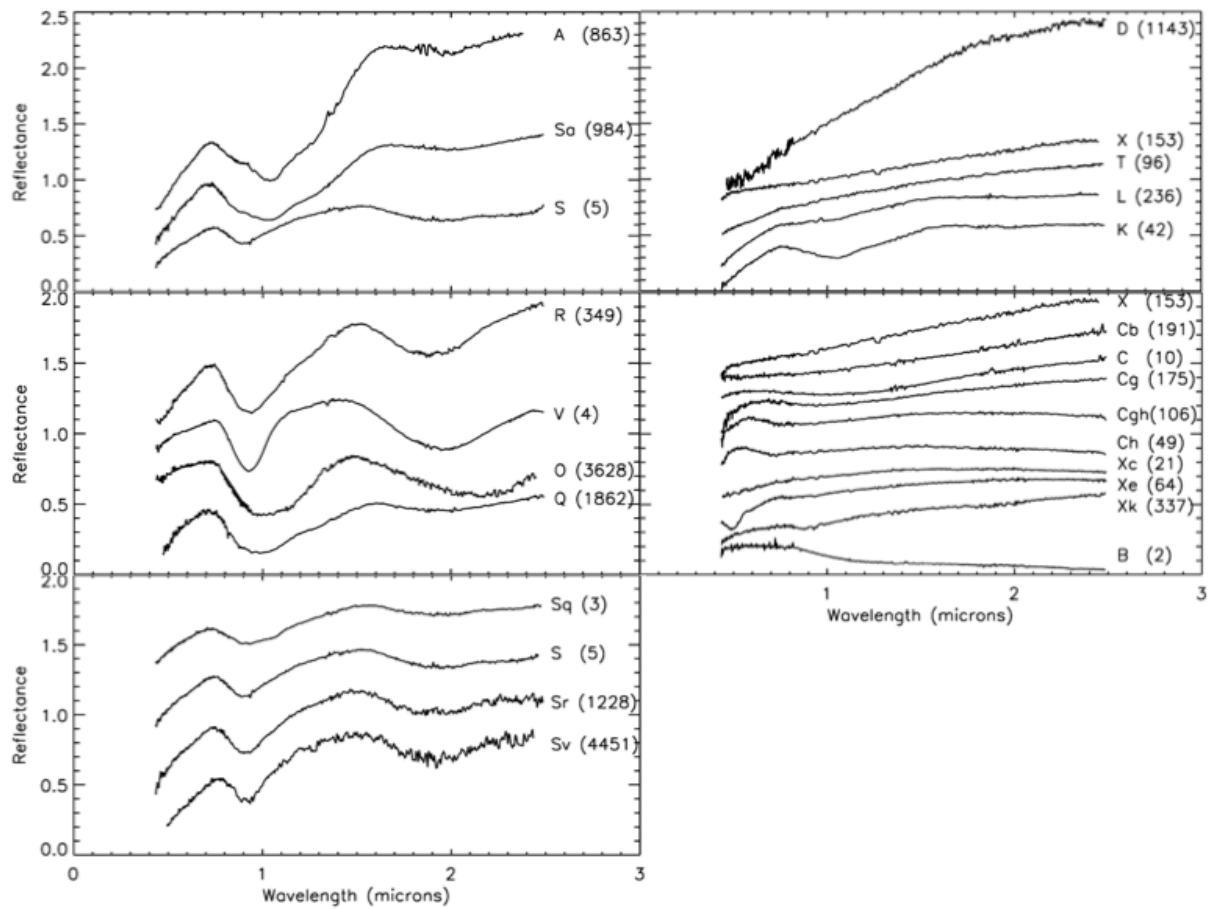

Figure 2. The range of asteroid relative-reflectances as seen in the red and near-infrared. These panels are from the work by DeMeo et al. (2009) and are their Figures 5, 6, 7, 11, and 14 (reprinted from Icarus with permission from Elsevier). There are 26 spectra here, representing the 24 taxonomic classes for the near-infrared devised by DeMeo et al. ("S" and "X" are shown twice); the taxonomic type is the letter at the right edge of each spectrum. The number after the type is the numbered asteroid that represents the prototype. Note that for clarity spectra within a panel have been shifted vertically; reflectance is set at unity for a wavelength of 0.55 µm. Asteroids on the left all show the classic two-band ("Band I" and "Band II") absorption features near 1 and 2 microns due to pyroxene and olivine. These types include the objects in the S-complex (S, Sa, Sq, Sr, Sv). Asteroids on the right lack these absorptions, including objects in the C-complex (C, Cb, Cg, Cgh, Ch, B) and X-complex (X, Xc, Xe, Xk); indeed many of the asteroids on the right lack distinguishing diagnostic absorption features, making determining their composition problematic.

One of the interesting results from the several hundred exoplanets that have been discovered is the fact that the architecture of our own Solar System seems to be relatively unusual (e.g. Udry and Santos 2007, Borucki et al. 2011). Our planetary system's layout -- with small terrestrial planets close to the home star, and large gas- and ice-giant planets farther out -- has not yet been seen in very many other systems. While observational biases and limitations have made it difficult to ascertain just how many terrestrial planets there are elsewhere, it is clear that there are a great many ways for giant planets to end up being distributed within a planetary system. This reality has motivated dynamicists to investigate possible modes of planetary migration. Some of this migration is influenced by the population of small bodies; essentially, the act of perturbing small bodies may transfer sufficient angular momentum (by the law of conservation) to change a planet's orbit. Thus the population of small bodies in the protoplanetary disk is a fundamental part of understanding the development of the planets' final layout in the Solar System. Furthermore, we are discovering other systems around older stars where clearly there are asteroid or comet belts with masses much larger than those we currently have in our Solar System (e.g. Lisse et al. 2007, Lawler et al. 2009). These disks of material are not protoplanetary, since the parent stars are usually hundreds of millions or billions of years old. Two questions stemming from these observations are why some systems have such massive disks, and whether our own Solar System went through a similar stage in the past.

Another area of relevance that pertains to the Solar System itself has to do with water and organic molecules. The existence and distribution of these substances are, of course, crucial things to understand since life on Earth (and so far, life as we know it) requires such compounds to exist. Both comets and asteroids contain water and organics in various forms and concentrations. Current thinking is that it is unlikely that Earth formed with its current water content, since it was likely too hot; most of Earth's water presumably was brought to it via small-body impacts (e.g. Drake and Campins 2006, Mottl et al. 2007). The original organic component of Earth was likewise probably miniscule, with small bodies perhaps providing the bulk of the organics (Anders 1989).

Lastly, small bodies have relevance today since they represent an impact hazard to Earth. The extreme case is the K-Pg mass extinction 65 million years ago, after which a large fraction of Earth's species became extinct in a geologically short time. Presumably an important factor in instigating this extinction was the impact of an approximately 10-km wide object striking what is now the Yucatan peninsula (Alvarez et al. 1980). Many other impacts have happened since that event, causing varying degrees of damage to local and regional ecosystems and in some cases whole biomes. A very recent and well-observed event was the arrival of the ~15-m wide Chelyabinsk bolide in February 2013. That asteroid, which was not known before the encounter, was possibly the largest object to strike Earth's atmosphere since the 1908 Tunguska event (Brown 2013). Somewhere out there is the next devastating small body that will eventually collide with us; the question is, how big is it, and is it coming in a few years, centuries, millennia, or longer? Unlike the dinosaurs, we can actually attempt to answer this question and try to guard against catastrophe.

**2. Origins**

2.1. Basics

The traditional, fundamental compositional distinction between comets and asteroids is the ice content. The zeroth-order explanation for that distinction involves the existence of a "snow line," a distance from the forming Sun within the protoplanetary disk beyond which water could condense. While the term "snow line" conjures up the idea of a sharp boundary in the protoplanetary disk, it appears (as mentioned in Section 1) that the reality is more complicated, with the disk having a trend of ice abundance that varied with time rather than a sharp fall-off (see e.g. the review by Encrenaz 2008). The most widely-accepted idea is that the comets and asteroids we see today are the remnants of the planetesimals that formed by accretion within the protoplanetary disk.

The specifics of the accretion process still provoke a major ongoing research effort, most strikingly epitomized by the so-called "meter barrier." While numerical modeling has been able to reproduce the accretion of solid-phase rock, metal, and ice grains into conglomerates up to submeter scales with the correct time scales, simulating the creation of larger bodies has been difficult. The main problem is that, in many simulations, a two-body collision between meter-size objects seems to result in a destructive collision rather than an accretion. One possibility is that the interaction between the meter-size boulders and turbulent gas in the disk can sufficiently affect the nature of the collisions so that km-scale bodies can actually be created (e.g. Johansen et al. 2007, Chiang and Youdin 2010).

There is evidence for a protoplanetary disk within which species aside from ice exhibited trends with distance from the Sun. The distribution of spectroscopic asteroid types is not randomly mixed, but instead follows a trend with heliocentric distance, which has been interpreted as a remnant of the solar nebula. This phenomenon has been known for decades (e.g. Gradie et al. 1989), but a recent demonstration of the situation is shown in Figure 3. Thousands of asteroids observed by the WISE spacecraft (Masiero et al. 2011) show that asteroid albedos (and similarly, taxonomic types and compositions) are not uniformly distributed in the Asteroid Belt; more primitive (lower albedo) objects dominate farther out, while more processed (higher albedo) objects dominate closer in. The correlation of spectral types with distance motivates the idea that the Asteroid Belt can be used to better understand the conditions in the solar nebula and the material that formed the terrestrial planets. However, discoveries of extra-solar planets and other planetary systems have greatly broadened our notions of what is "normal" and "typical". Also, increasing computation capabilities have allowed ever more realistic and detailed models of accretion, and are now challenging long-held ideas of what we thought we knew about Solar System formation. The Asteroid Belt and Kuiper Belt are important keys to constraining and testing these models.

The mass of the asteroids in the Asteroid Belt between Mars and Jupiter is about $6 \times 10^{-4}$ times the mass of Earth, extremely depleted compared to a protoplanetary disk extending smoothly from 1-5 AU and likely originally containing many Earth-masses of material (e.g. Morbidelli et al. 2012). The formation of Jupiter was long thought to be responsible for disrupting the formation of a planet in this region, and for scattering material around the Solar System. The asteroids remaining in the Asteroid Belt have dynamical signatures from Jupiter's gravitational effect, such as the Kirkwood gaps – evident as the gaps in semimajor axis as seen in Figure 1. These gaps are created by mean-motion resonances (MMRs) with Jupiter, where the orbital period of the asteroid is a small-integer-ratio of Jupiter's; they are denoted at the top of Figure 1. An MMR tends to enhance perturbations from Jupiter and so helps to kick the asteroid out of that region of orbital element space. The belt's inner edge is controlled partially by Mars. This is also seen in Figure 1; most of the asteroids now in the Asteroid Belt do not cross the Mars-crosser ("MC") line. The belt is also sculpted by the $\nu_6$ resonance – a secular resonance arising from when the asteroid's longitude of perihelion has a precession frequency matching that of Saturn (see e.g. Froeschlé and Morbidelli 1994). This resonance is most noticeable in a plot of inclination versus semimajor axis, as in Figure 3; the reason the asteroids look confined to low inclinations is because the $\nu_6$ resonance tends to kick out any asteroid with an inclination that is too high. The equivalent resonance with Jupiter, $\nu_5$, operates at higher inclinations, so in this case Saturn's influence is stronger than Jupiter's.

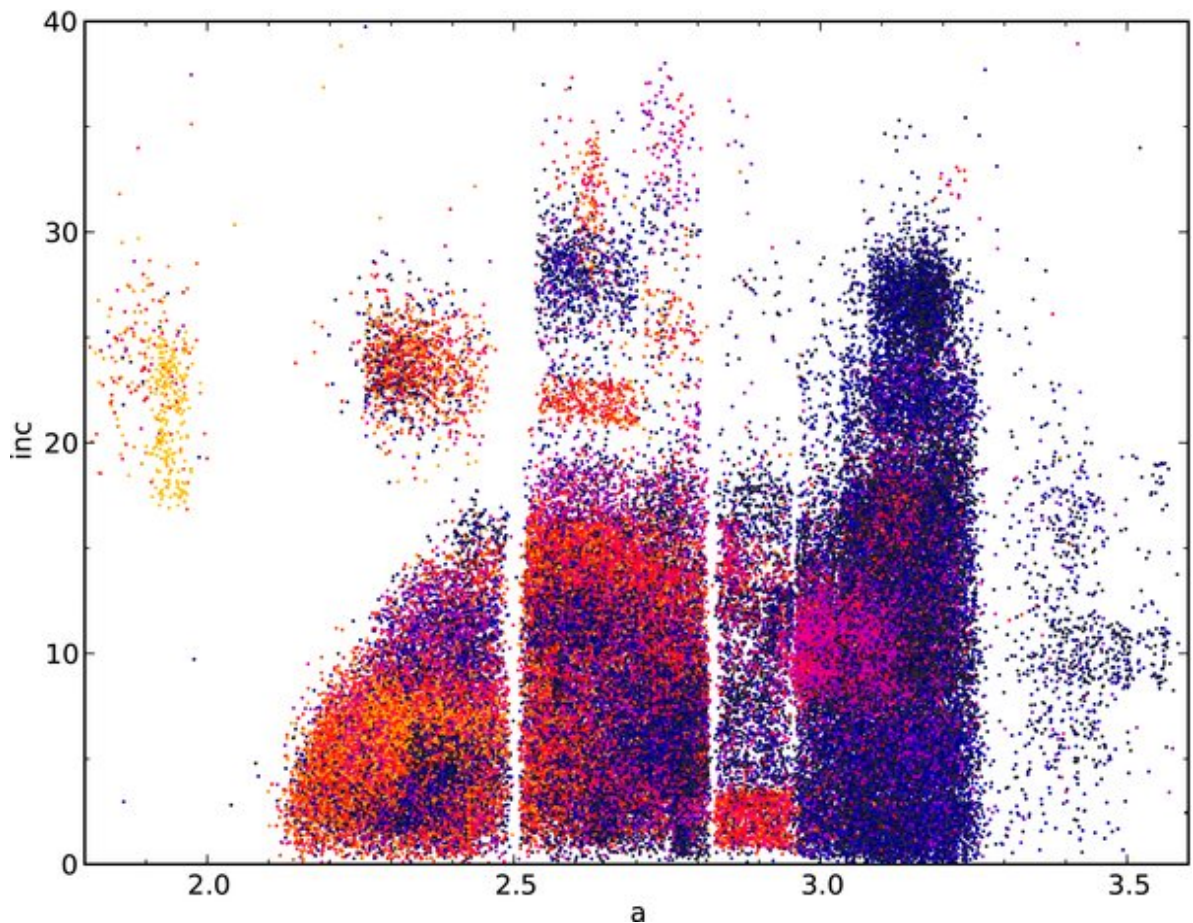

Figure 3. Scatter plot of main-belt asteroids in *a-i* space (where *i* is the orbital inclination) showing the albedos of the asteroids as derived by observations with the WISE spacecraft. This is Figure 12 from the work by Masiero et al. (2011; ©AAS, reproduced with permission). The color coding of the points gives the albedo, with black and grey as the lowest albedos (0.05 and below), increasing to blue, purple, pink, red, and orange, with yellow showing the highest albedos (above 0.35). There is a clear trend of albedo with distance from the Sun; the trend exists for composition as well.

2.2. Recent Models

Recent modeling efforts to reconcile formation timescales of the outer planets has led to consideration of large scale migration of the giant planets. The so-called Nice model (Tsiganis et al. 2005, Morbidelli et al. 2005, Gomes et al. 2005, Morbidelli 2010) – named after the city in France where many of the model's originators work – has come to be broadly accepted, although the details are still debated. Indeed the model, while explaining many observed orbital properties of small bodies in our Solar System, has only made limited observational predictions and could not account for the small mass of Mars (compared to Earth and Venus).

The basic premise of the model is that the outer planets (Jupiter, Saturn, Uranus, and Neptune) started out in orbits much closer to the Sun than they are now. There was also a disk of material – forerunners to today's comets – that was the remnant of the protoplanetary disk from which the four planets formed. Scattering of these small bodies (as mentioned in Section 1) caused the planets to migrate in semimajor axis, eccentricity, and inclination. Approximately ~800-900 million years after the planets formed, Saturn and Jupiter happened to fall into a 2:1 mean-motion resonance. This had a profound effect on the layout of the Solar System, strongly perturbing the remaining small bodies in the outer Solar System, as well as Uranus and Neptune themselves. There is even a possibility that Uranus and Neptune switched order; this falls out of some simulations and would nicely explain the long-standing mystery of why Neptune is more massive than the next planet closer in to the Sun. An example of the effects of the model are shown in Figure 4; small bodies in the vicinity are very rapidly scattered. The model can explain such other observables as the inclination distribution of the Jovian Trojan asteroids and the basic structure of the Kuiper Belt, and would be the cause of the Late Heavy Bombardment seen in the lunar cratering record, about 3.9 billion years ago.

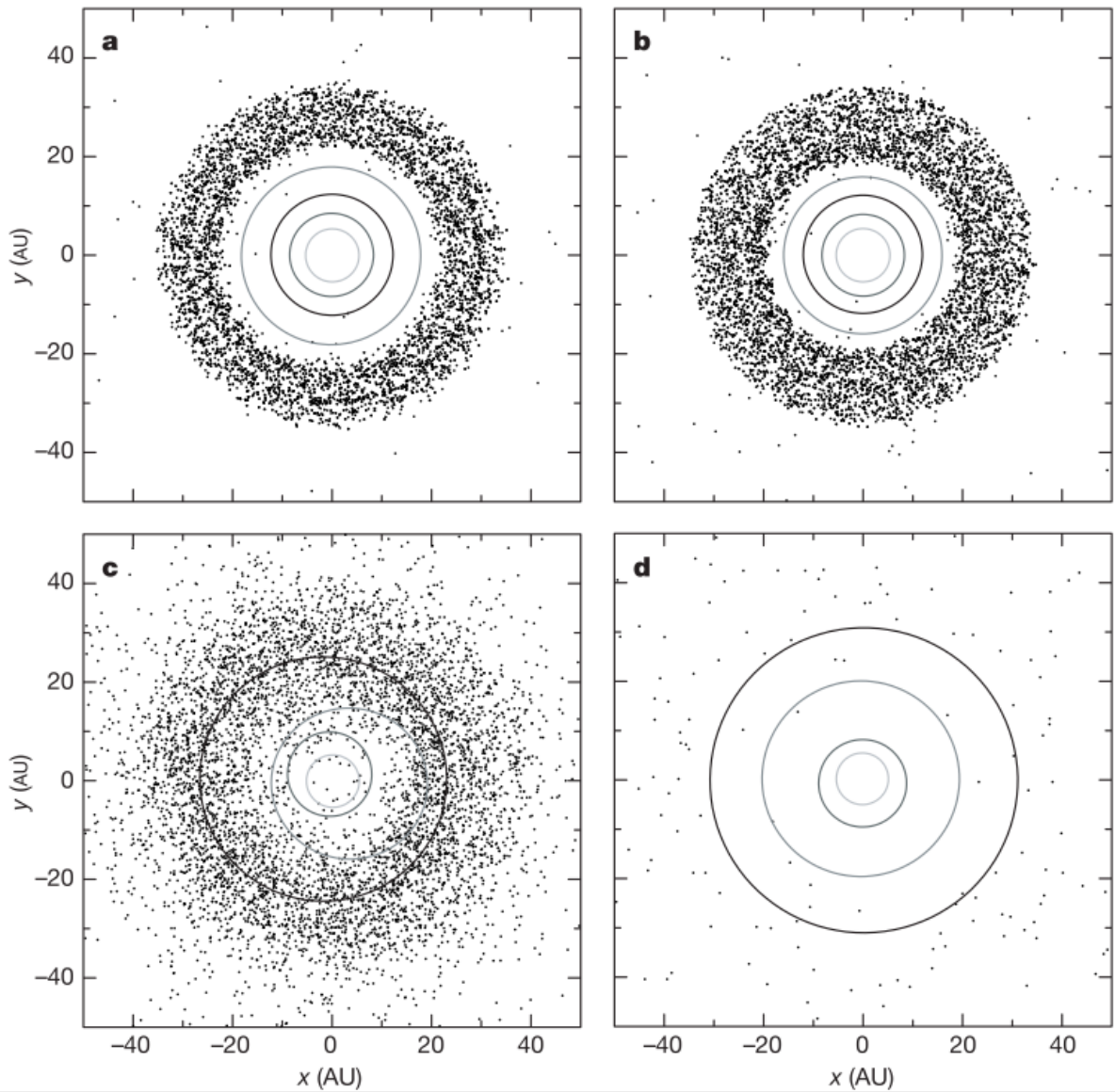

Figure 4. An example simulation of the Nice model; this is Figure 2 from the work of Gomes et al. (2005; reprinted by permission from Macmillan Publishers Ltd). The four panels show the development of the four outer planets plus thousands of massless test particles through the use of a numerical simulation. Panel a represents the Solar System 200 million years after the formation of the planets; panel b, 879 million years; panel c, 880 million years, and panel d, 1080 million years. Jupiter and Saturn enter the 2:1 mean-motion resonance at 880 million years, and the disk of test particles is quickly disrupted.

More recently, Walsh et al. (2011) have added an early, inward migration of Jupiter to 1.0 AU, which finally allows Mars to form with its small mass. However, this requires a complete re-working of our interpretations of the Asteroid Belt. The "Grand Tack," as this model has been called, is still being refined, but like the Nice model, it explains some long-standing mysteries in an appealing way. It too, may be overturned, but at the present it provides a useful framework for understanding the asteroids that is substantially different from before. Morbidelli et al. (2012) give a comprehensive review of the formation of the terrestrial planets, from nebular dust to present day bodies. The asteroids are an important constraint, as the Asteroid Belt as seen today is not consistent with early Jupiter migration in the gas disk (Walsh et al. 2012) or the later outer planet migration (jumping Jupiter) scenarios. Finding dynamical models that can match the current inclination and eccentricity distribution of the largest asteroids is surprisingly difficult, and serves as a sensitive measure of the dynamical evolution of the Solar System.

Morbidelli et al (2012) and references therein describe how the original material in the Asteroid Belt would have been disrupted by inward migration of Jupiter, leaving only the inner Asteroid Belt intact, and scattering some of these objects outward behind it as it moves. After Saturn moved into resonance with Jupiter, both began to migrate outward, and some of the scattered objects were scattered back into the Asteroid Belt region, along with many additional volatile-rich objects (C-type asteroids) that may have formed closer to 5 AU. This scenario solves the problem of having both high and low temperature materials side by side in the Asteroid Belt. However, it means that the compositional gradient with heliocentric distance is not primordial, and allows a larger amount of radial mixing than is usually considered probable. The problem of linking meteorites to their asteroid parent bodies also becomes nearly impossible, since the current position of the asteroids may not be at all related to where those bodies formed. However, the observations that asteroid families, collisional fragments of once intact parent bodies, are nearly all spectrally uniform, while many meteorites clearly show that differentiation has occurred, is no longer an impossible situation. Differentiated objects can form closer to the Sun, and be emplaced in the Asteroid Belt by Jupiter during migration. Undifferentiated objects form farther out, and are scattered inward by Jupiter. Both can now exist side by side, but neither is representative of material that condensed at 2-4 AU. We may have little hope to identify primordial material, if this accretion model is correct.

The topic of the origin of the Kuiper Belt, which is likely the original source of the Jupiter-family comets we see today, is discussed in detail in the chapter by McKinnon in this volume. Briefly, the orbital distribution of today's belt can be a useful constraint on the historical dynamical effects that sculpted it. For example, the slow outward migration of the outermost planet may have swept up some objects into mean motion resonances; Pluto, residing in Neptune's 3:2 resonance, is the most famous example. But the belt also holds a "scattered disk" of objects that were presumably scattered by Neptune into highly eccentric orbits. There is also a "hot" classical disk that maintains relatively low eccentricity but is dynamically "hot" in the sense of having higher inclination. In contrast, the "cold" classical disk has lower inclination objects. There seems to be some difference in origin of these various populations since there is some difference in their surface properties -- namely color (e.g. Sheppard 2012, Romanishin et al. 2010), which is for the most part the only measurable property available for the vast majority of known Kuiper Belt objects. As mentioned above, the Nice model would suggest that Saturn's 2:1 mean-motion resonance crossing disrupted the primordial Kuiper Belt, and it would imply that we may have a difficult time disentangling primordial and evolutionary effects.

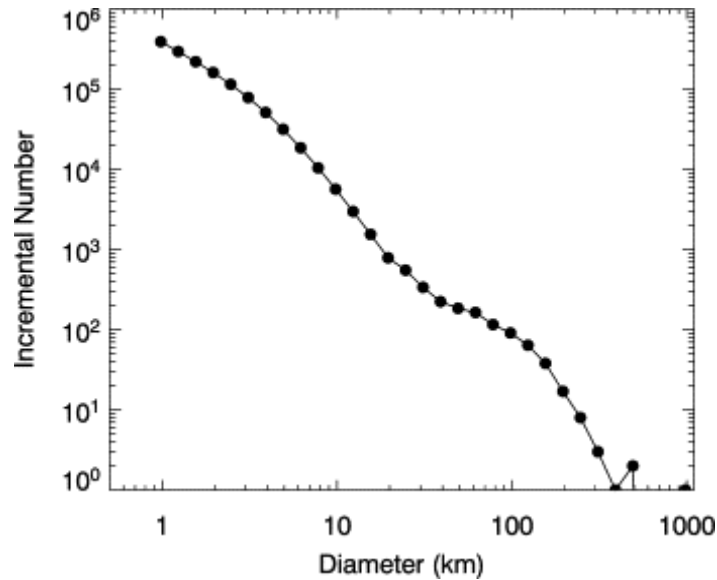

Figure 5. Main-belt asteroid size frequency distribution, as shown in Figure 1 from the work of Bottke et al. 2005 (reprinted from Icarus with permission from Elsevier). This distribution is computed from the absolute magnitude H distribution, transformed into diameter bins assuming an average geometric albedo $p_v$ of 0.092. For the fainter objects, those with H>12 (i.e. a diameter less than 18 km for the average albedo), the data are from the Sloan Digital Sky Survey. For the brighter objects, with H<12, the observational census is virtually complete and so the data are from the list of known asteroids.

2.3. Size Distributions

The size distribution of the asteroids has been measured to increasingly smaller sizes, but is still not linked to the dust distribution with any confidence. Models and observations have finally come together and agree that below about 10 km in diameter, all main belt asteroids are collisional fragments (e.g. Bottke et al. 2005). Asteroids bigger than 30-50 km are likely to be fractured from multiple collisions, but have lifetimes of 4.5 Gyr or more, and could be survivors of the formation of the planets. However if the source regions for the present day Asteroid Belt are as broad as 1—6 AU or more, these limits may change. There is some observational evidence for waviness in the cumulative size distribution – i.e., for a distribution that follows a power-law but with a bump at certain sizes (e.g. O'Brien and Greenberg 2005). The cumulative size distribution as shown in the review by Bottke et al. (2005) is shown in Figure 5. This waviness could be interpreted as being controlled by the strength-size relationship of the asteroids. However the existence of the waviness, and its location in the size distribution, is still under debate (e.g. Gladman et al. 2009).

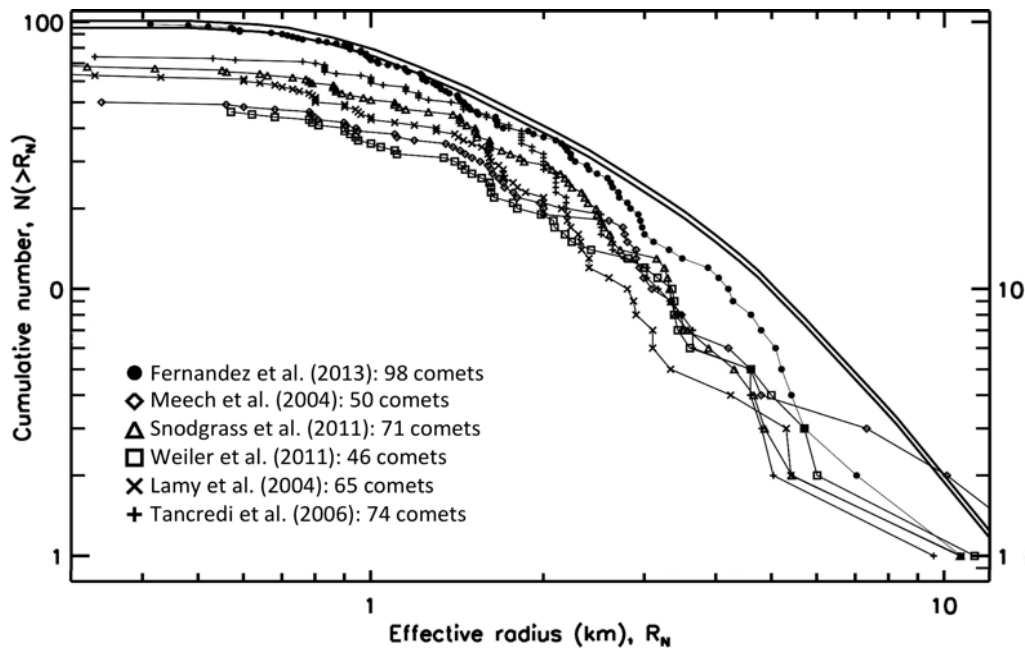

Figure 6. Cumulative size distributions of the Jupiter-family comets as reported by various workers. This plot is adapted from the work by Fernandez et al. (2013; reprinted from Icarus with permission from Elsevier). The legend indicates how many JFCs were used in each result. All results used primarily visible-wavelength measurements of nuclei, except for Fernandez et al., who used exclusively infrared measurements. There is broad agreement in the distributions. The double-solid lines indicate the intrinsic distribution model by Meech et al. (2004) that best reproduces their observed distribution; this model requires that there be a lack of sub-kilometer JFCs to account for the small number of such comets that have been discovered.

The current size distribution of comets is, compared to the asteroids, poorly known. Work on the Jupiter-family comets has been done by several teams (e.g. Meech et al. 2004, Fernandez et al. 2013), and suggests that there is currently a lack of sub-kilometer size comets. Some recent estimates of the distribution are shown in Figure 6. Note that there is no expectation that the current, observed distribution of comet sizes reveals the primordial distribution; the collisional environment in the Kuiper Belt is likely too violent (Farinella and Davis 1996). The observational situation is even worse for long-period comets. Using the current size distribution of comets to work backward and tease out the original size distribution is a very long-term project that will require having a much better understanding of collisions, of cometary mass loss, and of fragmentation.

## 3. Surfaces

### 3.1. Background

Twelve asteroids and five comets have been imaged at close range by spacecraft through flybys and/or rendezvous, returning a huge number of disk-resolved images of these small bodies. Table 1 lists all small-body encounters within $10^6$ km. Some views of these flyby targets are shown in Figures 7, 8, 9, and 10.

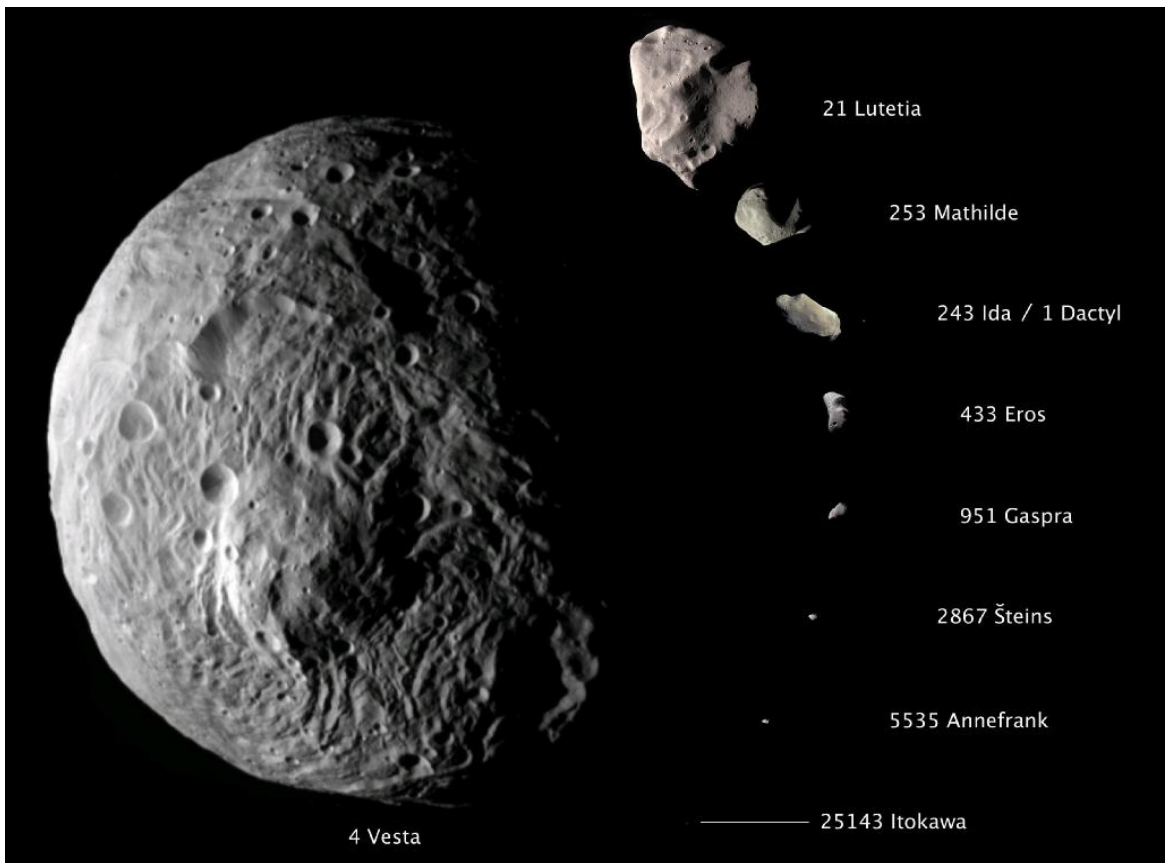

Figure 7. Montage of many of the asteroids that have had close spacecraft flybys, shown to scale. Note that the diameter of the largest object, Vesta, in this view is about 560 km, while that of the smallest object, Itokawa, is under 1 km. Courtesy NASA/JPL-Caltech/JAXA/ESA (Photojournal image PIA14316).

These planetary exploration missions have fundamentally changed Solar System small bodies from astronomical objects to geological objects. Our knowledge about the surface morphology and cratering characteristics has been greatly expanded as a result of those planetary missions. We will discuss the surface morphology of comets and asteroids separately; both kinds of objects are affected by cratering, but cometary surfaces are also changed on much shorter timescales by outgassing and erosion due to sublimation.

3.2. Surface Morphology on Asteroids

The first ever spacecraft flyby of an asteroid was in 1991, when the Galileo spacecraft visited (951) Gaspra while en route to Jupiter (Belton et al. 1992; Veverka et al. 1994). This flyby marks the start of observational studies of asteroid geomorphology. Since then, eleven more asteroids have been visited either with flybys or rendezvous missions (Table 1). Among those asteroids, (9969) Braille, (5535) Annefrank, and (132524) APL were observed at relatively low resolutions that do not allow for detailed studies of their surface morphology. The remaining asteroids represent a considerable sample of the general morphological properties of asteroids, from a small size of about half a kilometer in the maximum dimension for (25143) Itokawa (Demura et al. 2006), to the second largest of all main-belt asteroids, Vesta (Jaumann et al. 2012), and from near-Earth population – (433) Eros and Itokawa – to main-belt asteroids.

Table 1. List of small bodies visited in close-proximity (<$10^6$ km) by spacecraft, listed in chronological order. Under "Type," "HTC" refers to Halley-type comet; "JFC," Jupiter-family comet; "MBA," main-belt asteroid; "NEA," near-Earth asteroid. The letter before "MBA" and "NEA" refers to the taxonomic type; the types are discussed in sections 1.5 and 3.3. Note that flybys of Comets 21P/Giacobini-Zinner and 26P/Grigg-Skjellerup returned no imaging, hence we have put a '?' in the dimension column.

| Object | Type | Year | Spacecraft | Closest pass (km) | Best res. (m/pixel) | Dimension of object (km) |
|---|---|---|---|---|---|---|
| 21P/Giacobini-Zinner | JFC | 1985 | ICE | 7,800 | N/A | ? |
| 1P/Halley | HTC | 1986 | Vega 1 | 8,889 | ~300 | 15x7 |
| 1P/Halley | HTC | 1986 | Vega 2 | 8,030 | ~300 | 15x7 |
| 1P/Halley | HTC | 1986 | Giotto | 596 | 130 | 15x7 |
| 1P/Halley | HTC | 1986 | Suisei | 151,000 | 37,000 | 15x7 |
| (951) Gaspra | S MBA | 1991 | Galileo | 1,600 | 54 | 18x11x9 |
| 26P/Grigg-Skjellerup | JFC | 1992 | Giotto | 200 | N/A | ? |
| (243) Ida | S MBA | 1993 | Galileo | 2,400 | 25 | 60x25x19 |
| (253) Mathilde | C MBA | 1997 | NEAR | 1,200 | 200 | 66x48x46 |
| (433) Eros | S NEA | 1998 | NEAR | 3,827 | 363 | 34x11x11 |
| (9969) Braille | Q NEA | 1999 | Deep Space 1 | 26 | 182 | 2x1x1 |
| (433) Eros | S NEA | 2000-2001 | NEAR | 0 | cm to m | 34x11x11 |
| (5535) Annefrank | S NEA | 2002 | Stardust | 3,079 | 184 | 7x5x3 |
| 19P/Borrelly | JFC | 2002 | Deep Space 1 | 2,171 | 47 | 8x2 |
| 81P/Wild 2 | JFC | 2004 | Stardust | 237 | 14 | 6x4x3 |
| 9P/Tempel 1 | JFC | 2005 | Deep Impact flyby | 700 | 10 | 8x5 |
| 9P/Tempel 1 | JFC | 2005 | Deep Impact impactor | 0 | 3 | 8x5 |
| (25143) Itokawa | S NEA | 2005 | Hayabusa | 0 | cm to m | 0.5x0.3x0.2 |
| (132524) APL | S MBA | 2006 | New Horizons | 101,867 | ~2000 | 2.3 |

| (2867) Steins | E MBA | 2008 | Rosetta | 800 | 80 | 7x6x5 |
| --- | --- | --- | --- | --- | --- | --- |
| (21) Lutetia | C or M MBA | 2010 | Rosetta | 3,162 | 60 | 121x101x75 |
| 103P/Hartley 2 | JFC | 2010 | Deep Impact | 700 | ~7 | 0.7x2.3 |
| 9P/Tempel 1 | JFC | 2011 | Stardust | 181 | 12 | 8x5 |
| (4) Vesta | V MBA | 2011-2012 | Dawn | 210 | 20 | 573x557x446 |
| (4179) Toutatis | S NEA | 2012 | Chang'e 2 | 3.2 | 10 | 4.6x2.3x1.9 |

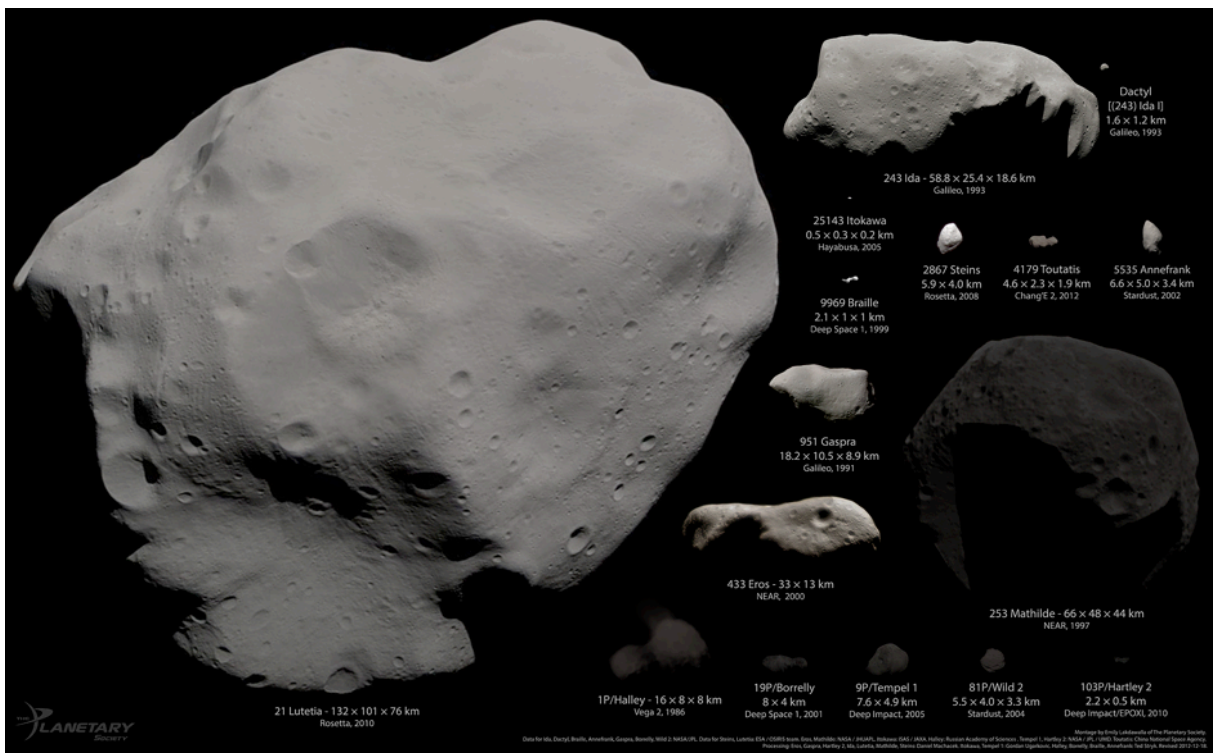

Figure 8. More complete montage of all small bodies that have had spacecraft flybys, except for Vesta, shown to scale. Each asteroid is as close to true color and albedo as possible. Note that the diameter of the largest object here, Lutetia, is about 100 km. Montage by Emily Lakdawalla. Data from NASA / JPL / JHUAPL / UMD / JAXA / ESA / OSIRIS team / Russian Academy of Sciences / China National Space Agency. Processed by Emily Lakdawalla, Daniel Machacek, Ted Stryk, Gordan Ugarkovic. Used by permission

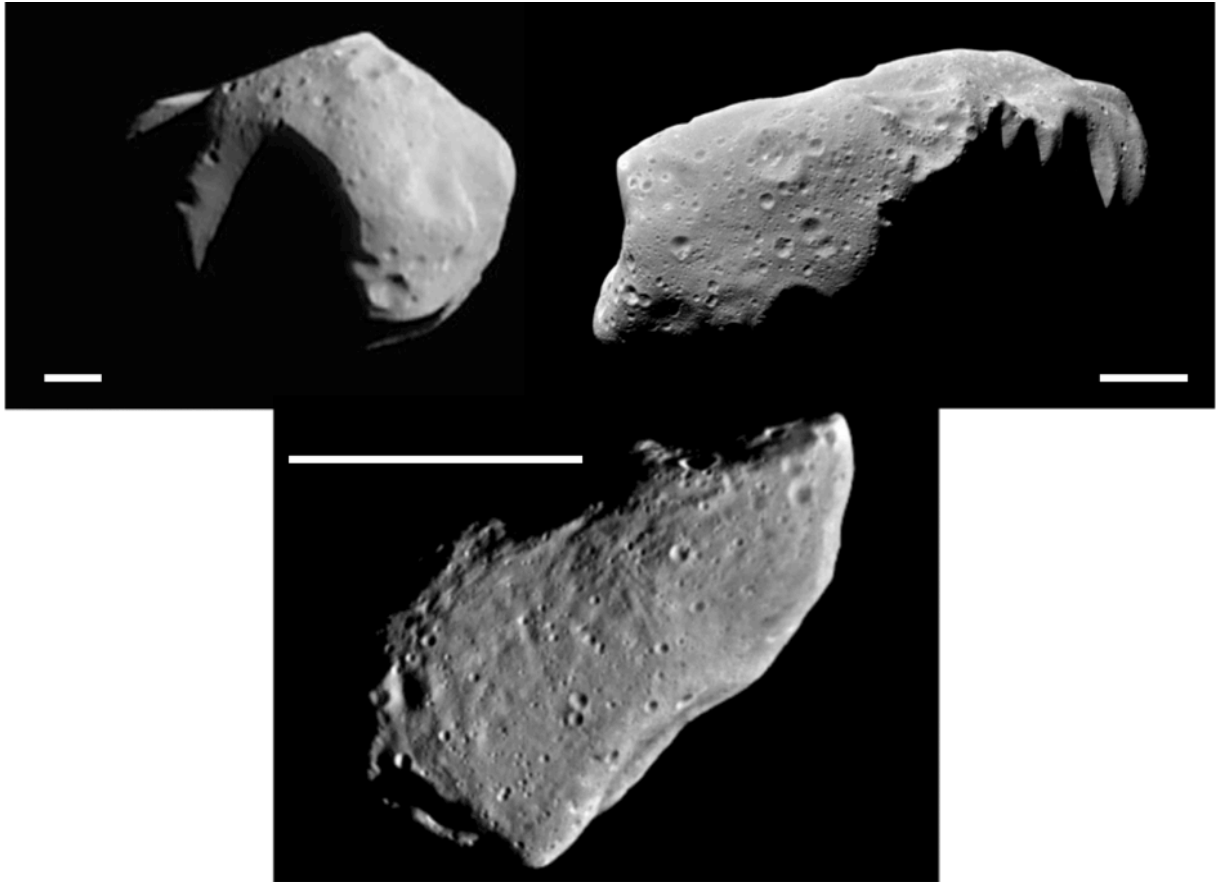

Figure 9. Higher-resolution views of some of the main-belt asteroidal flyby targets shown in Figures 7 and 8 (clockwise from top left): Mathilde, Ida, and Gaspra. White horizontal bar in each panel measures 10 km. All pictures courtesy NASA (Photojournal images PIA02477, PIA00135, PIA00118).

Despite their small sizes and therefore much weaker surface gravities compared to planets, asteroids display a large variety of surface features, including craters, boulders, lineaments (grooves, troughs, ridges), rough and smooth terrains, regolith, and landslides. Being geologically inactive, the surfaces of asteroids are usually dominated by impact craters. The surface features reflect the mechanical responses of the target asteroids to various impact energies under various geometrical settings, and therefore reveal their physical properties. We summarize the major geological features found on asteroids in this section, compare different asteroids, and discuss the implications to asteroids' evolutionary history and properties.

3.2.1. Craters

The dominant geomorphological features on the surfaces of almost all larger asteroids are craters, most of which are of impact origin, although this is not entirely certain for the small asteroid Itokawa due to the different morphology of the "circular depressions" (Hirata et al. 2009). After formation, craters on asteroids are subject to size-dependent degradation and removal processes, such as regolith movement, slope failure, and micrometeorite bombardment. Three collective properties of crater populations on asteroids are the most important and widely studied. First, the spatial density is a critical indicator of surface age since the last resurfacing (e.g., Ivanov et al. 2002, Chapman 2002). Second, the size-frequency distribution is related to the size-frequency distribution of impactors via cratering scaling laws (e.g., Holsapple et al. 2002 for a review; Holsapple and Housen 2007; Melosh 1989). The gravity and strength of asteroid material affects the scaling law, along with porosity. Third, the depth to diameter ratio (d/D) is affected by many factors including impact energy, target strength, and the degradation processes and trend (e.g., Melosh 1989, Carr et al. 1994, Sullivan et al. 1996), and is also an indicator of regolith depth (e.g., Vincent et al. 2012a, 2012b). The d/D on small bodies, however, is not well measured due to the uncertainties in reconstructing 3-D topography.

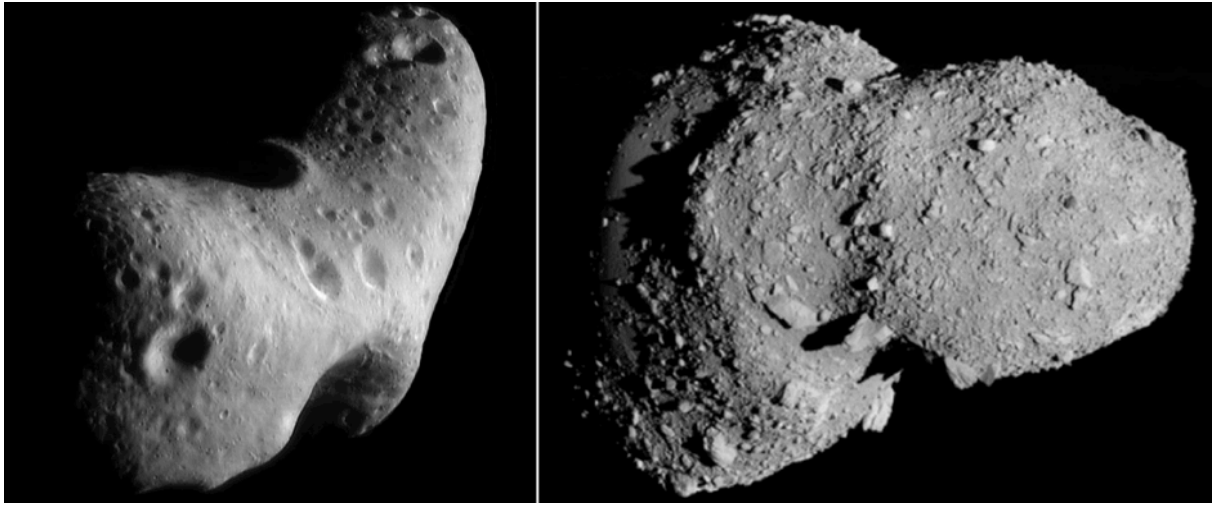

Figure 10. A comparison of Eros (left) and Itokawa (right), the two NEAs with detailed spacecraft observations. Note that Eros's long dimension is about 70 times longer than Itokawa's; Eros is a much bigger object. Also note the dramatic difference in crater density and surface texture. Eros picture credit: NASA/JHUAPL (Photojournal image PIA02923). Itokawa picture credit: ISAS/JAXA (image P-043-12079, used by permission).

The cumulative size-frequency distribution (SFD) of craters on asteroids generally follows a power law with an index between -2 and -4, which is consistent with the SFD of main-belt asteroids from both observations (Jedicke et al. 2002) and theoretical modeling (e.g., Bottke et al. 2005); see Figure 5. This particular range of the power law index is the expected result of asteroid collisional evolution (Davis et al. 2002). Starting from an asteroid population model, relative velocity, and the assumption of scaling laws, one can fit the size-frequency distribution and density of craters on an asteroid or a region to derive its crater retention age, which is the time since the the most recent resurfacing event (e.g., O'Brien and Greenberg 2005; O'Brien et al. 2006). While generally this approach matches the crater size frequency on asteroids and allows us to derive these relative ages satisfactorily on (e.g.) (243) Ida, Dactyl (Ida's moon), parts of (21) Lutetia, and parts of Vesta (Chapman et al. 1996a; Davis et al. 1996, Marchi et al. 2012a, 2012b), most asteroids show obvious deviations from the models at various crater sizes.

Gaspra (Figure 9) has few, if any, large craters, but is peppered with fresh, small craters of several tens to hundreds of meters that are saturated, i.e. in an equilibrium state (Chapman et al. 1996b; Carr et al. 1994). The number of craters on an object is said to reach equilibrium when for every new crater that forms, another is lost or covered up, so that the total number cannot increase. On Gaspra, the lack of large craters is inconsistent with the abundance of smaller craters. This property is so far unique to Gaspra. Gaspra possesses a few flat facets, which might be the remnants of large impact craters of sizes approaching the size of the asteroid itself (Thomas et al. 1994; Greenberg et al. 1994).  The lack of large craters on Gaspra indicates a young age of ~250 Myr (Chapman et al. 1996b).  However, Greenberg et al. (1994) and Chapman (1997) argued that the surface of Gaspra could be either much older or younger, if one assumes rubble pile structure (nearly strengthless) or monolithic structure (strong), respectively.

Unlike Gaspra, Eros (Figure 10) shows an extreme paucity of small craters of <100 m, while large craters >200 m are in an equilibrium state, similar to that of Ida (Veverka et al. 2000; Chapman et al. 2002).  Richardson et al. (2004) and Michel et al. (2009) showed that the most likely reason for the paucity of small craters on Eros is seismic shaking erasure.  The possibility that small impactors are preferentially removed by a size-dependent mechanism – i.e. by the Yarkovsky effect, a thermal radiation force that causes objects to undergo semimajor axis drift due to asymmetric radiation (Bottke et al. 2006) – has been considered but rejected (O'Brien 2009).

Itokawa (Figure 10) and (2867) Steins also show obvious depletion of small craters (Besse et al. 2012), probably also due to seismic shaking. The YORP (Yarkovsky-O'Keefe-Radzievskii-Paddack) effect (Bottke et al. 2006) – analogous to the Yarkovsky effect except it changes the asteroid's spin state due to that asymmetric radiation of thermal emission – might be responsible for regolith movement on Steins (Marchi et al. 2010a), effectively "reshaping" the asteroid in that sense. Alternatively, for Itokawa, the high abundance of boulders on its surface could have protected it from forming small craters, with impactors simply fragmenting the boulders rather than creating craters (Michel et al. 2009).

The most striking features on (253) Mathilde (Figure 9) are several large craters with sizes comparable to the size of the asteroid itself (Cheng and Barnouin-Jha 1999).  The number of those large craters exceeds the saturation density, suggesting to several authors that Mathilde may be composed of weakly bonded, porous, crushable material, that absorbs impact shock efficiently without disrupting pre-existing large craters (Housen et al. 1999; Asphaug 1999).  The population of small craters on Mathilde also saturates its surface, consistent with an old surface of perhaps several billion years (Davis 1999), although the estimate varies depending on the internal strength.

Vesta and Lutetia are both large and each shows a complicated cratering record on their surfaces.  Vesta is ~560 km in diameter (Jaumann et al. 2012), and is large enough to trap sufficient radiogenic heat during its accretion to lead to differentiation of its interior. Vesta is thus a differentiated protoplanet (e.g., Keil 2002; Russell et al. 2012).  On the other hand the internal structure of  Lutetia (~100 km across) is uncertain, but it is unlikely to be differentiated, so it represents an ancient planetesimal (e.g., Sierks et al. 2011; Coradini et al. 2011).  The cratering morphology on both asteroids includes all crater preservation states and different crater densities from region to region, suggesting very different crater retention ages from hundreds of Myr to >3.6 Gyr (Thomas et al. 2012; Massironi et al. 2012; Marchi et al. 2012a, 2012b; Schenk et al. 2012). On the oldest terrain on Lutetia, a region called Achaia, the size-frequency distribution shows a distinct flexure point at 4-7 km diameter, attributed to the presence of a stratified surface with fractured material overlying a more competent interior (Marchi et al. 2012a), similar to the case for Mercury (Marchi et al. 2011). The full analysis of cratering properties on Vesta from Dawn data is still ongoing as of the writing of this chapter. Readers are referred to, e.g., Yingst et al. (2013) and Williams et al. (2014) for the geology of Vesta.

The degradation states of craters can be gauged with the depth to diameter ratio ($d/D$), and the ratio can be also an indicator of the depth of regolith. Typically, the crater $d/D$ on these asteroids is between 0.12 and 0.15 (Gaspra, Ida, Eros) (Carr et al. 1994; Sullivan et al. 1996; Veverka et al. 2000), although for large asteroids Mathilde, Lutetia, and Vesta, this value could have a large range of <0.1 to >0.2 (Veverka et al. 1999; Vincent et al. 2012a, 2012b; Jaumann et al. 2012). The small asteroid Steins also has a large range of $d/D$ of 0.04 to 0.25 (Besse et al. 2012). The smallest asteroid Itokawa has much shallower craters with $d/D$ of 0.08 on average and a relatively narrow range of ±0.03, possibly due to the small size of the asteroid, the small number of craters with which to make the analysis in the first place, the incomplete formation of raised rims, and/or the infilling of crater by pebbles (Hirata et al. 2009, Michel et al. 2009). In contrast, the Moon, Mars, and Phobos typically have a depth to diameter ratio of about 0.2 (Shingareva et al. 2008). The wide range of $d/D$ values among all these bodies indicates that crater creation and degradation depend on the detailed properties of the targets themselves. A comprehensive understanding of the variety of depth-to-diameter ratios – and of asteroid cratering in general – is a project for the future; the study of cratering on asteroids is complicated by both observational and modeling limitations. The most obvious factors include the identification of secondary craters (formed by re-impact of ejecta), the existence of extremely degraded craters underlying new craters, and the often irregular shapes of asteroids with complicated local topography. Crater chronology is almost always limited by our knowledge of the impactor populations, the mechanical properties of the surface (affecting scaling laws), and the removal process. The result is that surface ages are often strongly model-dependent.

3.2.2. Boulders

The visiting spacecraft imaged boulders on the surfaces of Eros, Lutetia, Vesta, and especially Itokawa. The surface of Itokawa is in fact dominated by boulders rather than craters (Fujiwawa et al. 2006; Barnouin-Jha et al. 2008; Michikami et al. 2008). This is dramatically shown in Figure 10. The boulders on Vesta have not yet been thoroughly studied as of the writing of this chapter. Therefore we will focus on the other three asteroids to discuss boulders. The shape of boulders on Eros and Itokawa are consistent with laboratory impact experiments (Michikami et al. 2010). The size-frequency distributions of boulders on all of them, as well as on Phobos, follows a power-law index of nearly -3 within a few meters to 80 m (Thomas et al. 2000; 2001; Michikami et al. 2008), a steep slope of power index of -6 for <10 cm size on Eros near its landing size (Thomas et al. 2001), and steep slopes of indexes -5 for boulders > 80 m for both Eros and Lutetia (Thomas et al. 2001; Küppers et al. 2012).

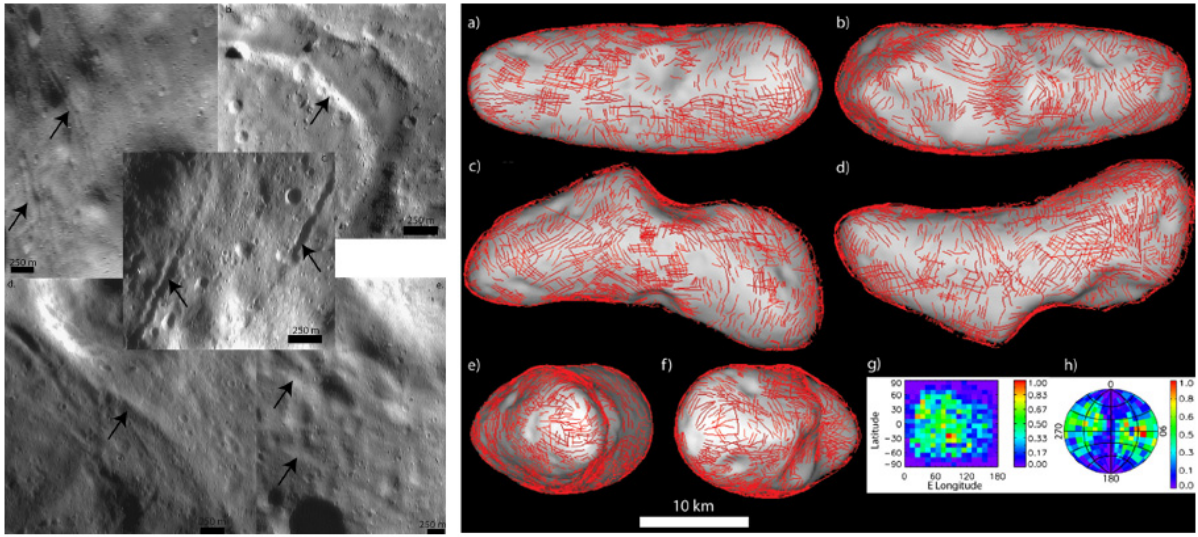

Figure 11. Lineaments on Eros. These are Figures 4 and 5 from the work by Buczkowski et al. (2008; reprinted from Icarus with permission from Elsevier). The left panel demonstrates various lineament morphologies, with the arrows indicating the relevant features, as follows: top left, grooves; top right, flat-floored troughs; bottom left, ridges; bottom right, shallow troughs; center, pit chains. The right panel shows the extensive network of 2141 lineaments (red marks) identified by software. Eros is shown in six orientations. The maps at bottom right (in two different projections) show the distribution of the poles of all the lineaments; the lineaments seem to be randomly distributed.

The origin of boulders is generally associated with impact breaking of the target material. Since boulders are subject to destruction by the bombardment of small impactors, the existence of a large number of boulders usually suggests relatively young ages of the associated impact craters. For example, Geissler et al. (1996) noted that the distribution of large blocks on Ida is consistent with ejecta from the impact that created Azzurra crater; Thomas et al. (2001) studied the spatial distribution of boulders and their characteristics on Eros, and established the formation of Shoemaker crater as the probable source of most of its boulders; a similar study for Lutetia by Küppers et al. (2012) also showed that most boulders on Lutetia were generated by the formation of the central crater in the youngest region, Baetica. The boulders on Itokawa, however, show a global distribution. There are no large craters on this small asteroid, and the total volume of boulders accounts for ~25% of the total excavated volume of craters, much higher than that on Eros (<1%) and on the Moon (~5%) (Thomas et al. 2001; Cintala et al. 1982). These arguments led Fujiwara et al. (2006) and Michikami et al. (2008) to conclude that the bulk of boulders on Itokawa most likely originated from the disruption of the parent body, and re-accumulated on the surface of Itokawa during its formation.

3.2.3. Lineaments

Lineaments were first seen quite prominently on the global scale on Eros (Thomas and Prockter 2010). The term includes grooves, troughs, and ridges, and they can be local or global in scale. Examples of lineaments are shown in Figures 11, 12, and 13. All the asteroids that are resolved at sufficiently high resolution – except for Itokawa, which is a rubble pile – show either local or global lineaments on their surfaces. They seem to be caused by propagation of stress or tension, and indicate to some extent a coherent interior. Local-scale lineaments appear to be associated with strong impacts that formed large craters. Global-scale lineaments could be associated with either the global response to seismic waves during large impacts or the stress environment of being broken up from the parent bodies (Veverka et al. 1994; Thomas et al. 1994; Sullivan et al. 1996; Prockter et al. 2002; Thomas et al. 2002; Buczkowski et al. 2008; 2012; Thomas et al. 2012). However an intriguing alternate hypothesis for the Eros lineaments was given by Greenberg (2008), who suggested instead that the lineaments are near a vein of stronger rock (perhaps created by partial melt) that was more resistant to impact erosion. Grooves seen in the regolith are probably expressions of fracture within a more coherent substrate (Thomas et al. 1979), and may provide additional clues about the interior structure.

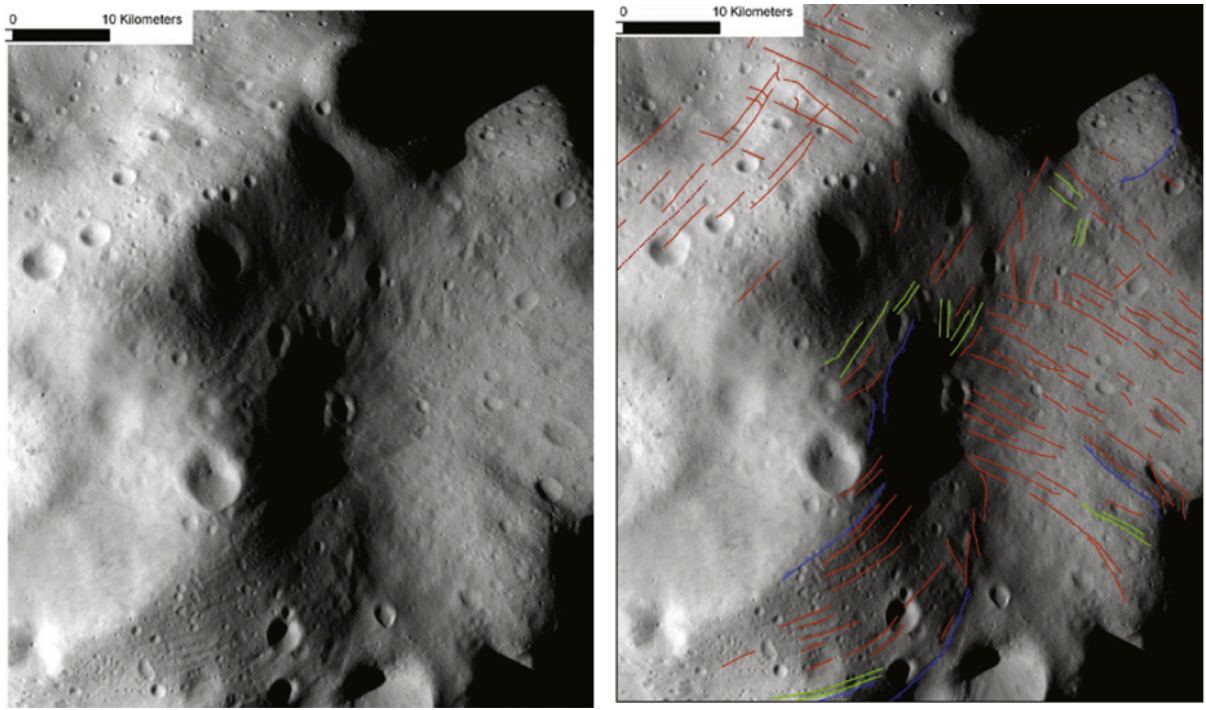

Figure 12. Lineaments on the surface of asteroid Lutetia. This shows panels a and b of Figure 6 from the work of Massironi et al. (2012; reprinted from Planetary and Space Science with permission from Elsevier), with the actual image from Rosetta on the left, and the lineaments marked with colored lines on the right. Note the scale bar at top.

The most interesting lineament systems exist on Eros, Lutetia, and Vesta. Buczkowski et al. (2008) mapped >2000 lineaments on Eros ranging up to tens of km in length (Figure 11). They found that, except for the lineaments that are radial or circumferential to ~10 craters and obviously associated with those impacts, the majority of lineaments belong to three sets: parallel to meridian; encircling one end of Eros, and towards the other end. While interpreting the first set as due to impacts on the long side of the body, they proposed that the second and third sets could be due to the stress environment in its parent body. Nevertheless, the large number of lineaments on Eros on a global scale is not consistent with a rubble pile internal structure.

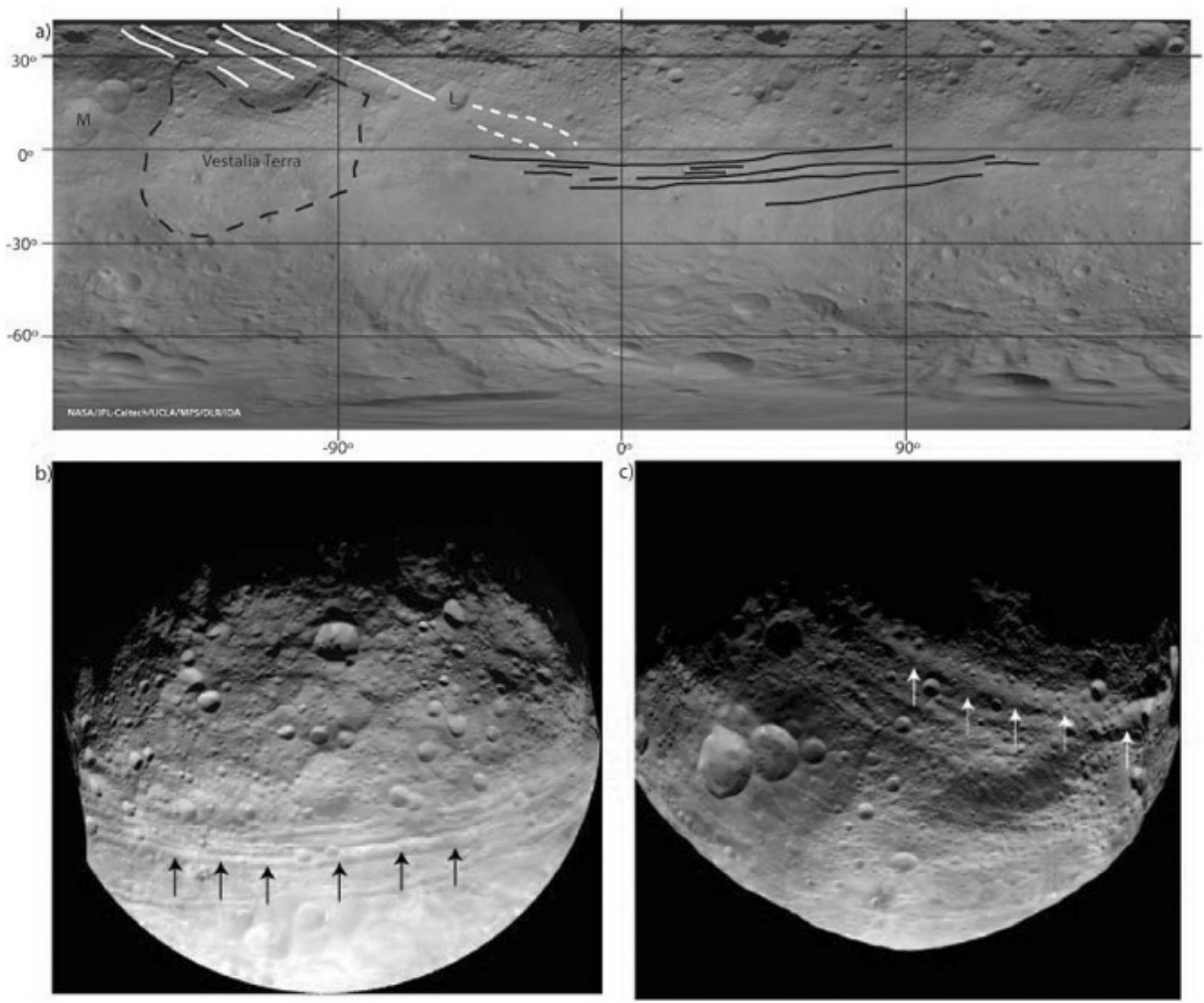

Figure 13. Lineaments on Vesta. This is Figure 1 from the work by Buczkowski et al. (2012; reprinted from Geophysical Research Letters with permission from John Wiley and Sons). The top panel is a cylindrical-projection map of Vesta; various systems of troughs are marked in black solid, white solid, and white dashed lines. The bottom left panel shows the equatorial troughs (the black solid lines in the top panel), while the bottom right panel shows the northern troughs (the while solid lines in the top panel). Both images are from the Dawn spacecraft.

Lutetia shows an extremely complicated system of lineaments on the full globe except for the geologically youngest region near the north pole (Thomas et al. 2012), and not as systematically distributed and organized as on Eros. An example of the lineament system is shown in Figure 12.

While the full analysis of lineament systems on Vesta is still ongoing, the most striking features are two sets of long linear structures near the equator and on the northern hemisphere that almost encircle the whole body (Figure 13) (Jaumann et al. 2012). Buczkowski et al. (2012) categorized these linear structures as grabens. These grabens appear to be co-planar with the respective poles coincident to the centers of two large basins near the south pole, Rheasilvia and Veneneia. They concluded that the properties of the grabens are consistent with impact-induced, global scale deformation of a fully differentiated body, completely distinct from the responses of small, undifferentiated asteroids to impacts.

### 3.2.4. Smooth Deposits

Both NEAs visited by spacecraft, Eros and Itokawa, display apparently smooth regions, termed "ponds" on Eros (Robinson et al. 2001; Cheng et al. 2001; Abe et al. 2006; Barnouin-Jha et al. 2008). The smooth areas on both asteroids are located in gravitationally low areas, indicative of movements of regolith with cm to mm scale particles. Cheng et al. (2002) proposed that the ponded deposits on Eros are consistent with fluid-like motion of dry regolith, possibly induced by seismic agitation from impacts, the same process that is responsible for preferentially removing small craters on both of these objects (Richardson et al. 2004; Michel et al. 2009).

### 3.2.5. Regolith

Regolith (defined as loose, fragmental debris, regardless of origin) was not originally predicted before the Galileo encounter with Gaspra because of the low gravity (e.g., Housen et al. 1979a; b), although other early work, including that of Veverka et al. (1986) indicated that regolith would accumulate on small objects. The surfaces of asteroids are now known to possess a substantial amount of regolith derived from impacts, and this regolith demonstrates considerable mobility. The depth of regolith is generally inferred from crater morphology and ejecta, distribution of smooth deposits, the morphology of geologic features such as boulders and lineaments, and spectrophotometric variations of the surface. In general, meters to hundred of meters thick regolith are inferred on asteroids, depending on the size of the objects (Carr et al. 1994; Chapman et al. 1996a, 1996b; Sullivan et al. 1996; Barnouin-Jha et al. 2008; Prockter et al. 2002; Besse et al. 2012; Vincent et al. 2012a, 2012b). The regolith on Vesta displays global variations from tens of meters to >1 km (Denevi et al. 2012; Jaumann et al. 2012) due to the complicated impact history.

Movement of regolith is evident from the localized smooth areas (as discussed in the previous section) and by mass wasting due to slope failure, as commonly seen on steep slopes of crater walls (e.g., Miyamoto et al. 2007; Thomas et al. 2012; Jaumann et al. 2012; Krohn et al. 2014). The mobility and evolution of regolith is generally inferred from both geomorphological evidence and the optical maturity of regolith (i.e. from space weathering; see next section). For NEAs, the causes of mass movement are generally considered to be (a) seismic shaking induced by impact, (b) YORP spin-up, and (c) tidal disturbance from close encounters with planets. In binary asteroids, tidal effects of one body on the other can similarly move regolith around.

### 3.3. Space Weathering

Our knowledge about the mineralogy of asteroids is dominated by remote sensing observations of the spectral reflectance properties of asteroidal surfaces, supplemented by laboratory experiments of meteorite samples. It has long been recognized that the optical properties of silicate rocks are subject to change due to exposure in the space environment, caused by solar wind irradiation and micrometeoroid sputtering (Hapke 1973). For the case of the Moon, the exposure causes the surface to darken, the spectral slope to redden, and the mineral absorption bands to weaken. Such optical phenomena and the associated processes are broadly termed "space weathering". The space weathering process on the Moon is well understood (see, e.g., Pieters et al. (2000); Hapke (2001) for comprehensive reviews). Laboratory measurements and experiments of lunar samples showed clear evidence that vaporization of silicate minerals and redeposition of nanophase iron (npFe$^0$) coat the surface of regolith particles (cf. Hapke 2001). Because the sizes of metallic Fe particles are smaller than visible wavelengths, they strongly absorb incident light, and their absorption coefficient decreases as the wavelength increases, resulting in darkening and reddening of the regolith particles, and masking out mineral absorption bands (cf. Hapke 2001). However, this process cannot be easily extended to asteroids in general (Gaffey 2010). Chapman (2004) presents an excellent historical review on this topic. Our understanding of space weathering on asteroids has been severely limited by both the lack of direct samples and the limited high spatial resolution spectroscopic data. Because mineralogical identification on asteroids almost entirely relies on the spectral reflectance properties of asteroidal surfaces, understanding the space weathering processes is an important aspect of asteroid science. As we shall see in this section, the surfaces of asteroids display different manifestations of space weathering from that on the Moon. For the convenience of our discussion, we refer to the space weathering process on the Moon and its associated optical effects as "lunar-type".

Space weathering on asteroids was first proposed to explain the spectral difference between the most abundant samples in our meteorite collections, the ordinary chondrites (OCs), and the most abundant type of main-belt asteroids, S-types (cf. section 1.5). Presumably the S-types were the sources of OCs, yet the spectral differences remained problematic. Under the assumption that no mechanical or mixing process should change the mafic band centers and their relative strength, Gaffey et al. (1993) studied the mineralogy of S-type asteroids, and found that the spectra of a particular subclass (known as "S(IV)") implies mineralogy similar to the known OCs, therefore providing a generic link between S-types and OCs. However, the S-type asteroids show much weaker absorption bands, and much redder spectral slopes than do OCs. On the other hand, in the NEA population, a small class of asteroids, labelled the Q-types, with sizes of a few km, show similar spectral characteristics to those of OCs (McFadden et al. 1984; Binzel et al. 2002). This relationship is a good reality check because the OC meteorites were NEAs before falling to Earth, so one would expect there to be a closer spectral match. Binzel et al. (1996; 2002) showed that non-primitive NEAs show a continuous spread in their spectra from OC-like to S-like, implying a process that is continuously operating on the surfaces of one class and changing their optical properties to the other. Both observational (Binzel et al. 2010) and dynamical (Nesvorny et al. 2010) evidence suggested that tidal encounters to within ~5 planet radii of Earth and Venus could reset the weathered surfaces on small S-type asteroids, by exposing a relatively fresh unweathered surface and changing them to Q-types.

To put it another way, S-type main-belt asteroids have highly-weathered surfaces because they have been exposed to the space environment for ostensibly tens of millions to billions of years. Q-type surfaces, i.e., those with little or no weathering, on the other hand would appear in the NEA population because NEAs are collisional fragments and relatively recent escapees from the Asteroid Belt, thus their surfaces have only been exposed for millions of years or perhaps much less.

Recent observations of small S-type asteroids in the main belt that are of comparable size to NEAs also show spectral variation, which is logical because these objects have fresher surfaces and may be on their way to becoming NEAs. Some of these small objects have spectra that match the OCs (Mothé-Diniz et al. 2010). Further studies of the colors of many S-type dynamical families in the main belt have shown that their red spectral slopes are clearly correlated with dynamic ages since breaking up, consistent with continuous weathering process on S-type asteroids (Nesvorny et al. 2005).

Recent asteroid missions have both confirmed the existence of space weathering on asteroids, and raised more questions. The Galileo spacecraft flybys of two S-type main-belt asteroids Gaspra and Ida showed that the surfaces of Gaspra and Ida both have large variations in colors and at least a factor of two variation in the strength of mafic bands. While the overall spectra of Gaspra and Ida are typical of S-type asteroids, small, fresh craters and ejecta appear to be bluer, and the ridges with steep slopes tend to be bluer than other areas (Belton et al. 1992; 1994; Geissler et al. 1996; Sullivan et al. 1996). Close examination of the spectral trend on Ida revealed that the geomorphologically freshest units have spectra intermediate between OCs and the S-type asteroids (Chapman 1996). However, because the albedo variation on Ida is small, the space weathering process does not seem to have darkened the surface of Ida as much as it has of the Moon.

The clearest evidence of space weathering on Eros was identified in the large crater named Psyche, where brighter materials are exposed on the wall and darker materials are deposited on the bottom (Clark et al. 2001). The albedo variation from wall to bottom is a factor of several. The color of the bright crater wall is slightly bluer than that of the dark crater floor, consistent with, but much subtler than, lunar-type space weathering (Murchie et al. 2002; Izenberg et al. 2003). Therefore, the wall of Psyche crater was recently exposed due to mass wasting of the original regolith on the slope, and is much less space weathered than the crater floor. However, there are a number of inconsistencies in terms of space weathering on Eros. First, the color and spectral variations on Eros are remarkably uniform to within 2% (McFadden et al. 2001; Izenberg et al. 2003). The lack of color contrast between the supposedly fresh crater wall and the weathered crater floor certainly suggests different weathering processes on Eros than on Gaspra, Ida, and the Moon; this is represented schematically in Figure 14 (Gaffey 2010). Second, the downslope movement of Psyche crater wall material had to be driven by, e.g., other impacts after the formation of the slope. However, Psyche crater appears to be the youngest on Eros; none of the small craters on Eros appear fresh, i.e., brighter and bluer than average surface. Cheng et al. (2002) considered that the shaking of Eros, probably induced by tidal forces during close approach with terrestrial planets, could induce mass wasting on slopes and the formation of ponds. In addition, ejecta blocks on Eros, mostly excavated by the impact which formed Shoemaker crater, show similar albedo and spectra to other areas on Eros, a relation that also indicates relatively mature surfaces.

The Hayabusa mission to Itokawa provided evidence of space weathering on this small (550 m) NEA from both spectroscopic observations and the first samples directly returned from known sites on a S-type asteroid. Hiroi et al. (2006) showed that a dark region on Itokawa has a spectrum clearly consistent with significant space weathering with accumulated npFe$^0$ (nanophase metallic Iron) with about 0.03-0.07 vol.%. Ishiguro et al. (2007) analyzed the spectral data of the whole surface of the asteroid to map the space weathering maturity on Itokawa, and found that fresh materials are generally observed in regions of steep slopes and craters, whereas mature materials are ubiquitously distributed. Laboratory measurements of the Itokawa samples showed surface modifications on the rims of some particles, with sulfur-bearing Fe-rich nanoparticles, interpreted as a consequence of alteration by vapor deposition (Noguchi et al. 2011; 2014).

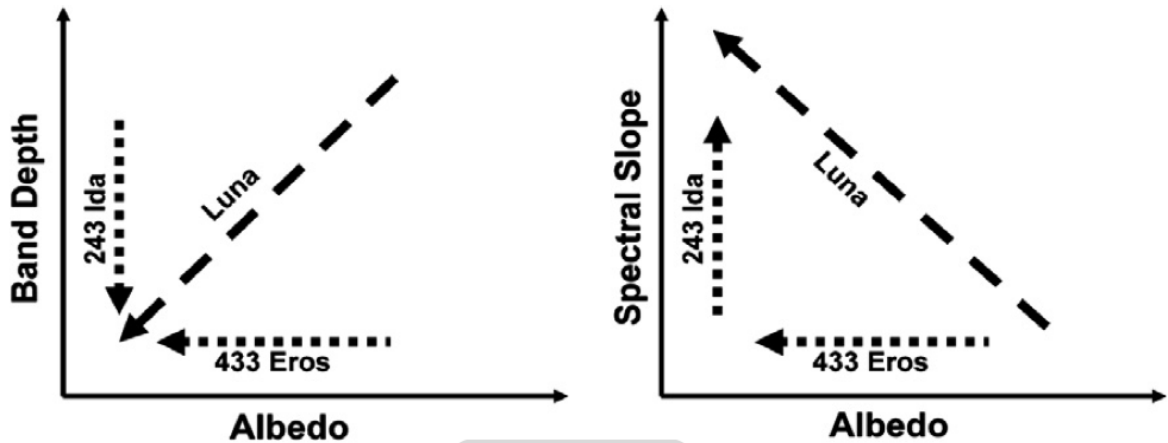

Figure 14. Comparison of space weathering across asteroids and the Moon. This is Figure 4 from the work by Gaffey (2010; reprinted from Icarus with permission from Elsevier). There are many manifestations of "space weathering," and the trends of albedo vs. depth of an absorption band vs. spectral slope can be different depending on the properties of the surface under study.

The presense of npFe$^0$ is one consequence of space weathering that appears to be the same on both the Moon and on asteroids. In addition to the evidence of npFe$^0$ found in Itokawa regolith samples (Noguchi et al. 2011; 2014), nano-size metallic particles are also found in some OC meteorites (Moretti et al. 2005; 2006; Longobardo et al. 2011) and in the howardite (i.e. likely Vesta fragment) Kapoeta meteorite (Noble et al. 2010). Furthermore, extensive laboratory experiments of space weathering have been conducted to simulate the effects of micrometeoroid impacts through laser impulses and solar wind bombardment through plasma irradiation (e.g., Hapke 1973; 2001; Dukes et al. 1999; Yamada et al. 1999; Strazzula et al. 2005; Brunetto and Strazzulla 2005; Marchi et al. 2005; Brunetto et al. 2006; 2007; Noble et al. 2007; Loeffler et al. 2008; 2009). The most significant result from these experiments has been to reproduce the lunar space weathering effects by generating npFe$^0$ coating on mineral grains (Keller and McKay 1993). The time scale of space weathering due to micrometeoroid impacts has been estimated to be of order $10^8$ years (Sasaki et al. 2001).

For non-S-type asteroids, such as the more primitive objects in the C-complex, the effect of space weathering is not obvious. This lack of spectral detection might be because the dark, opaque, carbonaceous materials (which are thought to be the dominant composition of low albedo asteroids like those in the C-complex) may obscure the effects of space weathering. Or alternatively, a low abundance of Fe might suppress the formation of npFe$^0$, and slow down the process of lunar-type space weathering.

Finally, we point out the puzzling case of Vesta, which resides on the opposite end of the traditional space weathering problem, i.e., Vesta shows a lack of weathering, even though it is expected. McCord et al. (1970) found an exact match between the spectrum of Vesta and the spectra of the most abundant achondrite meteorite samples, the howardite-eucrite-diogenites (HEDs), leading to the conclusion that Vesta is the source of HEDs. Compelling evidence from almost every aspect confirms the HED-Vesta connection. Early evidence includes the identification of small (few km) spectrally similar asteroids (now known as taxonomic type V) that provide a dynamical pathway to transport meteorites from the Asteroid Belt to Earth (Binzel and Xu 1993), and the existence of a large basin near Vesta's south pole (Thomas et al. 1997a, 1997b). Recent observations by the Dawn mission confirmed the mineralogical connection between Vesta, HEDs, and V-type asteroids (De Sanctis et al. 2012; 2013; Ammannito et al. 2013), and confirmed a highly impacted surface with two large, overlapping basins near the south pole, with the youngest one perhaps just ~1 Gyr old (Marchi et al. 2012b; Schenk et al. 2012). However, laboratory studies suggest that ion irradiation on HEDs darkens and reddens the samples rapidly, corresponding to a space weathering time scale of $10^5$ to $10^6$ years (Fulvio et al. 2012). The lack of space weathering on Vesta itself is thus a mystery (see summary by Pieters et al. 2006). To explain it, Vernazza et al. (2006) postulated that Vesta has a magnetic field that can shield it from solar wind irradiation. The discovery of magnetism in the Millbillillie eucrite (Fu and Weiss 2011) seems to support this idea. But magnetic shielding does not completely protect the whole surface of Vesta from weathering, and Dawn should have identified local weathered areas with high-resolution imaging and spectroscopy. Alternatively, the whole surface of Vesta could have been recently resurfaced by a major impact. However, the age of the largest impact basin on Vesta is much older than the time scale generally accepted for space weathering, as mentioned above. On the other hand, the comparisons of red slopes among V-type asteroids and HEDs support the existence of space weathering on V-type asteroids, although the reddening trend with solar wind exposure age is difficult to interpret (Marchi et al. 2010b). Pending the solution of whether the traditional view of space weathering on the Moon and S-type asteroids occurs on Vesta, Pieters et al. (2012) proposed an alternative view of the space weathering based on Dawn observations, in which a physical process on the surface such as regolith mixing, rather than chemical alteration, dominates the space weathering. Still, the most significant question for Vesta is that we do not know how to fit the lack of traditional space weathering into the whole paradigm of such processes on the Moon, Eros, and Ida.

3.4. Surface Morphology on Cometary Nuclei

The encounters with comet 1P/Halley by ESA's Giotto spacecraft (Reinhard et al. 1986) and the Soviet Union's Vega spacecraft (Sagdeev et al. 1986) marked the first closeup imaging of Solar System small bodies. During the close encounter, the Halley Multicolour Camera (HMC) onboard the Giotto spacecraft centered on the brightest part of the inner coma, showing the silhouette of a large, solid and irregularly shaped nucleus, and jet-like dust activity that was much brighter than the nucleus (Keller et al. 1986). Due to the bright jets in the foreground of the nucleus, it was difficult to see the surface morphology of this cometary nucleus from these images. Nevertheless, the solid nature of cometary nuclei as icy conglomerates as proposed by Whipple (1950) was directly proved correct. And for the first time, the albedo of a cometary nucleus was directly measured to be ~4%.

Since the P/Halley flybys, there have been five additional cometary missions returning disk-resolved images of four more cometary nuclei (Table 1). A montage (not to scale) of the comets is shown in Figure 15. The imaging data from those cometary missions has formed the foundation of our current understanding of the surface morphology of cometary nuclei. While the comets have a variety of overall morphologies, there are many common features, including pits and pitted terrains, smooth areas, icy patches, and numerous small, bright or dark spots.

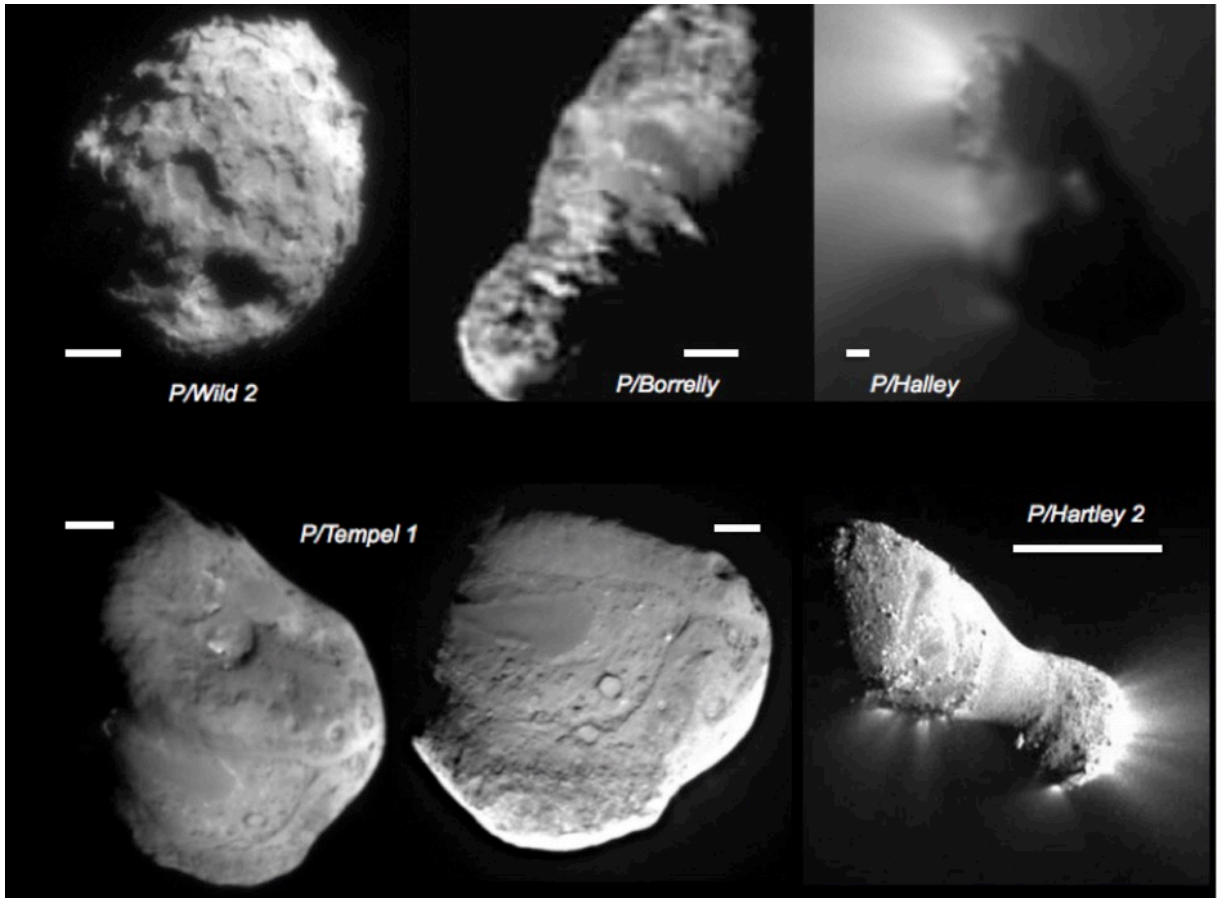

Figure 15. Montage of the five comets that have had spacecraft encounters (with imaging). P/Tempel 1 has been visited twice. P/Halley was visited four times (see Table 1) but for clarity we only show a Giotto picture. White horizontal bar in each panel represents 1 km. Note the variety of surface morphologies. Image sources: P/Wild 2 is courtesy NASA/JPL-Caltech (Photojournal image PIA05571); P/Borrelly is courtesy NASA/JPL (Photojournal image PIA03500); P/Tempel 1 (left image) is courtesy NASA/JPL/UMD (Photojournal PIA02142); P/Tempel 1 (right image) is courtesy NASA/JPL-Caltech/Cornell (Photojournal image PIA13860); P/Hartley 2 is courtesy NASA/JPL-Caltech/UMD (Photojournal image PIA13570); P/Halley is from the Giotto archive of data in the NASA Planetary Data System (Keller et al. 1992).

Comets have numerous (nearly) circular depressions with or without raised rims on their surfaces, generally termed "pits" due to their completely different morphology from impact craters on asteroidal surfaces. Pits are observed on all four cometary nuclei with various sizes and morphologies (Figure 15; Britt et al. 2004; Brownlee et al. 2004; Thomas et al. 2007; 2013a; b). The associated pitted terrains usually occupy a large fraction of a comet's surface, except in some regions where there are smooth terrains. The cumulative size distributions of pits on comets 9P/Tempel 1 and 81P/Wild 2 have slopes between -1.7 and -2.1 (Thomas et al. 2013a), significantly different from the size distribution of typical impact craters on the Moon or asteroids (between -2 and -4). Therefore pits on comets either do not all originate from impacts or have been modified after formation. The pits on comet P/Tempel 1 are up to a few hundred meters in diameter and up to 25 m in depth, mostly without flat floors. Those on P/Wild 2 have larger sizes up to ~1.5 km, mostly with flat floors and some with central peaks. The most distinctive feature on comet P/Wild 2 is the nearly vertical walls of the pits, sometimes with overhangs. While some pits with raised rims on some comets could have impact origins (such as the two circular depressions bracketing the Deep Impact Mission (DI) impact site on comet P/Tempel 1), most of them are probably cryo-volcanic collapse features associated with cometary activity (Belton et al. 2008; Belton and Melosh 2009). Belton et al. (2013) ascribe >90% of the pits to cometary outburst activity.

The apparently smooth areas on comets P/Tempel 1 (Thomas et al. 2007; 2013a) and 103P/Hartley 2 (Thomas et al. 2013b) are probably one of the most intriguing features observed; such areas are visible in Figure 15. These terrains have smooth texture up to 30-m scale, but most likely also at a 5-m scale, with slightly lower reflectance than the surrounding terrains, and they are usually laterally confined. They occupy topographic lows with slopes up to a few degrees, suggestive of flows. The "mesas" on comet 19P/Borrelly as termed by Britt et al. (2004) appear to be similar to the smooth areas on comets P/Tempel 1 and P/Hartley 2. Belton and Melosh (2009) proposed a fluidized multiphase transportation of dust resulting from the sublimation of material with higher volatility than water, such as CO or $CO_2$, from beneath the surface as the origin of the smooth areas on comet P/Tempel 1. No such smooth area was observed on comet P/Wild 2. Belton (2010) used the smooth areas to hypothesize an evolutionary sequence of the four comets, where the surface of comet P/Wild 2 represents the early stage in the sequence and has not developed smooth areas that are large enough to be visible, while the surface of comet P/Hartley 2 represents the latest stage with the largest fraction of surface covered by smooth areas.

From the Deep Impact flyby spacecraft images of comet P/Tempel 1, Sunshine et al. (2006) for the first time unambiguously discovered water ice deposits on the surface of a cometary nucleus. Similar water ice concentration was again observed on comet P/Hartley 2 (Sunshine et al. 2011). The icy patches cover less than 1% of the total surface areas of comets, and only contain about 3-6% water ice, with typical particle sizes of ~30 µm, much larger than those in ejecta observed by Deep Impact (~1 µm) (Sunshine et al. 2007), and in comet P/Hartley 2's coma (Protopapa et al. 2014). Temperature measurements and thermal modeling suggest that the water ice deposits are thermally decoupled from the refractory dust on the surface (Groussin et al. 2007; 2013). The locations of the identified ice deposits on comets P/Tempel 1 and P/Hartley 2 are both near the morning terminators, suggesting that they are probably condensation of water vapor in diurnal cycles as opposed to ice exposed from the interior (which would have much smaller particle sizes, too; Sunshine et al. 2007). It is quite possible that the reason for water ice being definitively discovered on only two cometary nuclei is that only the Deep Impact spacecraft, among those having visited comets, is equipped with a spectrometer capable of detecting water ice. Therefore, it is reasonable to speculate that water ice patches are likely common on cometary nuclei.

In addition to the large-scale features discussed above, there are numerous bright and dark spots on all well-imaged cometary nuclei (Figure 16). Some of them appear to be albedo features, such as some bright spots on comets P/Wild 2 and P/Tempel 1. The spots could be water ice concentrated in areas due to topography, but it is not possible to determine their origins conclusively because of the limited resolution or the unavailability of spectroscopic data. On the other hand, some dark spots could either be small-scale albedo features, or small, deep pits or holes (Nelson et al. 2004).

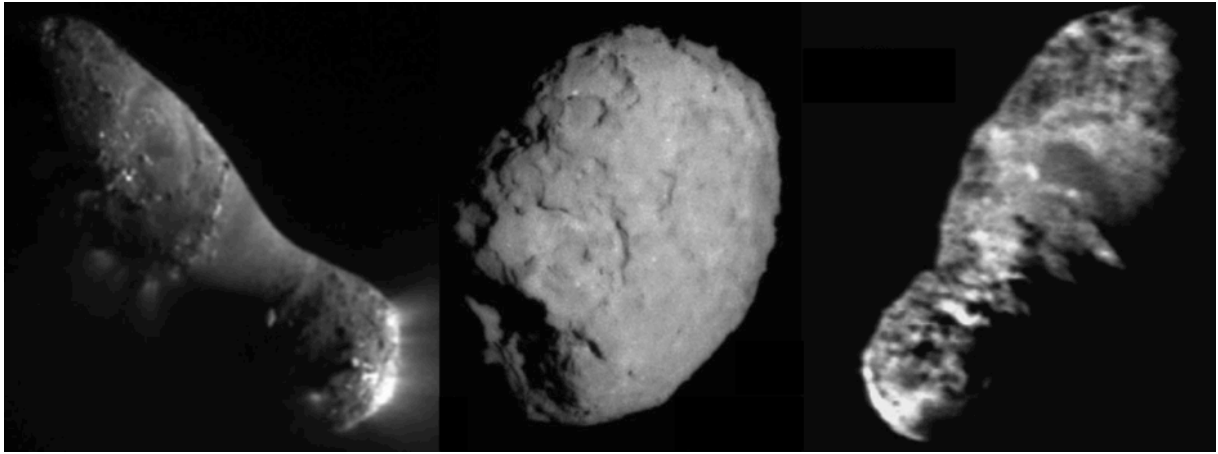

Figure 16. Examples of the bright and dark spots seen on cometary surfaces. Images are of P/Hartley 2, P/Wild 2, and P/Borrelly, from left to right. The P/Hartley 2 and P/Borrelly images are the same as those seen in Figure 15, but with better contrast to enhance the features. P/Wild 2 image is from the Stardust archive of data in the NASA Planetary Data System (Newburn and Farnham 2008). Some of the bright spots could be footpoints to jets. The dark spots could be low albedo features or actual pits.

Cometary surfaces are under constant change due to outgassing activity. The second flyby of comet P/Tempel 1 by the NExT mission in 2011 just one perihelion passage after the Deep Impact flyby in 2005 showed clear change in surface morphology (Figure 17; Thomas et al. 2013a), although no obvious change in photometry was identified (Li et al. 2013). The most significant change observed was the retreat of the scarp bounding the smooth area near the south pole by at least 50 m. In addition, at least two crudely triangular areas evident in 2005 disappeared by 2011, representing a backwasting along at least 1000 m of the boundary. It is estimated that the total volume loss was about $2-4 \times 10^5$ m$^3$, corresponding to $8-16 \times 10^7$ kg assuming an average density of 0.47 kg/m$^3$ (Richardson et al. 2007). It has also been noted that there is concentrated jet activity originated near the scarps (Farnham et al. 2013). In addition to the boundary of the smooth features, several small bright albedo spots (<30 m) in the region have changed in contrast and extent. However, due to their small sizes, the changing viewing geometry, and the different imaging instruments, it remains uncertain whether these are real changes or are due to the effect of different illumination and viewing geometries, or instrument effects. Interestingly, no obvious ejecta blanket produced by the DI impact could be identified on the surface of comet P/Tempel 1 (Schultz et al. 2013).

The high-resolution images of comets P/Borrelly, P/Wild 2, and P/Tempel 1 reveal another common feature, i.e., ubiquitous surface layering that possibly extends into the interior. Based on these layers, Belton et al. (2007) hypothesized a "talps" model, or "layered pile" model to describe the internal structure of JFCs. In this model, the nuclei interiors are composed of layers of different thicknesses, sizes, and possibly compositions, that were accumulated during the primordial accretion phase of comets through low speed collisions between cometesimals. This hypothesis presents a completely different internal structure from the classic rubble pile model for asteroids (cf. Richardson et al. 2002, and references therein). While the evolution of asteroids mostly occurred in the inner Solar System and was dominated by intensive collisions after their formation, leading to their rubble pile nature, the evolution of comets probably took a completely different path. JFCs we see today were frozen for ~4 Gyr in the Kuiper Belt, i.e. where they are though to have formed, before being gravitationally perturbed into the inner Solar System (e.g., Duncan and Levison 1997; Duncan et al. 2004; Morbidelli and Brown 2004). Comets in JFC orbits are active for only 7% of the orbit (Duncan et al. 2004). Therefore the evolution of JFCs is first dominated by the collisional environment in the Kuiper Belt, and then on their surfaces by volatile sublimation during repeated perihelion passages. If talps are primordial in JFCs, then their preservation seems to indicate a much more benign collisional history than the asteroids have undergone. Future cometary missions, especially the CONSERT experiment on ESA's Rosetta mission that will use ground-penetrating radar to study the internal structure of Comet 67P/Churyumov-Gerasimenko (Schulz 2009), should provide definitive tests on the talps model.

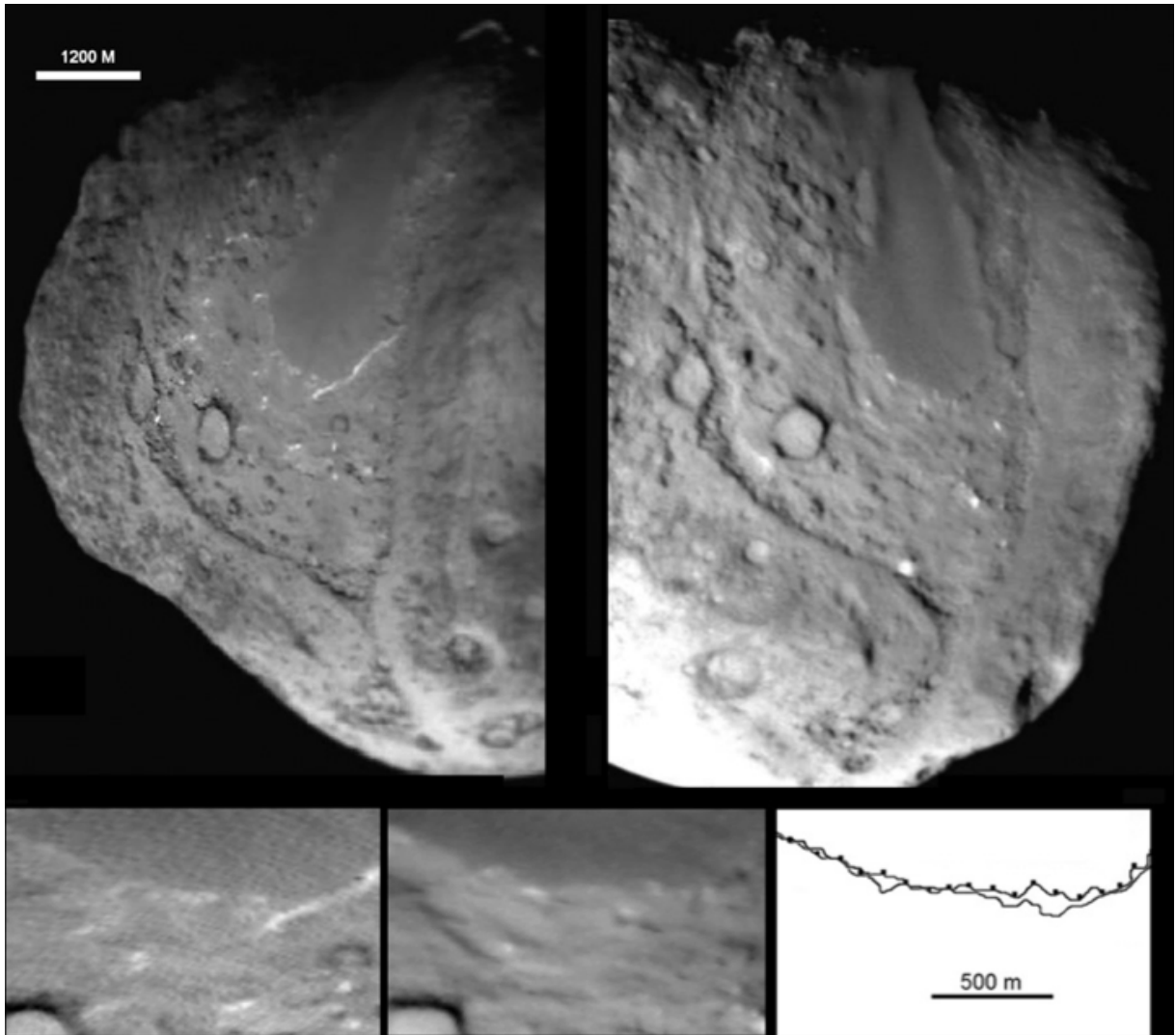

Figure 17. Changes in the surface of comet P/Tempel 1 from one orbit to the next. This is Figure 9 in the work by Thomas et al. (2013a; reprinted from Icarus with permission from Elsevier). The left image is from the Deep Impact flyby in 2005, and the right image is from the Stardust flyby in 2011. Various bright and dark spots have appeared and/or disappeared in the intervening six years. The edge of the scarp has clearly changed, losing up to 50 meters of extent in some locations. This change is shown in the bottom three panels, with the right panel showing the traces of the scarp.

**4. Structure, Interior, Spin**

4.1. Structure of Asteroids

From the discovery of the largest asteroid, (1) Ceres, in 1801 by Piazzi (Foderá Serio et al. 2002) until the 1990s, asteroids were largely unresolved point sources. Reflected light can reveal surface composition, and perhaps some level of structure through the thermal properties, but the interiors of asteroids were largely inaccessible to study. As described in the previous section, spacecraft missions can reveal exquisite detail of a body's surface structure and in some cases interior by means of gravity measurements. However, the number of missions will by necessity be limited, so studies of the population must rely largely on groundbased remote sensing.

Groundbased radar does reveal the shapes and surface details of near-Earth asteroids at few-meter resolution, and has led to some surprising discoveries about their structure. Ostro et al. (1989) summarize the early efforts to study the shapes and outer contours of NEAs. The two planetary radar facilities, Arecibo Observatory and Goldstone Deep Space Network, are complementary in their capabilities and ability to study solid Solar System bodies, primarily NEAs (Ostro et al. 2002 and references therein). Starting in 1998, the increased sensitivity of the Arecibo planetary radar system, along with the huge increase in the discovery rate of NEAs, began to reveal their detailed shapes. Some were the expected collisional fragments, with irregular facets, craters and ridges. However, some were surprisingly spheroidal, even at sizes of only a few hundred meters. The self-gravity of such small bodies is not nearly enough to overcome the strength of the rock, so these must be strengthless rubble-piles. However, other objects in the same size range were seen to be elongated, or rapidly rotating, which required a coherent rock. The diversity of NEAs is remarkable, and indicates a variety of formation mechanisms (Nolan et al. 2005, Ostro et al. 2002).

Radar imaging is accomplished by sending a monochromatic signal to the asteroid and receiving the signal which reflects from the solid surface. Because the precise timing and characteristics of the transmitted signal sent are known, the return signal reveals much about the body, depending only upon knowing the time and speed of light. We can use the time sampling and rotational velocity information to reconstruct an image of the asteroid, with some projection effects, and ambiguity in that several points on the asteroid surface may have identical distance and apparent velocity, and thus map to the same location in the radar image (see Ostro et al. 1993 and Magri et al. 2007 for more details).

The resolution of a radar image depends on the signal-to-noise ratio (SNR) of the signal reflected from the asteroid. The power of the transmitter, size of the telescope, and gain contribute to the SNR, but the dominant factor is usually distance from Earth. The SNR depends on the inverse fourth power of the distance, so NEAs dominate the sample of high-resolution images. About 300 objects have been imaged as of 2014, with about 50 of those at the highest resolution available at Arecibo, 7.5 meters per pixel. Before 2011, the highest resolution at Goldstone was 19-m per pixel, but currently for very close approaches to Earth, images from GSSR can reach 3.75-m per pixel as in asteroid 2005 YU$_{55}$. Examples of some radar images are shown in Figure 18. As described earlier, near-Earth asteroids are collisional fragments of larger objects that are then perturbed into planet crossing orbits (e.g. Bottke et al. 2002a). While some NEAs do indeed look like irregular collisional shards, many others do not. About 25% of the NEAs between 200-800 m are spheroidal, and another 10-15% are two spheroidal lobes joined together. Elongated objects, with axial ratios of 2:1 or more, comprise about 10% of the sample, while binary systems are another 15% of NEAs less than 10 km in diameter (Taylor et al. 2012). An asteroid less than 10 km can only be spherical from its own gravity if it is strengthless, although surface features such as craters and ridges can be supported by friction in the surface material. Surface features seen in radar images must be interpreted with caution, because the image is not a spatial image, but it does map into a spatial image. A radar image is really distance from the observer in one dimension, and rotational velocity in the other dimension, or distance from the rotation axis. By convention, radar images always appear as if the object is viewed from the north pole, illuminated from 90 degrees phase angle. From a single image, the location of a feature on the surface cannot be determined uniquely. However, a series of images over several days can be used to derive a shape model (e.g. Nolan et al. 2012, Magri et al. 2011, Ostro et al. 2005). More limited data sets can be used to identify surface textures and features which often reveal internal structure.

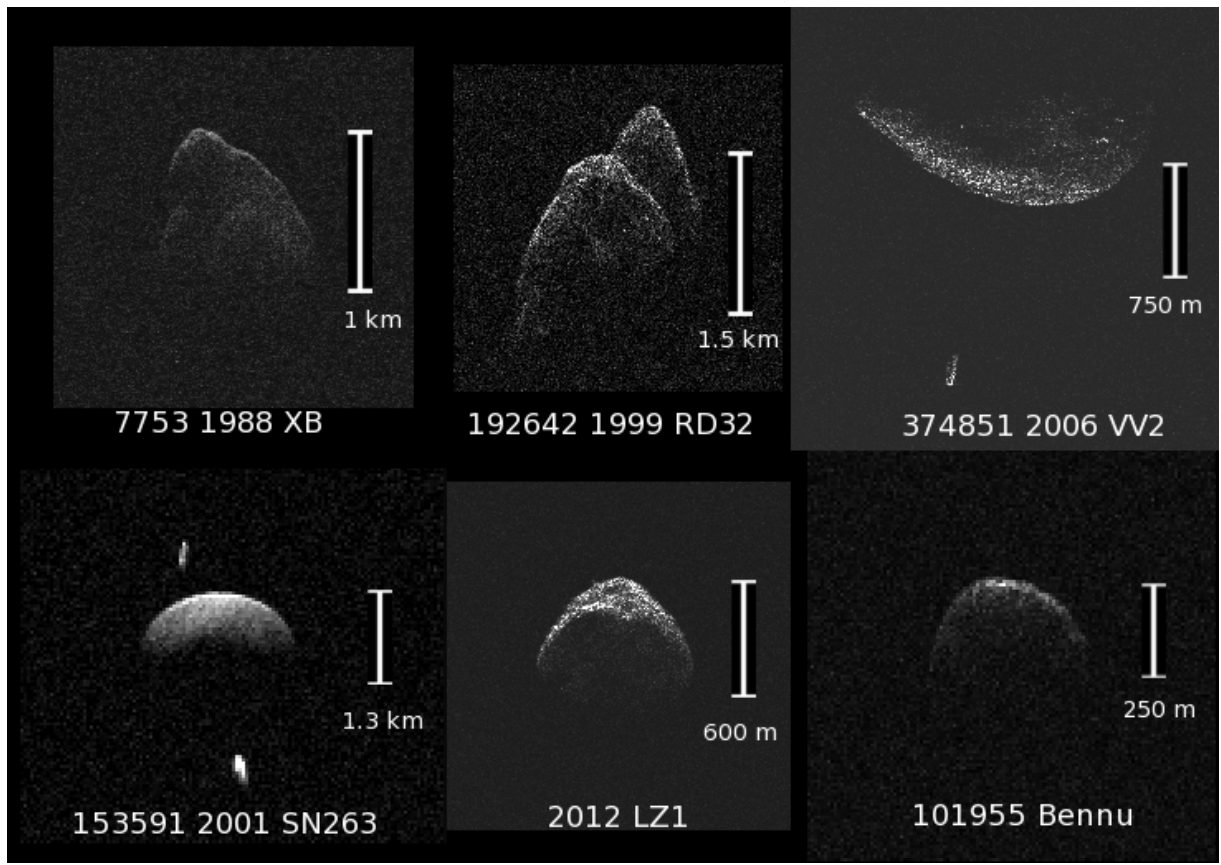

Figure 18. Arecibo radar images of a variety of asteroids can reveal surface features as small as 7.5 meters. Scale bars show the approximate size in the vertical direction. In each of these images, the distance from Earth increases from top to bottom (except for 2006 VV2, where it is the opposite), and Doppler frequency, or object rotation velocity in the line of sight increases from left to right. These are not spatial images, but for a rigid body, they map into a spatial image, and surface features can be interpreted readily with some projection effects.

The sample of objects shown in Figure 18 includes some notable objects. Asteroid (153951) 2001 $SN_{263}$ is the first triple asteroid system discovered among the NEAs, where the outer satellite is larger (below the primary in this image) orbiting in about 6.2 days, and the smaller inner one (above primary) orbits in about 16.5 hours. The OSIRIS-REx mission will visit asteroid (101955) Bennu (formerly 1999 $RQ_{36}$), a 550-m spheroidal asteroid, which although rotating relatively quickly (4.2 hours) does not appear to have a satellite larger than 10 meters at the present time. Both (374851) 2006 $VV_2$ and Bennu have one or more boulders on the surface larger than 10m, perhaps similar to those seen on Itokawa and Eros. The population of NEAs is varied in size and shape, composition and rotation, suggesting a variety of possible outcomes for the same formation conditions.

4.2. Spin of Asteroids

The rotation rate of an object can put constraints on its internal structure. Optical lightcurve observations of asteroids and comets have been carried out by both amateur and professional astronomers for many decades, and provide an extensive database with which to study the population of asteroids. The most reliable data from the asteroid lightcurve database compiled by Warner et al. (2009) – and which is frequently updated at http://www.minorplanet.info – is plotted in Figure 19 for main-belt asteroids, with H magnitude on the x-axis and rotation frequency (rotations per day) on the y-axis. The H magnitude is the brightness of an asteroid at 1 AU from both Earth and Sun, and at zero phase angle, so is a proxy for the asteroid diameter. The plot shows that for about H<20, a rotation frequency of 10, or about 2.4 hours rotation period is a limiting value for all asteroids. This corresponds roughly to the fastest rate at which a strengthless rubble pile structure can spin without flying apart (e.g. Pravec and Harris 2000). Smaller asteroids, (H>20) can break this spin barrier, and can rotate as rapidly as 30 seconds. This certainly requires some tensile strength, although some internal fractures and porosity can still be present. The spin limit depends on the material density, so it is slightly slower for icy bodies. No asteroids appear to be both small and rotating slowly, although these are difficult to measure using groundbased optical telescopes. Radar measurements should see small objects rotating slowly if they exist, but the distribution from radar observations matches the lightcurve distribution very closely, despite having very different observational biases.

4.3. Binaries

The discovery of the first binary near-Earth asteroid, (185851) 2000 $DP_{107}$, in 2000 (Margot et al. 2002) opened up a new window on studying interior structure of asteroids. Radar images allow the sizes of both components to be observed so the mass of the primary body can be derived from the orbit. Searches for binary asteroids had been inconclusive prior to this time, even leading to theoretical arguments as to why binaries did not exist and would not be stable (e.g. Gehrels et al. 1987). Lightcurve measurements suggested binary systems existed for decades, but the possibility of two-lobed, or contact binary objects could not be ruled out for most of the cases. Clear evidence of occultations and eclipses in asteroid lightcurves had not been confirmed prior to the discovery of 2000 DP$_{107}$. However, after the first radar detection of a NEA binary system, showing the clear separation of the components, the lightcurve evidence was then considered much more convincing.

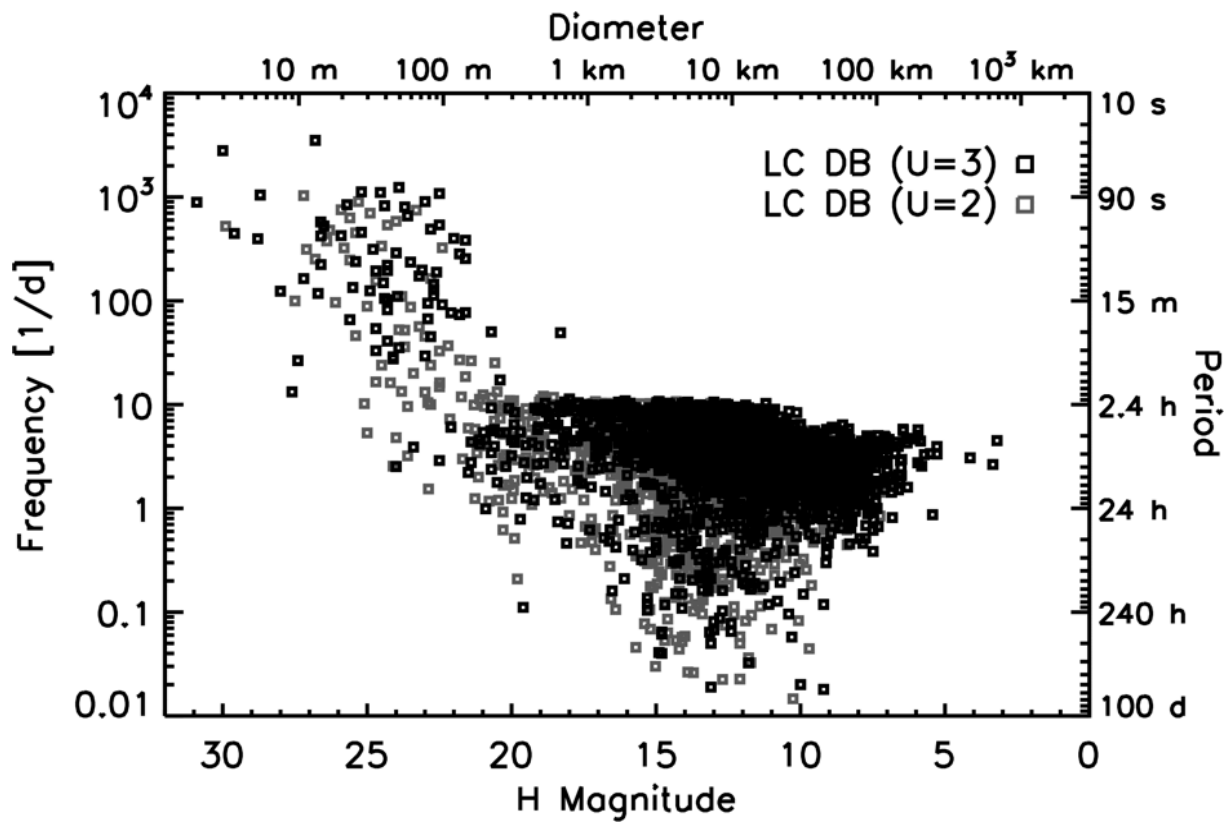

Figure 19. Distribution of the rotation rates of asteroids. Rotation rates are given as frequencies (left axis) and periods (right axis). The rates are shown as a function of diameter (top axis), which is estimated from the absolute magnitude (H; bottom axis) using an assumed geometric albedo of 0.20 (cf. Figure 5). Plotted here are the higher quality observations ("U" parameter of 2 or 3) taken from the asteroid lightcurve database (Warner et al. 2009), based on many nights of observation for each asteroid to determine the rotation rate from changes in the brightness of the object as it rotates. Note the sharp cut in the distribution at a rotation period near 2.4 hours. This rate seems to be a limit for asteroids larger than H magnitude of about 21, which has been interpreted as being the fastest that a strengthless, rubble-pile body can spin without breaking apart. Figure is courtesy Patrick Taylor.

As of 2014, two thirds of the approximately 50 known NEA binary systems have been discovered using radar, still the most powerful means of identifying binary systems. Lightcurve detection continues to be an important contributor, and is now showing binary systems are equally common in the main-belt population in the few km size range (Pravec et al. 2006). Yet another technique that works quite well for binary discovery in the main belt is direct imaging through the use of telescopes that have adaptive optics. Normally, the atmosphere distorts the wavefronts from an astronomical source that are collected by a telescope; turbulence in the atmosphere leads to the light coming through many cells of air that each have slightly different indices of refraction. When the telescope tries to bring all the light to a focus, the resulting image is blurry, usually far blurrier than would be expected from just considering the diffraction limit of the telescope's optics. For example, a 10-meter telescope at a good astronomical site, without the aid of adaptive optics, might normally achieve images with ~0.5 arcsecond resolution, whereas the diffraction limit (Rayleigh criterion) at visible wavelengths for such a telescope is about 0.01 arcsecond. "Adaptive optics" is the term for the technological feat of having a deformable mirror in the optical path (after the secondary mirror) that de-blurs the light by changing the light's reflection just enough to cancel out the effects of the turbulent cells of air. The mirror is deformed and re-deformed many times per second. An example of the incredibly good resolution that can be achieved through adaptive optics is shown in Figure 20, where one can clearly see a triple asteroid system, (216) Kleopatra (Descamps et al. 2011) .

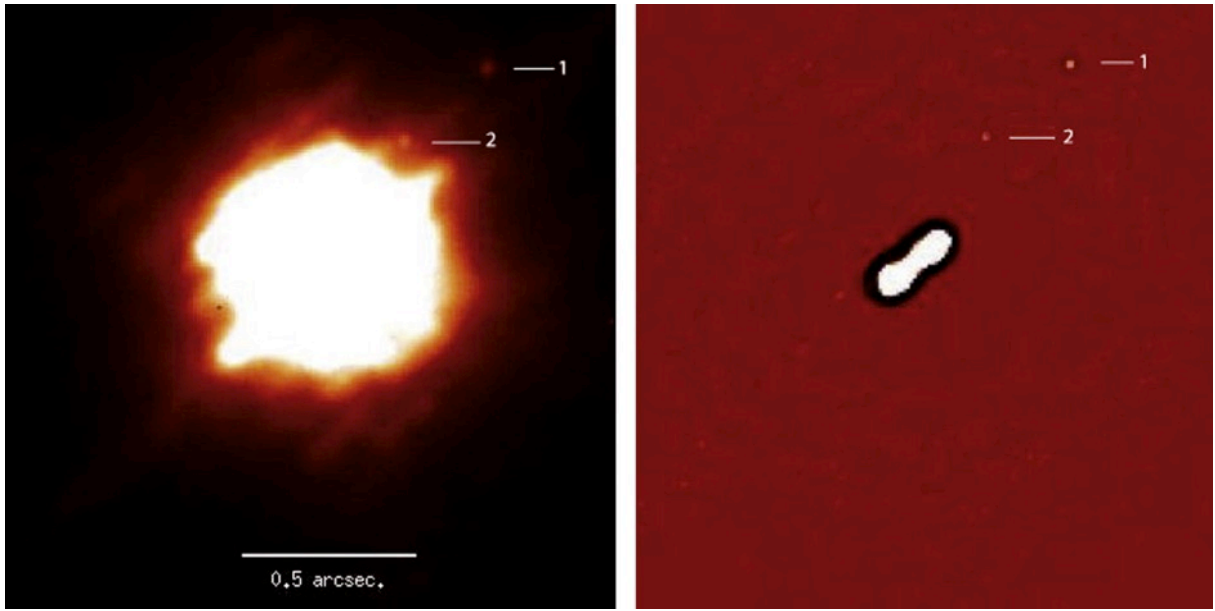

Figure 20. Adaptive optics image of asteroid Kleopatra as taken by Descamps et al. (2011) with the Keck II telescope atop Mauna Kea, Hawaii. This is their Figure 1 (reprinted from Icarus with permission from Elsevier). The panel on the left shows the original image, which has been heavily stretched to show the effects of diffraction. The panel on the right shows the image after some deconvolution and filtering and has a resolution of about 0.035 arcsecond. The elongated nature of the asteroid is evident, and Kleopatra's two moons are also clearly revealed. Note that the effective resolution in the right panel rivals the resolution obtained by (e.g.) the Hubble Space Telescope.

Following the discovery of binary 2000 $DP_{107}$ several formation theories were put forth. Tidal effects due to planetary encounters, collisional debris, disruption by collisions and re-formation as a binary, and rotational fission all seemed viable possibilities (Merline et al. 2002). However, after several more binary systems were observed, and their common features compared, it was clear that some of these formation theories could be ruled out.

All NEA binary systems have a fast rotating primary body, usually close to 2.5 hours, near the theoretical spin barrier for a rubble-pile structure (see Figure 19). The angular momentum of the system was well above the stability threshold, suggesting that spin-up and fission had occurred. The size of the secondary is remarkably consistent, between one-third to one-fifth of the primary diameter, and the orbit falls within 3-5 primary radii away. Many of the satellites are in synchronous rotation, and most are in very circular orbits, suggesting substantial tidal evolution. However, a few systems are not in synchronous rotation, and some satellites have orbital eccentricities of up to 0.15; one system, (153958) 2002 $AM_{31}$, has eccentricity near 0.45 (Taylor et al. 2013).

Ostro et al. (2006) and Scheeres et al. (2006) discuss the shape and orbital dynamics of the binary asteroid, (66391) 1999 $KW_4$, in exquisite detail from radar observations. The elongated shape of the satellite demonstrated libration, and material shifting on the surface due to the extremely low gravity explains the smooth central band of the primary. Many other asteroids have now been shown to have this same basic shape, not all of them binary systems, which suggests either that a satellite may have formed and was subsequently lost, or that this shape can be the result of spin up alone, and does not require the presence of a satellite to form initially. Future spacecraft missions to such objects may help answer this question.

The diversity of binaries among asteroid taxonomic types suggests that mechanical structure, and not composition, is the most important factor in forming binaries. Binary systems are found (e.g.) among S-complex objects, X-complex objects, and V-types (e.g. 2006 $VV_2$). In particular, some binaries share characteristics that point to formation by spin-up due to the YORP effect, in particular some V-types and Xe-types, especially those Xe-types that are part of the Hungaria family (the high inclination, low semimajor axis, high albedo group seen in the upper left of Figure 3). Note that Xe-types used to be known as E-types in the Tholen system, and may have enstatite surfaces.

The densities of NEA binaries, 1.0-2.0 $g/cm^3$, implies porosity of 30-50% if the grain density is 2.5-3.2 $g/cm^3$ for ordinary or carbonaceous meteorites assumed to be the parent body analogs for at least some of the S-complex and X-complex objects. Britt and Consolmagno (2003) measure porosities of meteorites and find that cracks and fissures on all scales give moderate porosities (20-30%) for seemingly solid samples. Asteroid porosities of 50% or even more are not inconsistent with many meteorite samples we have. A thorough compilation of asteroid porosities is given by Carry (2012) and shown in Figure 21.

The first asteroid shape model resulting from radar observations, that of (4769) Castalia (formerly 1989 PB), is a two-lobed structure, reminiscent of a molar tooth (Hudson and Ostro 1994). Many more asteroids have since been seen to have this same basic shape, not only NEAs, but also comet nuclei (Magri et al. 2011, Harmon et al. 2010, 2011). Unlike near-Earth binary systems, these contact binaries tend to be slowly rotating, and have nearly equal mass lobes (Benner et al. 2008). Jacobson and Scheeres (2011) suggest that when an object splits into nearly equal mass lobes, a stable binary system is difficult or impossible to form. The equal mass systems, (69230) Hermes (Margot et al. 2003) and (90) Antiope, are rare in the near-Earth and main-belt zones. In the Kuiper Belt, binary systems of nearly equal mass are more common, but the discovery biases are quite different from those in the other regions, as are the size ranges observable. The conditions that would lead to formation of a contact binary rather than a rapidly rotating binary with mass ratio 1:3 or more may not really be fundamentally different, but may instead be a stochastic process.

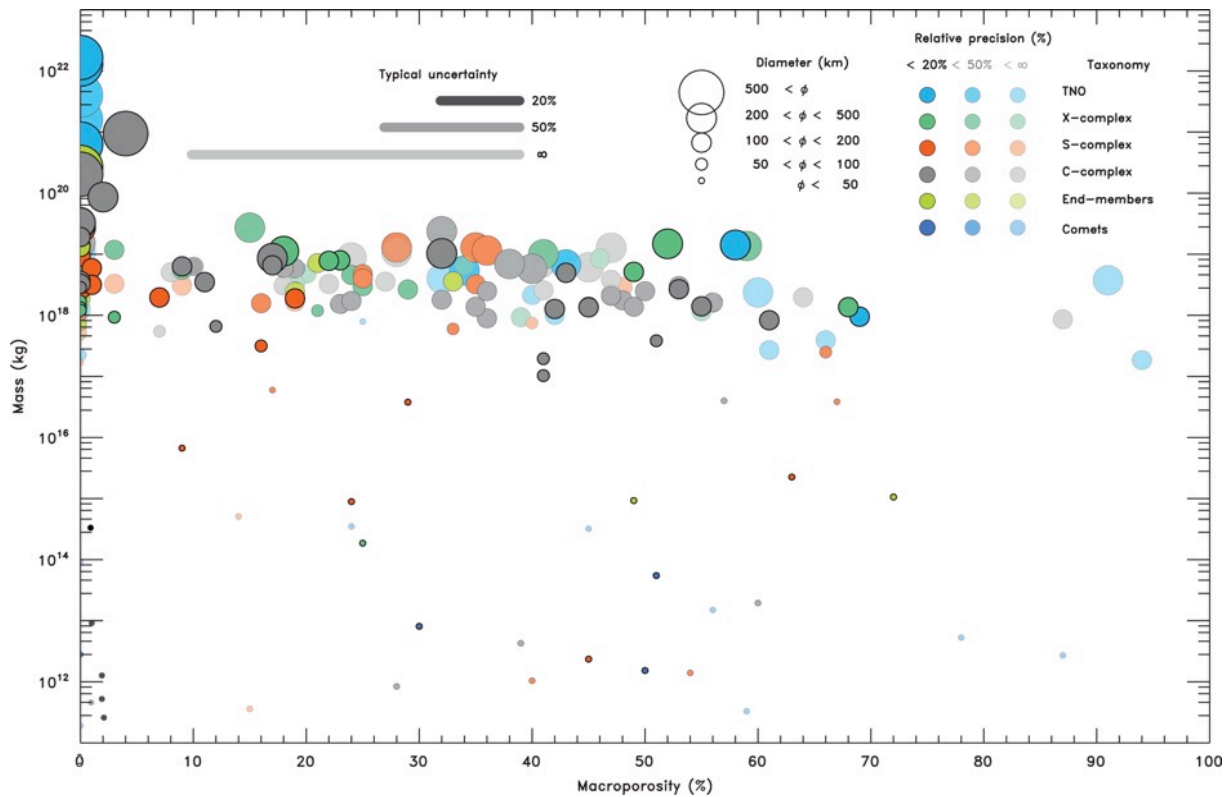

Figure 21. Scatter plot of macroporosities and masses of various small bodies as compiled by Carry (2012). This is his Figure 8 (reprinted from Planetary and Space Science with permission from Elsevier). Shading of each circle indicates the approximate uncertainty of the porosity measurement; Carry (2012) quantized the uncertainty as 20%, 50%, and "infinity" (i.e. very large). Most of these results are from the 1990s and later; it is only since then that we have come to realize that many small bodies have large porosity.

4.4. Cometary Properties

Comets lag far behind asteroids in terms of our understanding ensemble properties of structure and rotation. There are fewer available targets for study by radar, and rotational studies at visible wavelengths can be complicated by the fact that often the coma is much brighter than the nucleus. Radar delay-Doppler imaging of the sort that appears in Figure 18 has only been obtained on four comets: P/2005 JQ$_5$ (Catalina) (Harmon et al. 2006), 73P/Schwassmann-Wachmann 3 (Howell et al. 2007), 8P/Tuttle (Harmon et al. 2010), and P/Hartley 2 (Harmon et al. 2011). This last comet is shown in Figure 22, and one can compare the radar image of P/Hartley 2 to those shown in Figures 15 and 16.

In addition to the delay-Doppler imaging, basic radar spectra have been obtained from 11 other comets total (Harmon et al. 2004, Harmon et al. 2011), yielding constraints on nuclear sizes and on rotation rates. In some cases, the radar echoes contain a wide (in velocity space) skirt of flux. Since the wavelength of the radar is cm-scale, this skirt is thought to show echoes from very large grains (>2 cm diameter) in the near-nucleus coma. Typical cometary activity can only lift (through the drag force) grains up to cm-size; larger grains are usually too heavy and do not reach escape velocity (Grün and Jessberger 1990), although in the context of fragmentation, larger grains can be liberated (e.g. Howell et al. 2007). In any case the radar echo skirt is showing us the large end of the comet's dust coma, and can provide an important constraint on the total mass loss.

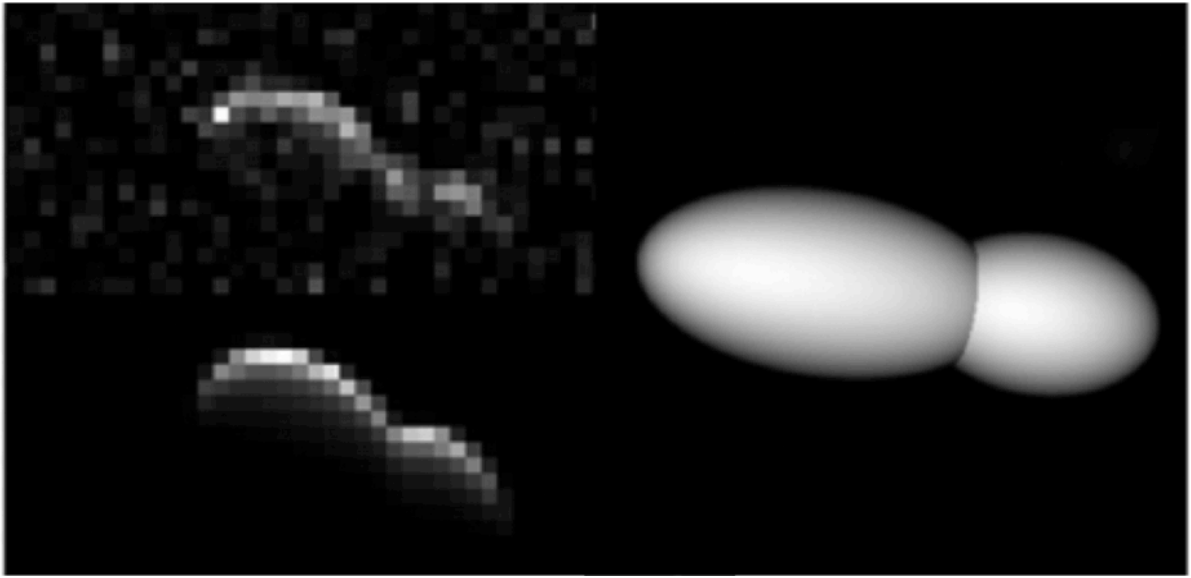

Figure 22. Radar echoes from and shape modeling of comet P/Hartley 2. This is Figure 2b from the work by Harmon et al. (2011; ©AAS, reproduced with permission). One of the delay-Doppler images of the nucleus is shown at top left, along with a simulated image from a shape model of two joined ellipsoids at bottom left. That shape model itself for the nucleus is shown at right. The long axis of the nucleus is about 2.3 km. Compare the model to the spacecraft imaging shown in Figures 15 and 16. Radar observations prior to the fly-by of the comet by the Deep Impact spacecraft helped to refine the position of the comet nucleus, and were critical to the success of the mission. The radar image sequence also helped to refine the rotation state. This is an excellent example of how groundbased observations provide important context for spacecraft missions.

A review of cometary rotation is provided by Samarasinha et al. (2004), and from 2004 to 2014 there have been only a few new rotation periods found for comets. The existence of excited (non-principal axis) rotation states has been noted in a few comets (e.g. Comet P/Halley; Belton et al. 1991), but it is not always easy to confirm such spin states. One important advance since the mid-2000s however has been the discovery that several comets change their rotation periods significantly on orbital timescales. This phenomenon was first confirmed for 10P/Tempel 2 (Mueller and Ferrin 1996, Knight et al. 2012), although it might have been seen in other comets earlier (e.g. Schleicher et al. 1991). Recently, changes in the rotation period have been noticed in comet 2P/Encke, P/Tempel 1, and P/Hartley 2 (Mueller et al. 2008, Chesley et al. 2013, Meech et al. 2011, Samarasinha et al. 2011). The change is on the order of 0.1% to 1%. For P/Encke, the change is a few minutes per orbit; for P/Tempel 1 and P/Hartley 2, it is on the order of 1 hour per orbit. The primary driver is a net torque exerted on the nucleus by the reaction force of its outgassing. The physical importance of understanding such short-term changes in the rotational state is that one can then try to understand how long it takes for the excited rotation state to damp out, an effect which depends on the bulk solidity of the nucleus itself – i.e. how much different pieces of the low-density nucleus rub up against each other. In other words, studies of cometary rotation states provide a clever way to try to understand the physical nature of cometary interiors.

The rapid changes that a comet can make to its own rotation period mean that in general a comet has no memory of its primordial spin state. An intriguing study – but likely only feasible far in the future – would be to look at the distribution of rotation states among long-period comets that are each on their first trip in from the Oort Cloud and before they actually turn on. In other words, how are the long-period comets spinning before they feel any torques from their own outgassing? Such observations would likely give us a window into a more primordial spin state distribution, or at least certainly a glimpse of what very long-term, evolutionary dynamical processes might affect the comets while they are in the Oort Cloud, tens of thousands of AU from the Sun. The difficulty is that there are virtually no long-period comets discovered that are new in the Oort sense that are inactive; such objects are simply too faint to detect with current technology.

Even such fundamental quantities as the mass and the density of a cometary nucleus have been exceedingly difficult to constrain. Spacecraft have in general been too far from each comet under study to have their trajectories gravitationally deflected. The ejecta cloud produced by the impactor experiment with the Deep Impact mission suggested that the density of comet P/Tempel 1 is very low, about 400 kg/m$^3$ (Richardson et al. 2007) although the quoted range is 200 to 1000 kg/m$^3$. The shape of comet P/Hartley 2's nucleus, like two ellipsoidal lobes, suggests that it might be showing an equipotential surface; if that is the case, Thomas et al. (2013b) derived a density of 200-400 kg/m$^3$. Before these missions, the most common way to estimate a comet's mass was through the use of its non-gravitational parameters, i.e. by making use of the changes to a comet's orbit caused by the reaction force of its own outgassing. Not only does outgassing change the spin state (as discussed above), but the orbit then deviates from a purely gravitational solution (Marsden et al. 1973). Some recent estimates have been made for P/Wild 2 (upper limit of about 600-800 kg/m$^3$; Davidsson and Gutierrez 2006) and P/Borrelly (100-300 kg/m$^3$; Davidsson and Gutierrez 2004). Other estimates using the fact that the density must be sufficient to keep a comet from splitting apart due to its rotation are in the range of about 200 – 500 kg/m$^3$ (Davidsson 2001). While it is wise to keep in mind that some of these analyses are model-dependent, the results are consistent with the idea that cometary nuclei are underdense and very porous bodies.

**5. Composition**

5.1. Basic Premises

It is commonly assumed that because the comets we see today have spent most of their lifetimes in an environment too cold for most geochemical processes to occur, and are too small to differentiate or retain heat of formation, that to first order the gas that comes off pristine comets will tell us about the compositional and chemical environment of the young Solar System's protoplanetary disk. The situation is more complex for asteroids which may have suffered more processing due to internal heat and in some cases differentiation, but some primordial information can still be inferred.

As discussed earlier, the comets we see today are vestiges of the icy planetesimals that were formed during the era of planetary formation. When such a planetesimal was created, while the primary component was water, many icy species ($CO$, $CO_2$, $CH_3OH$, $CH_2O$, etc.) were incorporated into its body. The relative amounts of these species were a direct sample of the chemical and thermal properties of the protoplanetary disk at that time and at that location (or perhaps the locations sampled by the planetesimal if scattering occurred while it was accreting).

Since the volatile content of asteroids in general does not outgas, but is locked in the minerals, it is much more difficult to ascertain compared to comets. Except in the case of sample return, asteroid composition is determined either by analysis of meteorites (which may have lost some of their volatiles during atmospheric entry or due to weathering after reaching the ground), or by modeling the compositional diagnostics in the reflectance and emission spectra of the surfaces. These multi-component fits are highly dependent on the typical grain size of the regolith on the surface, making it difficult to distinguish between, for example, adsorbed water in layer phyllosilicates, OH groups on hydrated silicates, or actual water on clay grains. Basically, most remote sensing observations – even by a spacecraft that might be considered to be "in situ" – are only sensitive to the top most microns of an object's surface, leaving most of the bulk of the object itself unsampled. Radio and radar observations can penetrate deeper – several wavelengths, so roughly a few decimeters – but there are fairly restrictive detectability limits in that wavelength regime.

As mentioned in Section 1, some asteroids have been observed to have outbursts of dust, showing something like a cometary coma and tail. While some of these objects have merely suffered a collision and thrown off dust as debris from an impulsive impact, several objects have been observed to have sustained dust production over weeks or months. Presumably such activity is driven by water (although this has yet to have observational confirmation) and rather than the sharp classifications of icy comets and rocky asteroids it seems likely there is a spectrum of compositions between the two endpoints. The active objects with asteroid like orbits have been dubbed the "Main Belt Comets". The prototype MBC, with the comet designation 133P/Elst-Pizarro but also the asteroid designation (7968), is shown in Figure 23. Dynamical modeling of the long tail shows that the cause of the ejection of dust from the object could not have been impulsive; it must have lasted for months, and therefore is likely to be water-sublimation driven. Furthermore, P/Elst-Pizarro has shown activity consistently now at nearly the same orbital longitudes for four orbits (Hsieh et al. 2013). There are currently less than ten known MBCs, but surveys are on-going and it is expected more will be discovered. Statistical analysis of asteroid populations has suggested there may be on the order of a few hundred to a few thousand MBCs (Gilbert and Wiegert 2010; Hsieh 2009; Bertini 2011). The existence of these objects has already transformed the way we think about the formation of planetesimals and mixing in the early Solar System (see e.g. Bertini 2011). As we move from the discovery phase to determination of composition of these objects they should prove to be a valuable tool in constraining protoplanetary chemistry.

5.2. Volatiles

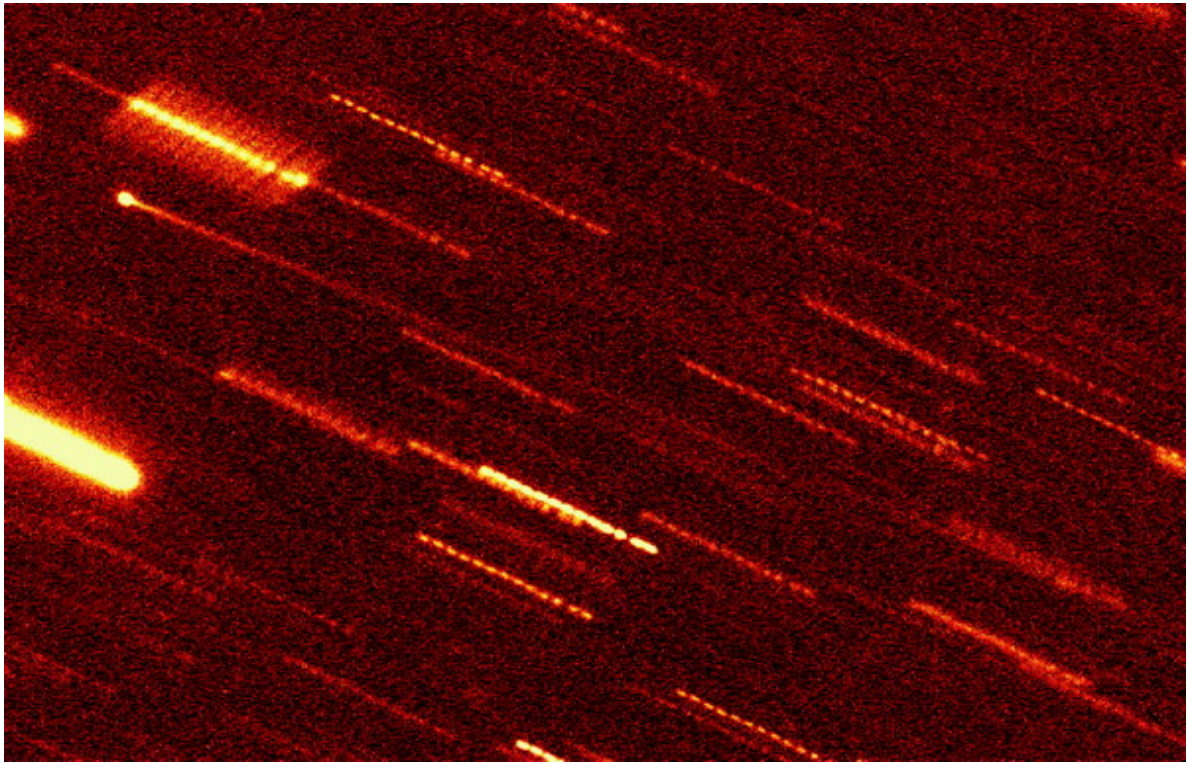

Figure 23. Image of Comet 133P/Elst-Pizarro, a.k.a. Asteroid (7968), a.k.a. the first "Main Belt Comet," as observed by Hsieh et al. (2004). This is their Figure 4 (©AAS, reproduced with permission). The comet is the continuous object with the bright spot (the comet's head) in the upper left and the long tail extending down toward lower right. Other dotted sequences of spots in the image are background stars; note that the comet's tail is in a different orientation from the star trails. The comet's tail extends several arcminutes, and is continuous, giving strong evidence that the object was continuously active for months on end, and thus being driven by cometary activity. While about ten objects in the Asteroid Belt have shown extended emission from dust, about seven (including P/Elst-Pizarro) show evidence for comet-like activity; the others were likely the result of impulsive events such as from an impact event. The persistence of ice to drive such activity has significant implications for our understanding of small-body evolution.

As discussed in section 1, the primary difference between comets and asteroids is that the composition of comets is dominated by volatiles, while asteroids are dominated by rock forming minerals but should have volatiles in increasing proportion with distance from the Sun. As comets approach the Sun their volatiles begin to sublimate, likely from the outer 1-2 m (Prialnik et al. 2004). This produces a coma that is usually bright enough to make it difficult to discern the nucleus itself from Earth-based observations. Thus, frequently the nucleus's properties are inferred from studies of the coma. From observations of gas in cometary comae, we can examine the distribution, composition and rate of gas production from the nucleus. Modeling of these properties can yield for example the rotational state of the nucleus and lower limits on its size (e.g. Schleicher and Woodney 2003, Samarasinha et al. 2011). Some gases sublimate directly from the nucleus (parent) and others are produced in the coma (daughter), either by chemical reactions or, more commonly, by photodissociation of the original molecules. As water is the most abundant volatile, all other volatiles in comets are typically measured by abundance relative to water or its direct photo-fragment OH.

The various volatile species found in comets have a wide variety of sublimation temperatures, which means they have the ability to begin outgassing at a variety of heliocentric distances. CO is the most volatile with a sublimation temperature of 25 K, though the somewhat more abundant $CO_2$ with its sublimation temperature of 80 K and ability to begin outgassing around 13 AU, is believed to be the likely driver of most distant comet activity (Meech and Svoren 2004). Water gradually takes over as the primary driver of cometary activity as heliocentric distance decreases. This variation in onset sublimation temperature leads to variations in abundance with heliocentric distance for some species (Figure 24, Biver et al. 1997, 2002). This is not the only reason abundances can vary with heliocentric distance, for instance the variation in the HNC/HCN ratio has been shown to be due to chemistry in the coma (Irvine et al. 1998, Rodgers and Charnley 1998).

Chemical abundances of volatiles vary from comet to comet, but when large numbers of comets are observed, statistically significant groupings appear, thus much work from the 1980s to the 2010s has gone into developing and understanding compositionally based taxonomies. The first of these was done with daughter fragments OH, NH, CN, $C_3$, and $C_2$ as these are readily observed in the optical, often with smaller aperture telescopes (~1-2 m), owing to their electronic transitions at visible wavelengths. A'Hearn et al. (1995) observed 85 comets, discovering that a significant fraction of Jupiter family comets (those originating in the Kuiper Belt) were depleted in carbon-chain molecules – $C_2$ and $C_3$ – relative to comets originating in the Oort cloud. This is shown in Figure 25, which is taken from the work by A'Hearn et al. As of 2013 this work has been extended to 101 comets, and principal component analysis continues to show that that the carbon-chain depletion is not associated with evolution, but is primordial in nature (D. Schleicher, private communication). Additional optical surveys have been done spectroscopically. Those done by Fink (2009) and Langland-Shula and Smith (2011) also divide their comets into taxonomies and agree with the A'Hearn et al. results where they overlap on carbon-chain depletion.

To determine composition one would ideally observe the species that sublimate directly from the surface – i.e. the parent species -- rather than the chemical fragments. Most of these species, however, only have spectroscopic transitions in the infrared (vibrational and rovibrational) and radio (rotational). Some important species – most significantly, $CO_2$ -- have transitions that occur at wavelengths where the atmosphere is effectively opaque. Such species can only be observed from space. Emissions from many parent species at radio wavelengths tend to be faint enough that they can only be observed in the brightest comets. Despite these challenges the database of observed comets at these wavelengths continues to grow and taxonomies of parent species are beginning to emerge. Figure 26 from the work of Mumma and Charnley (2011) shows all of the species thus far observed. Radio surveys of 14 different species, 10 of them primary, in more than 40 comets have revealed no significant groupings related to dynamical origin (Biver et al. 2002, Crovisier et al. 2009).

The late 2000s and early 2010s have seen a burst of detections of $CO_2$, in over 40 comets, thanks to spacecraft observations, both in-situ and remote (e.g. Combes et al. 1988, Crovisier et al. 1997, Feaga et al. 2007, Ootsubo et al. 2010, Reach et al. 2013). The abundance of $CO_2$ relative to water varies from ~3-30% when measured within 2.5 AU of the

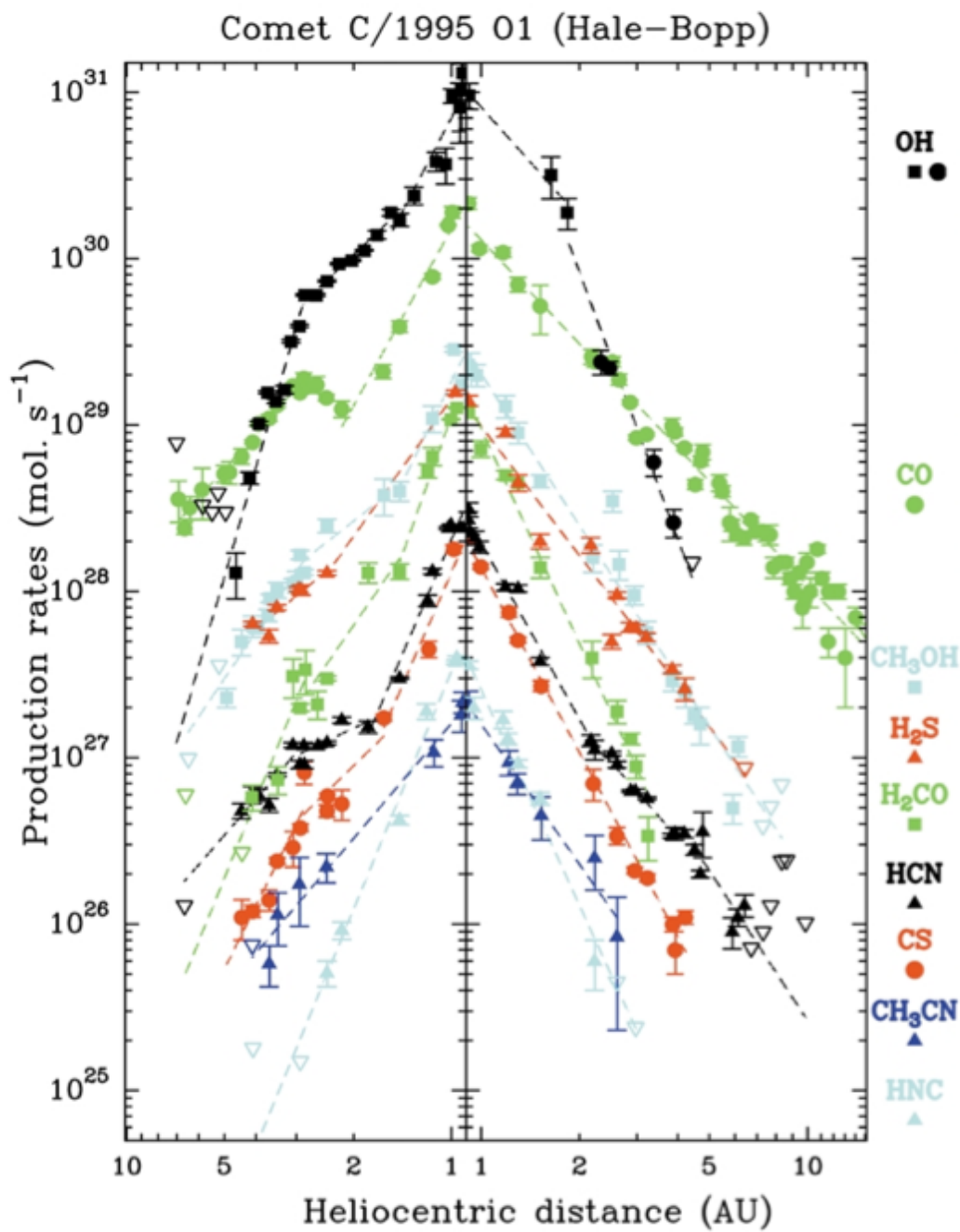

Figure 24. The so-called ``Christmas tree diagram," i.e. the production rates from Comet C/1995 O1 (Hale-Bopp) as observed using radio telescopes over the course of several years in the late 1990s and early 2000s. This is Figure 5 from the work by Biver et al. (2002; with kind permission from Springer Science and Business Media). This was the first time that comet scientists were able to follow the production rates of so many species for such a long time interval, monitoring how their relative abundances changed as the thermal wave from the Sun penetrated to the interior and as various ices sublimated.

Sun (i.e., where water-driven outgassing is most vigorous), and no abundance variation based on primordial origin is observed (Mumma and Charnley 2011).

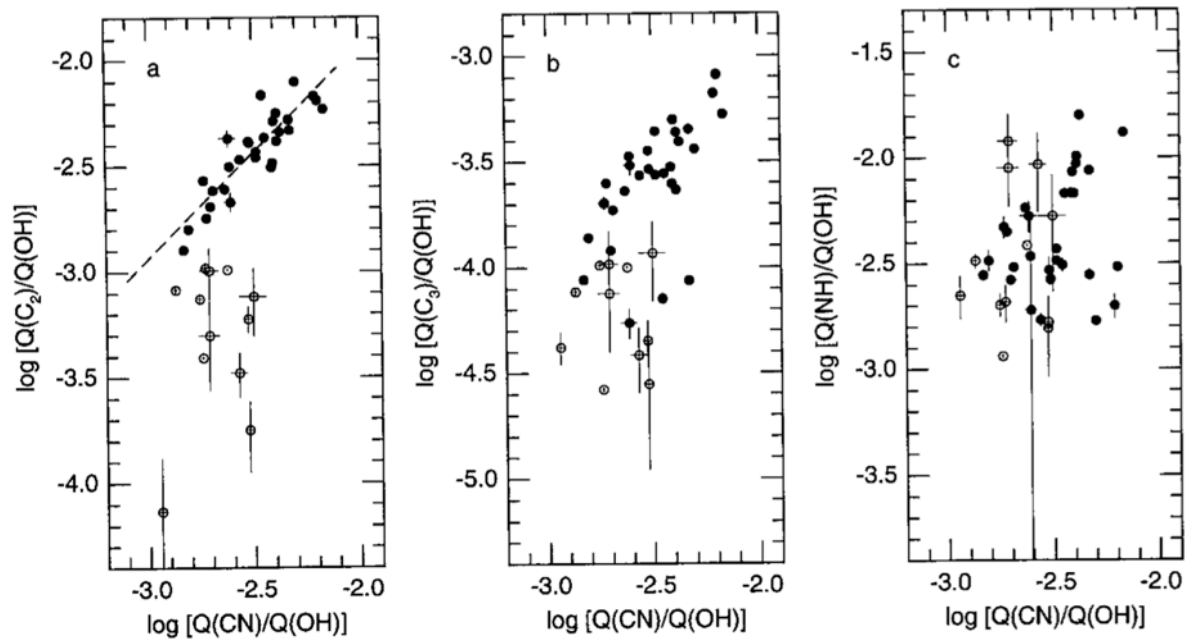

Figure 25. Correlations between relative abundances of daughter species in the comae of comets. This is Figure 10 from the work by A'Hearn et al. (1995; reprinted from Icarus with permission from Elsevier). *Q(X)* refers to the production rate of species *X*. The strong depletion of $C_2$ and $C_3$ in some comets – denoted by the open circles – indicates a significant heterogeneity in the comet population. For the "typical" comets (filled circles), the abundance of $C_2$ and CN (relative to water) is about the same (left panel), with the abundance of $C_3$ about a factor of 10 lower (middle panel). The dichotomy of abundance ratios does not extend however to NH (right panel).

Many of the organic species shown in Figure 26 are seen, as mentioned, at infrared wavelengths through the use of high-resolution, cross-dispersed echelle spectrometers. Such instruments can observe the vibrational transitions of many species at once. This has the advantage of there being no uncertainties in relative abundances due to rotational variation. In the infrared, water, the species to which all others are compared, can also be measured directly. Species included so far in surveys include $H_2O$, CO, HCN, $CH_3OH$, $H_2CO$, $CH_4$, $C_2H_2$, $C_2H_6$, OCS, and $NH_3$ (DiSanti and Mumma 2008, Bockelee-Morvan et al 2004, Mumma et al 2003). These surveys, which have included 26 comets so far, find three groupings based on organic abundances: organic-normal, organic-enriched and organic-depleted. Interestingly, unlike the optical photo-fragment based surveys mentioned above, there is no clear relation between composition and origin, since comets of both Kuiper Belt and Oort Cloud origin occur in all groups. Work to increase the sample size and improve statistics is on-going (Mumma and Charnley 2011).

The isotopic ratios of a number of elements in comets is of interest because they are considered signatures of the environment in which the Sun formed. The D/H ratio is one of the most sought after due to the question of how much comets contributed to Earth's water. The six Oort cloud comets measured before 2010 all had D/H ratios much higher than Standard Mean Ocean Water (SMOW). The first measurement of a Kuiper Belt comet was made by the Herschel space telescope and P/Hartley 2 was observed to have the same D/H ratio as SMOW to within the errors (Hartogh et al. 2011). This detection was followed by an upper limit for another JFC, 45P/Honda-Mrkos-Pajdušáková, which was consistant with the low D/H observed in P/Hartley 2 (Lis et al. 2013). The enigma of these observations is that the deuterium in comets originating in the Kuiper Belt was expected to be enriched relative to that of the isotropic comet population (Kavelaars et al. 2011). If the protosolar nebula began with enriched D/H similar to dense molecular clouds (~ $1 \times 10^{-3}$, Butner 2007), and the reactions that reduce that ratio in the protoplanetary disk decrease with distance due to decreasing temperature, density and turbulent mixing with distance (eg. Drouart et al. 1999, Mousis et al. 2000 and Aikawa and Herbst 2001), then it would be expected that the D/H ratio would increase with increasing distance. While this led Hartogh et al. (2011) to conclude that it was possible that either these models could be incorrect or that P/Hartley 2 could be an escaped Trojan (since Trojans possibly formed closer to the Sun where D/H ratios would be more depleted), the discovery of a second depleted JFC makes it even more important to measure the D/H ratio in a greater sampling of this population to determine whether this is a real taxonomic difference and also whether this can constrain our models of Solar Sytem formation. Additionally, this question has significant implications for the delivery of water to the inner Solar System. A thorough review of D/H observations and challenges can be found in work by Mumma and Charnley (2011).

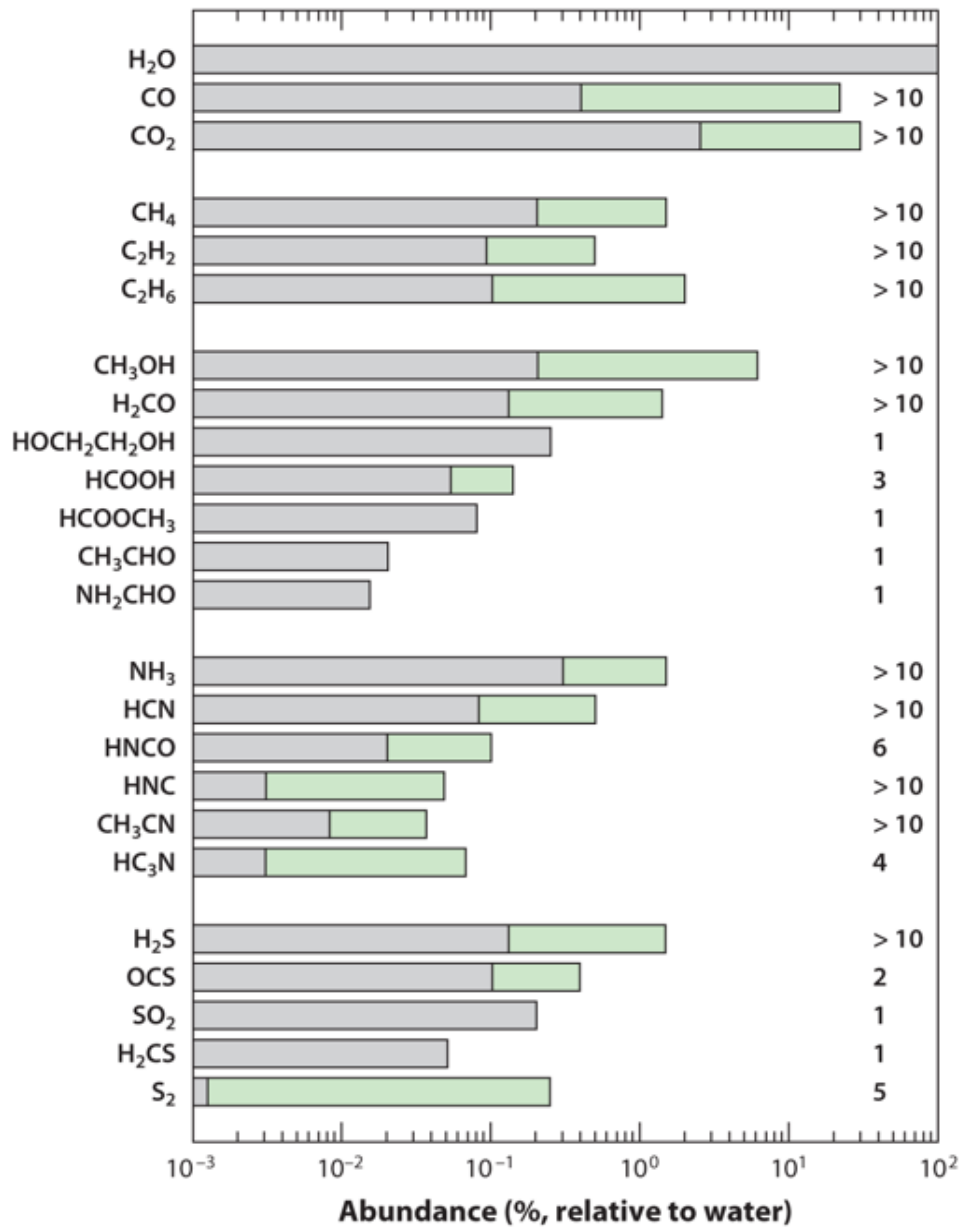

Figure 26. Relative production rates of various volatile species in comets. This graph is Figure 4 from the work by Mumma and Charnley (2011; reproduced from Annual Reviews of Astronomy and Astrophysics with permission from Annual Reviews). All abundances are scaled to water, which is set at 100%. The red segment shows the range among all measured comets; the numeral at the right edge gives how many comets in which the production rate of that species has been measured. After water, carbon dioxide, carbon monoxide, and methanol are the next most abundant species. Note the large variation in diatomic sulfur.

Since studies of comet composition make use of the coma and not the actual cometary nucleus itself, it is reasonable to wonder if the composition of the coma really can tell us the composition of the nucleus. More specifically, how do we know that the layer in the nucleus from which the coma gases originate is truly indicative of the interior composition? How much heterogeneity is there?

An important clue was provided by the Deep Impact spacecraft's visit to comet P/Tempel 1. The mission used an impactor to excavate a crater, yielding an opportunity to compare the composition of typical outgassing from surface layers, to that of freshly excavated deeper layers as much as 20 m below the surface (Richardson et al. 2007). Measurements from both the ground and the spacecraft showed to within the uncertainties no change in abundance in most chemical species between the typical near surface outgassing and the ejecta plume from the excavated crater. Parent species $CO_2$ (Feaga et al 2007), CO (Feldman et al. 2006), HCN (DiSanti et al 2007), as well as daughter fragments (photodissociation products of the larger parent molecules) NH, CN, C3, CH, NH2 and C2 (Cochran et al 2007) all retained the same relative abundance to water. So at the 0 to 20-meter depth level, the degree of heterogeneity in comets – at least in comet P/Tempel 1 – seems to be low.

The topic of heterogeneity came to the forefront again after the spectacular data returned by the Deep Impact spacecraft as it flew by P/Hartley 2. Heterogeneity is demonstrated in Figure 27, which shows various images of the comet as constructed from the extensive infrared spectroscopy that was obtained. By looking at specific wavelengths in the spatially-resolved spectra, spectral maps could be created. This particular figure is from the work of A'Hearn et al. (2011). The maps show that the sublimation of water was predominantly coming from the central region of the nucleus, as seen in the bottom left panel labeled "$H_2O$ vapor." However, the reflected sunlight off the dust (top left) and the emission from organic species (middle right) are almost entirely spatially uncorrelated with the water vapor. Instead, they are strongly correlated with the $CO_2$ emission as seen in the top right panel, indicating that $CO_2$ at one end of the comet is driving much of P/Hartley 2's activity. Indeed, the $CO_2$ is even pulling off chunks of water ice (bottom right panel) into the coma.

However the heterogeneity question is not yet fully understood, since there is good evidence that some comets are homogeneous perhaps all the way through – in particular comet P/Schwassmann-Wachmann 3, which was observed extensively in 2006 after it had broken into several large pieces during the previous two apparitions in 1995 and 2001. The gas and dust compositions in the comae of the comet's fragments were basically identical (Dello Russo et al. 2007, Sitko et al. 2011). There was no indication that the two main pieces of the comet (fragments B and C) had radically different makeup. Both comet P/Schwassmann-Wachmann 3 and P/Hartley 2 are Jupiter-family comets, but the lack of heterogeneity in P/Schwassmann-Wachmann 3 suggests that the two comets could not have been formed from material that had experienced the same amount of mixing in the protoplanetary disk. Of course, there are few comets with detailed enough measurements to make a claim about heterogeneity, so this topic requires more data before broad generalizations can be justified.

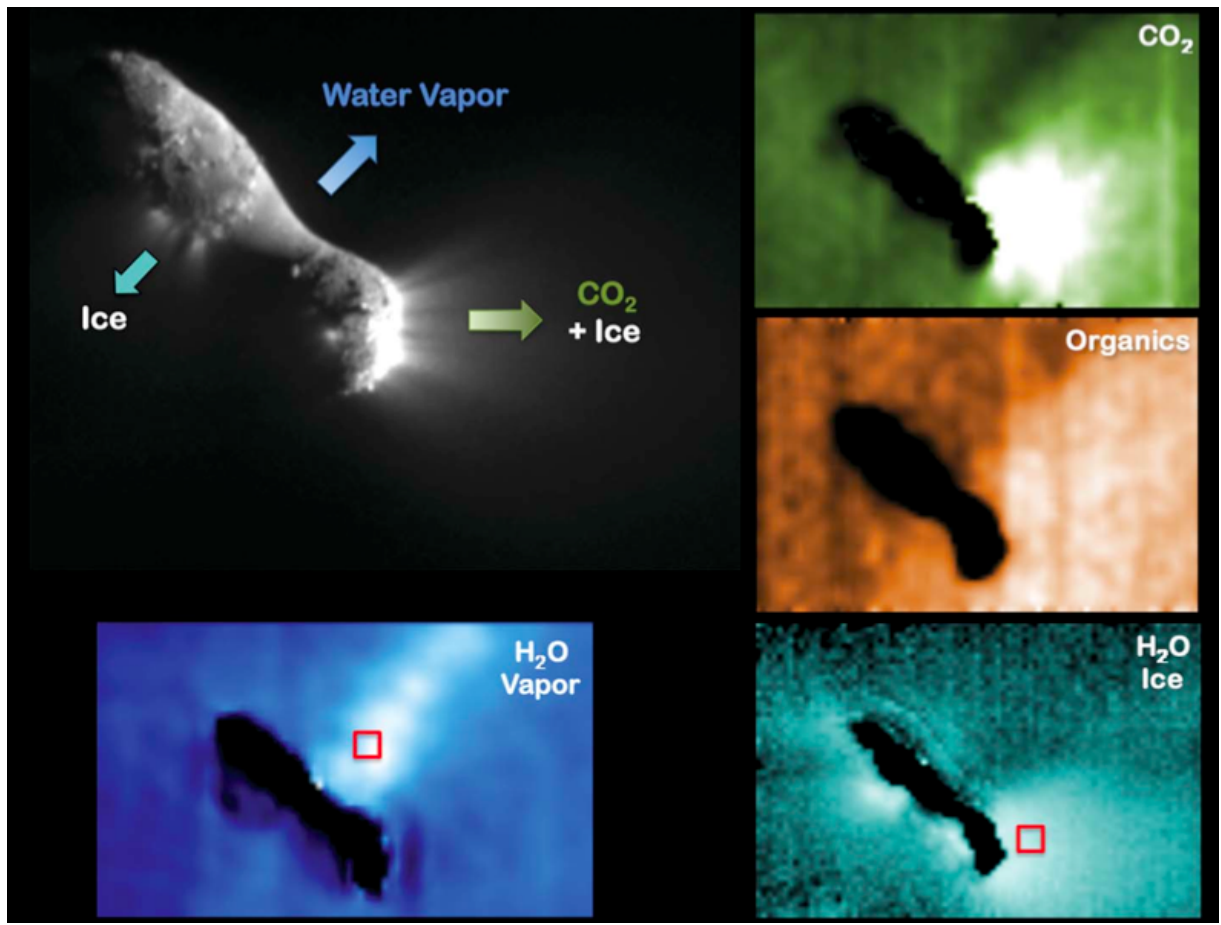

Figure 27. Heterogeneity in the nucleus of comet P/Hartley 2 as imaged by the EPOXI mission. This is Figure 5 presented by A'Hearn et al. (2011; reproduced from Science with permission from AAAS). The flyby of this comet revealed for the first time the significant role of carbon dioxide in driving cometary activity in a Jupiter-family comet in the inner Solar System. Top left panel: context image of the comet in visible-light continuum. Bottom left panel: Water vapor map, showing that most of the water is sublimating from the center of the comet's nucleus. Top right panel: $CO_2$ vapor map, showing that most of the carbon dioxide is sublimating from one of the ends of the nucleus. Middle right and bottom right panels: Maps of organic emission (C-H stretching bonds) and water ice absorption which clearly show the vastly stronger correlation with the $CO_2$-driven activity rather than the water-driven activity. The long axis of the nucleus is about 2.3 km.

The volatile content of MBCs is a new development that has taken on added importance with the claim of water ice on the surface of asteroids (24) Themis and (65) Cybele (Rivkin and Emery 2010, Campins et al. 2010, Licandro et al. 2011). While it was certainly true that the primitive bodies in the outer main belt were thought to have appreciable water content, the observational results provide confirmation. Note however that the existence of water-bearing minerals on asteroids – as evinced by the 3-micron absorption band – has been known since the late 1970s, when it was first seen on Ceres by Lebofsky (1978). Interestingly, there is evidence of transient or variable OH coming from Ceres, first reported by A'Hearn and Feldman (1992) using IUE, and then later by Küppers et al. (2014) using Herschel. In particular, the Herschel observations have shown evidence of OH while Ceres is near perihelion, and the amount is seen to vary with Ceres's rotation, suggesting specific source regions on the surface. Detection of water and/or ice in the outer part of the Asteroid Belt has also motivated some theoretical work on understanding how volatiles – in particular, water ice – can survive in the Asteroid Belt for billions of years (e.g. Schorghofer 2008). The key seems to be having low thermal inertia so that diurnal and annual heat pulses from sunlight simply cannot make it down deep into the interior of the bodies. Thus the natal water ice can survive for long periods of time, and is available when, for example, an impactor strikes the body, clearing away the topmost layers and exposing the underlying ice to space and to the telescopes of astronomers.

The longevity of water ice and the poor thermal coupling between volatiles and rocks is also exemplified by the findings of the Deep Impact spacecraft on both comet P/Tempel 1 and P/Hartley 2, where surface ice was seen to cover a small percentage of each nucleus's surface in locations where ostensibly the local temperature of the rock was well above the water-ice sublimation temperature (Sunshine et al. 2006, Groussin et al. 2013). While this ice is likely not primordial, but probably a recondensation of settled water that was formerly in the gas phase, it is remarkable that cold ice could so clearly be seen surrounded by much hotter rock. Furthermore, the fact that $CO_2$ was so abundant in as small a comet as P/Hartley 2 – one might have expected such a highly-volatile species to have been completely baked out of the comet – lends further credence to the idea of cometary ice being decoupled from cometary silicates, probably helped by high porosity.

Much of the water content of the asteroids is studied through its existence within hydrated minerals, largely phyllosilicates. Studying the hydration is important for understanding what was going on in the young Solar System such that liquid water could actually be in the physical presence of some of these minerals. In particular, the absorption band near 3-microns has been studied extensively in many asteroids, as mentioned above, going back over 30 years (e.g. Lebofsky 1980). The aqueous alteration of minerals is seen in many primitive asteroids, but it is important to note that not all of them have it (e.g. Barucci et al. 1998, Rivkin et al. 2002, Rivkin 2012). Interestingly, some of the asteroids that show evidence of hydration – e.g. the so-called M-types (under the Tholen taxonomy) – were once thought to be largely metallic asteroids, and would have to have been heated to quite high temperatures. The presence of hydrated minerals and the lack of FeO seem to contradict this scenario. Radar observations do not support most of the M-types having significant iron metal, at least near the surface (Shepard et al. 2010, 2015). However, iron-nickel meteorites must come from somewhere, and the parent bodies of these are still under debate (e.g. Emery and Lim 2011). In the Bus-DeMeo taxonomy, the former M-types now fall within the X-complex, distributed among several taxa, and are apparently a compositionally heterogeneous group.

5.3. Mineralogy

As stated in section 1.5, Figure 2 has a sampling of the spectral diversity among the asteroids. Note that the figure is not exhaustive; the overall slope of an asteroid's spectrum can change depending on how much weathering it has suffered, so in some cases a single type of asteroid can actually encompass a range of spectral slopes.

In any case, the figure demonstrates that many asteroids have absorption features that are diagnostic of the surface composition; the most prominent and commonly seen ones are the deep features near 1.0 and 2.0 microns, commonly referred to as "Band I" and "Band II," respectively. Both absorption bands are due to pyroxene, while olivine also contributes to Band I. The band centers and band areas can be used to place some constraints on the mineral abundances of Ca and Fe in the pyroxenes, of clino vs. ortho pyroxene, and of olivine (e.g. Gaffey et al. 2002). However, it is important to note that the mixing of various different particle sizes and compositions can lead to non-unique interpretations of observed spectra.

Another frequently seen absorption feature occurs near 0.7 microns, which seems to arise in asteroids with phyllosilicates on the surface (Vilas 1994). The oxidized iron within the hydrated mineral undergoes a charge transfer from doubly to triply ionized (i.e. $Fe^{2+}$ to $Fe^{3+}$). As might be expected, the existence of this feature is correlated with an absorption feature near 3 microns that is also indicative of hydrated minerals (e.g. Rivkin et al. 2002, Rivkin 2012); the visible-wavelength feature however is much easier to observe.

Direct sampling of asteroid mineralogy has now become possible with the success of the Hayabusa mission, which returned samples from the surface of Itokawa. In particular, one major result was that the strongly-suspected link between S-type asteroids and ordinary chondrites was confirmed (Nakamura et al. 2011). Furthermore, the isotopic oxygen abundance suggests that Itokawa itself is one of the sources of the ordinary chondrites (Yurimoto et al. 2011).

Comet mineralogy has largely been studied through the dust grains that are dragged into space by the sublimating gas. An excellent recent review has been given by Kelley and Wooden (2009). Typical grain sizes are on the order of 1 micron, which is small enough that emission features from silicates appear in the 10-micron region, a wavelength regime amenable to groundbased spectroscopy. One of the most intriguing results from the study of many comets is the existence of crystalline silicates. Such crystallinity is in contrast to interstellar silicates which may start out crystalline but are amorphized on million-year time scales (e.g. Kemper et al. 2004). Thus crystalline silicates within the comets must have been heated to crystallize before being incorporated into the comet. We discuss this further in section 6.3 below.

The silicates themselves are most commonly olivines and pyroxenes, as evinced by the shape of the 10-micron silicate feature. The shape also can be used to derive the degree of crystallinity, and also the relative amounts of Mg and Fe. Many comets seem to have Mg-rich olivines and pyroxenes, or at least roughly equal amounts of Mg and Fe. This is not universal however, and, among the comets in which it has been studied, there appears to be diversity in the Mg-to-Fe silicate abundance. (Pyroxenes with some fraction of calcium (e.g. diopside, hedenbergite) are not commonly seen in comets although there are some claimed detections, including in the Deep Impact ejecta.) This high average Mg-to-Fe ratio is interesting since in asteroids and meteorites it is more common to see the Fe-rich silicates. This dichotomy is presumably related to the compositional gradients in the protoplanetary disk. The cometary silicate composition is also somewhat dependent on dynamical class, perhaps again revealing something about formation regions of these objects.

The existence of hydrated minerals in comets, as claimed in the Deep Impact ejecta, would have significant implications for cometary formation, since presumably the hydration could only come from liquid water, i.e. water heated out of the solid phase. Yet we know that comets must have retained most of their water ice and indeed even the more volatile and highly-abundant species like CO and $CO_2$.

Our understanding of the mineralogy of comets was helped greatly by the success of the Stardust mission, which brought dust grains from comet P/Wild 2 back to Earth. A summary of Stardust's major results in this regard has been provided by Bradley et al. (2009), and implications of the results have been reviewed by Wooden (2008). One major result from the enabled laboratory studies is that the high degree of crystallinity in the silicates – mentioned above as a result from telescopic observations – was confirmed. This was a direct laboratory measurement and so a nice corroboration of the telescopic data. Such materials must have been heated and then brought out past the snow line, potentially many AU away, in order to be incorporated into the icy comets.

The result is even more significant since it appears that the fraction within the Stardust sample of so-called "presolar" grains – grains that condensed in the interstellar medium or around other stars before finding their way into our own protoplanetary disk – is lower than what was expected based on what is seen in primitive meteorites (Stadermann et al. 2008). At face value this implies that the rocky component of comets is quite heavily derived from material that was in the inner Solar System and flowing outward. However there is some evidence that presolar grains were preferentially destroyed by the collection process (Floss et al. 2013), which may complicate the interpretation.

Yet another interesting aspect to the Stardust sample is the low abundance of chondrules and phyllosilicates (Zolensky et al. 2006, Nakamura et al. 2008). With the disk evidently cycling material outward to the comet-formation region fairly efficiently, one might expect that collisions among young asteroidal bodies in ther inner Solar System could deliver chondrules and aqueously-altered materials to the region where comets could incorporate them. Chondrules and phyllosilicates are after all common, judging by the meteorite record and by asteroid spectroscopy. However the P/Wild 2 grains contain few of either, suggesting that the parent bodies to today's cometary nuclei are "pre-accretionary," i.e. formed before the epoch of chondrule formation (Wooden 2008). While the details of the thermodynamics, transport, and chemistry in the protoplanetary disk are important for getting the right answer, the result can place constraints on disk evolution models.

## 6. Big Questions and The Future

In this last section we summarize some of the significant problems with regard to small bodies that we think are the most pressing. We also provide some discussion of how such problems could be addressed. The study of small bodies has been recognized by the Planetary Science Decadal Survey (Committee on the Planetary Science Decadal Survey 2011) as a way to address some of the questions of the highest overarching importance in the field. What the small bodies lack in mass relative to other parts of the Solar System, they make up for in importance to understanding Solar System history and evolution.

There are several general points that are useful to keep in mind regarding observations of small bodies. (1) Discovery far outpaces characterization. There are over 600,000 known asteroids, almost 500 Jupiter-family comets, and a few thousand known Halley-type and long-period comets. It is far easier to discover such objects than to study them and discover their compositional, structural, and physical properties. (2) Many small bodies have no or only very weak spectroscopic features that can be used for compositional diagnostics. To put it another way, the jump from reflectance and thermal emission properties and taxonomy to composition and mineralogy is often not straightforward, nor unique. (3) Observations rarely sample an object deeper than a few microns; many datasets obtained on small bodies really are only addressing the topmost microscopic layer. This is especially unfortunate because it is that layer that is the most processed. Seeing deeper down into an object is very useful but often very difficult to do unambiguously. (4) While we know that the small body populations are a diverse group, we have only begun to understand the full variety of comet and asteroid properties. In other words, we know we are ignorant, but we are not yet quite sure just how ignorant! (5) The meteorite collection is indeed vast, but a significant barrier to advancing our understanding of small bodies is the relative lack of samples with known context. At the moment we only have samples from Itokawa and from P/Wild 2.

### 6.1. Where is the water?

As discussed above, we know that water makes up a significant fraction of the mass of a comet, but its contribution to the masses of the asteroids is much more uncertain. An important clue is the fact that CI and CM meteorites can be up to 15% water by weight (Salisbury et al. 1991). The water content of asteroids is a vital question since it is important to know what kind of objects actually provided Earth with its water. Part of this investigation requires studying how much of the water is bound up in hydration of minerals. This is a fundamental chemistry problem and needs to be addressed if we are to understand how much water there actually was available in the protoplanetary disk. Recent discoveries of water, or at least OH, in somewhat non-intuitive locations such as the poles of Mercury and the surface of the Moon suggest that it is crucial to know the details of the thermal environment in order to understand where water can survive.

Observationally, one approach that could vastly improve the state of the art is to have a large spectroscopic census of the 3-micron region in many asteroids. This wavelength region contains a broad absorption feature due to overlapping vibrational overtones of water and OH. The spectrum of each object should have high enough signal-to-noise to be able to use the shape of the absorption feature to distinguish between water itself and water that is bound in a hydrated mineral. Such a census would let us find surface water abundance as a function of taxonomic type and heliocentric distance, and thus give us some connection between water and the birth location of the object. To some extent, this project is already underway, but spectroscopy at these wavelengths is not trivial, requiring dry air with low precipitable water above the telescope. Preliminary results suggest that the distribution of hydrated asteroids is not related to asteroid size or orbital location in a simple way (e.g. Rivkin et al. 2002).

6.2. Where are the organics?

As with water, the key to understanding the organics is in spectroscopic studies of the 3.2 to 3.6-micron region. C-H bonds in organic species have stretching modes at those wavelengths. High signal-to-noise is needed to be able to differentiate between different sets of organics within an asteroid. Currently one usually just sees a broad feature that can be difficult to assign to specific species, since the phase of the material is solid. Organic materials in meteorites have been studied extensively, but the relationship to specific asteroids or locations is lacking (e.g. Pizzarello et al. 2006 and references therein).

Fortunately, comets can provide more insight about organic abundances, for the icy bodies at least, since the cometary activity naturally dredges up interior volatiles, making them visible in the coma as discussed earlier in this chapter. The difficulty is determining which volatiles are native to the cometary nucleus itself and which are products of chemistry that can occur within the collision zone of the coma, within a few radii of the nucleus's surface.

There has been a tremendous increase in the study of cometary organics since the late 1990s, owing to improvements in infrared spectrometers and to access to the large telescopes that are required. Thanks to these observations, a clearer picture is emerging of how various species -- $C_2H_2$, $CH_3OH$, $C_2H_6$, $H_2CO$, etc. -- are distributed among the comet population (Mumma and Charnley 2011). Continued observation of comets -- from all dynamical classes -- is certainly warranted. The number statistics are still fairly low, so a census of the organics in even 10-20 more comets, both short- and long-period, would be very helpful to discern trends. The difficulty is that there are only a few comets per year that are bright enough for this kind of observation. This suggests that JWST, representing potentially a tremendous jump forward in the number of faint targets that are accessible, could bring considerable data to the problem.

A sample-return mission to a comet could in principle provide us with cometary species that could be studied in exquisite detail in the laboratory. This would be a boon for our understanding of planetesimal accretion and cometary evolution. Such a mission would be technically complex, however, since one problem to be solved is preserving organics and volatiles at low temperatures during the journey back to Earth so that no chemistry occurs along the way. The Stardust mission unfortunately was not able to bring back many organics since most of the organics sublimated away as the dust grains hit the collector aerogel at high relative speed. One interesting result however was the discovery of glycine in some of the Stardust samples (Elsila et al. 2009). We can look forward to the day when more amino acids and other complex species will be measured in cometary materials.

6.3. What was the thermal and compositional gradient in the protoplanetary disk? How much mixing was there?

An apparent compositional gradient in the current Solar System has been identified for several decades (Gradie et al. 1989, Carvano et al. 2010), as was described above in Section 2. Recent dynamical modeling – in particular models like the Nice model and the Grand Tack model that have the giant planets migrating significantly and disrupting the orbital distribution of planetesimals – has suggested that the interpretation of this apparent, current gradient is not easy. The goal, after all, is to find the *primordial* compositional gradient. Two main problems in understanding what the primordial gradients were like remain: (a) there is uncertainty in what dynamical mixing has happened, and (b) we do not have enough compositional data to begin with. Detailed compositional studies of a wide range of small bodies will be necessary if we are to address the situation, and naturally the first step is understanding composition as a function of object type and object's location in the Solar System right now. We can then hope to make use of advances in our understanding of Solar System dynamics to constrain the initial conditions.

Although it had been suspected earlier, strong evidence for tremendous mixing in the protoplanetary disk comes from the fact that Stardust brought back silicate dust grains from comet P/Wild 2 with a degree of crystallinity that could only have been formed in high temperatures (Zolensky et al. 2006). This result was discussed in section 5.3, above. In fact, more than half of the Fe-bearing silicates show evidence of having been processed in the inner Solar System (Ogliore et al. 2009).

There are several models which could explain the transport of small heated grains to the outer Solar System. The Shu X-wind model (Shu et al. 1994, Shu et al. 2000) has been a leading hypothesis since it can explain astrophysical phenomena regarding disks and jets around young stellar objects. The model uses the magnetohydrodynamics of the star-disk interaction to fling material outward after it has had a chance to be heated close to the protostar. The magnetosphere of the accreting star transfers angular momentum to the disk, which induces an outward wind. While much of the wind blows along field lines out of the disk altogether, some grains that were heated in the stellar accretion region will be blown back into the disk at distances appropriate for cometary accretion. However, there are several significant problems with this scenario as the source of crystalline grains, including the possible lack of time to produce the inferred initial $^{26}Al/^{27}Al$ ratio (Desch et al. 2010, Boss 2012).

Other possible explanations come from one and two dimensional hydrodynamic models of turbulent, viscous disks. These models explain the outward transport of grains through diffusion and/or mid-plane gas flows in the first 1 - 2 Myr of disk formation. These grains would have to be incorporated into icy bodies very early on, before they fell back to smaller radii (e.g Hughes and Armitage 2010). Additionally, in full 3D models of "Marginally Graviationally Unstable" (MGU) disks, the disk appears to evolve through one or more phases of gravitational instability, and self-gravitational torques can transport material both inward and outward (e.g. Boss 2007). Each of these approaches have implications for mixing and heterogeneity in the disk at varying timescales.

In any case, one of the significant clues to the early protoplanetary disk lies in the solids with depleted initial $^{26}$Al abundance. In meteorites, both chondrules and a fraction of the calcium-aluminum-rich inclusions show evidence for zero initial abundance, indicating they formed before or several million years after the event(s) that infused the disk with $^{26}$Al. According to Boss (2012), better Pb-Pb isochron ages for the depleted inclusions, as well as tighter constraints on the initial heterogeneity of short-lived isotopes $^{60}$Fe and $^{26}$Al, will help us understand the initial conditions. Furthermore, measurements of the isotopic ratio of oxygen in molecules observed in protoplanetary disks will help us understand mechanisms for fractionation and spatial heterogeneity of oxygen isotopic abundances since these isotopes display evidence for mass-independent fractionation in our Solar System (Clayton 1993, Sakamoto et al. 2007, Boss 2012).

While meteorites continue to be a source of samples, sample return missions represent a promising way to make further progress, since parent bodies can be chosen for their known composition and dynamical properties.

6.4. What did small bodies do to Earth?

The dominant, current hypothesis is that Earth and the terrestrial planets likely formed too close to the young Sun in too warm a region to allow hydration at formation (Morbidelli et al. 2000). In simplistic terms this was thought to indicate that there was a "frost line" or "snow line" in the protoplanetary disk that marked the boundary of where water could condense out of the gas phase. Recent work has shown this is indeed an oversimplification and that condensation in the real protoplanetary disk was more complicated, as well as time-dependent (e.g. Encrenaz 2008, Bell 2010, Morbidelli et al. 2012, van Dishoeck et al. 2014). While it has long been assumed that that small bodies from the outer Solar System brought water and organics to what would have otherwise been a dry Earth (e.g. Matsui and Abe 1986), the lunar D/H ratio and Earth's mantle composition both suggest that it was *possible* for there to have been enough water inside the snow line for Earth to have formed wet, and thus that very little water was brought from impactors (e.g., Marty 2012, Saal et al. 2013). On the other hand the composition and isotopic ratios in the mantle have also been used to argue in favor of contributions by both asteroids and comets (e.g. Halliday 2013). Greater knowledge of isotopic ratios of small bodies and of the history of fractionation in the Solar System will be a key part of understanding the origin of Earth's water.

As discussed in section 5.2, until 2010, the water D/H ratio in six Oort-Cloud comets showed that the D/H ratio was much higher than observed in Earth's oceans (i.e. in standard mean ocean water, SMOW). This made having comets as a source of Earth's water problematic. Modeling suggested that if Jupiter were in a nearly circular orbit then water could be delivered to the terrestrial planets by chondritic material from beyond the frost line (Raymond et al. 2004). The discovery that the JFC P/Hartley 2 has a water D/H ratio similar to SMOW (Hartogh et al. 2011), revives the possibility that comets could have delivered a significant fraction of Earth's water. Furthermore, two other recent results on the D/H ratio – (a) the D/H measurement of less than 1.5 times SMOW in the Oort cloud comet C/2009 P1 (Garradd) by Bockelée-Morvan et al. (2012), and (b) the finding of an upper-limit to the D/H ratio in comet P/Honda-Mrkos-Pajdušáková that is consistent with that of P/Hartley 2 (Lis et al. 2013) – demonstrate a possibile continuum in cometary D/H. Clearly the D/H ratio in comets is a more complicated issue than was previously thought. In any case, given the success of the asteroid models at delivering water, the answer may be that Earth's volatiles were delivered by some combination of asteroids and comets – unless it turns out that the wet Earth hypothesis is correct. It is possible that we will need to develop and fly the technology to sample small body noble gas isotopes before we are reasonably certain of the origin of Earth's water.

It is clear from both this question and several of the others posed in this section that understanding the dynamical evolution of the small bodies is critical to being able to reconstruct the protoplanetary disk. The current dominant model of large scale planetary migration in our Solar System, the Nice model mentioned in earlier sections, has done an admirable job of explaining many observed phenomena, but the difficult part is using it to make a prediction that is both (a) observationally feasible, and (b) would make a strong test of the idea's validity. Compositional information on the small bodies is one possible way -- since for example the model would predict that one would see a similar primordial composition among the outer main-belt asteroids, Trojans, and trans-neptunian objects. The difficulty is that it is still hard to interpret the surface composition due to a lack of diagnostic features. What is unknown is whether the surfaces would really be the same because of similar primordial condition, or simply because of similar subsequent evolution. Sample return missions may be required before we know that answer.

Non-gravitational forces are an important aspect to the dynamics of many objects, as it is now clear that over fairly short dynamical time scales they can move asteroids and comets significantly around the Solar System. In particular, for asteroids, the Yarkovsky and YORP effects need to be better understood – and this means understanding the thermal properties as well as the structural properties in the interior. This is important since the Yarkovsky effect affects the delivery of asteroids from the Asteroid Belt to near-Earth space, including those that could someday strike Earth. For comets, outgassing torques and forces – on the whole much stronger than Yarkovsky effects – can measurably change a comet's orbital elements on month- or year-long time scales, thereby making it more difficult to gauge just how significant the cometary impact risk is for Earth.

6.5. Forthcoming breakthroughs

The next decade should bring tremendous insight into our understanding of comets and asteroids. First, the Rosetta spacecraft will rendezvous with comet P/Churyumov-Gerasimenko in 2014 and observe it close-up for several months. Rosetta is primarily a European Space Agency mission but NASA has contributed some of its instruments. For the first time, we will be able to watch a comet turn on as its activity gets started. The spacecraft also has a lander, Philae, that will be able to study the surface directly, our first such opportunity to do so on a comet. The spacecraft will arrive at comet P/Churyumov-Gerasimenko in May 2014 -- when the comet is 4.0 AU from the Sun, outside the zone where water could sublimate -- and will stay with the comet at least until December 2015. The comet's perihelion occurs in August 2015, when it will be 1.24 AU from the Sun.

Asteroid sample return should figure prominently in the next several years, with both the U.S., Europe, and Japan all potentially sending sample-return missions to primitive asteroids. The Japanese mission, Hayabusa 2, is a follow-up to its successful Hayabusa mission that brought back samples from asteroid Itokawa in 2010. Hayabusa 2 will travel to (162173) 1999 JU3, which is a primitive asteroid. Launch would happen in 2014 or 2015, with asteroid arrival in 2018. The spacecraft would collect a sample and leave the asteroid in 2019, returning to Earth in 2020. The U.S. mission, OSIRIS-REx, which stands for "Origins Spectral Interpretation Resource Identification Security Regolith Explorer," will retrieve samples from asteroid Bennu. This object is likewise a primitive asteroid. Launch will happen in 2016, with arrival to the asteroid in 2019. The spacecraft would study the asteroid for many months, departing likely in 2021 and getting back to Earth in 2023. Even if only one of these missions is successful, we can expect to have tremendous new insights into our understanding of primitive bodies in the 2020s.

6.6. Longer-term missions

In the farther future significant scientific return can be obtained from robotic missions that visit Trojan asteroids, primitive main-belt asteroids, and dormant/extinct comets. In the U.S., NASA has plans for human visitation of asteroids in the relatively near future, mid-2020s. However it is unclear exactly what path toward that goal will be taken. Nonetheless, there are significant engineering obstacles to a successful manned mission to an asteroid that need to be overcome. Some of these obstacles – sometimes known as "strategic knowledge gaps" – can be answered through telescopic observation and robotic visitation of the asteroids in question. Some of these questions include: could astronauts actually operate in the nearly-zero gravity environment around an asteroid? How would the spacecraft rendezvous? What are the surface and immediate subsurface layers like? Can one grab onto the surface or is it too powdery? Could one anchor to it? What thermal variation would there be as the asteroid rotates in and out of sunlight? Is the asteroid in a principal-axis rotation state, or is it tumbling? How far down would one have to dig to get to more pristine material, if at all?

While these are difficult problems to solve, one cannot help but be excited about what secrets of the small bodies and of our Solar System will be revealed in the coming years. The synergy of Earth-based telescopic observation (to provide the big picture about many objects) and spacecraft missions (to provide the detailed picture of a few key objects) is sure to continue to provide astronomers and geophysicists with new, thrilling insights.

**Acknowledgements**
We greatly appreciate the assistance of Michael Nolan and Patrick Taylor in preparing this manuscript. We also thank Peter Thomas and an anonymous reviewer for their careful reading and suggestions. Some of our figures made use of data archived by the Small Bodies Node of the NASA Planetary Data System. Y.R.F. and E.S.H. acknowledge support from NSF (through Grant No. AST-1109855 to The Johns Hopkins University). L.M.W. acknowledges support from NASA (through Grant No. NNX09AB71G issued through the Planetary Atmospheres program). The Arecibo Planetary Radar Program is supported by the National Aeronautics and Space Administration under Grant No. NNX12AF24G issued through the